\documentclass[aps,prb,twocolumn,superscriptaddress,showpacs,floatfix]{revtex4-2}

\usepackage{graphicx}
\usepackage{dcolumn}
\usepackage{bm}
\usepackage{color}
\usepackage{lineno}
\usepackage{gensymb}
\usepackage{amsmath}
\usepackage{amsfonts}
\usepackage{color}
\usepackage{amssymb}%
\usepackage{blindtext}
\usepackage{xspace}
\usepackage{siunitx} 
\usepackage{miller}

\usepackage[version=3]{mhchem}
\usepackage{natbib}
\usepackage[]{hyperref}

\providecommand{\U}[1]{\protect\rule{.1in}{.1in}}

\newcommand{\tbmno}{TbMnO$_3$}
\newcommand{\lamno}{LaMnO$_3$}

\newcommand{\DM}{Dzyaloshinskii-Moriya}
\newcommand{\cryopad}{\textit{CRYOPAD}}
\newcommand{\spinw}{\textit{SpinW}}

\newcommand{\matlab}{\textit{MATLAB}}
\newcommand{\aaxis}{$\vec{a}$}
\newcommand{\baxis}{$\vec{b}$}
\newcommand{\caxis}{$\vec{c}$}
\newcommand{\abplane}{$ab$}

\newcommand{\bcplane}{$bc$}
\newcommand{\qvec}{$\vec{Q}$}
\newcommand{\etal}{\textit{et al.}}

\newcommand{\REmno}{$RE$MnO$_3$}
\newcommand{\dymno}{DyMnO$_3$}

\newcommand{\Hparab}{$\vec{H}\parallel\vec{b}$}

\newcommand{\mubohr}{\,$\mu_B$}

\begin{document}
\title{Spin-wave dispersion and magnon chirality in multiferroic TbMnO$_3$}

\author{S. Holbein}
\affiliation{$I\hspace{-.1em}I$. Physikalisches Institut, Universit\"at zu K\"oln, Z\"ulpicher Str. 77, D-50937 K\"oln, Germany}
\affiliation{Institut Laue-Langevin, 71 avenue des Martyrs, F-38042 Grenoble CEDEX 9, France}

\author{P. Steffens}%
\affiliation{Institut Laue-Langevin, 71 avenue des Martyrs, F-38042 Grenoble CEDEX 9, France}

\author{S. Biesenkamp}
\affiliation{$I\hspace{-.1em}I$. Physikalisches Institut, Universit\"at zu K\"oln, Z\"ulpicher Str. 77, D-50937 K\"oln, Germany}

\author{J. Ollivier}%
\affiliation{Institut Laue-Langevin, 71 avenue des Martyrs, F-38042 Grenoble CEDEX 9, France}

\author{A. C. Komarek}%
\affiliation{$I\hspace{-.1em}I$. Physikalisches Institut, Universit\"at zu K\"oln, Z\"ulpicher Str. 77, D-50937 K\"oln, Germany}
\affiliation{Max Planck Institute for Chemical Physics of Solids, N\"othnitzer Stra\ss e 40, D-01187 Dresden, Germany}

\author{M. Baum}
\affiliation{$I\hspace{-.1em}I$. Physikalisches Institut, Universit\"at zu K\"oln, Z\"ulpicher Str. 77, D-50937 K\"oln, Germany}

\author{M. Braden}
\email{[e-mail: ]braden@ph2.uni-koeln.de}
\affiliation{$I\hspace{-.1em}I$. Physikalisches Institut, Universit\"at zu K\"oln, Z\"ulpicher Str. 77, D-50937 K\"oln, Germany}
\date{\today}

\begin{abstract}

Inelastic neutron scattering experiments combining time-of-flight and polarized techniques yield
a comprehensive picture of the magnon dispersion in multiferroic TbMnO$_3$ including the dynamic chirality.
Taking into account only Mn$^{3+}$ moments, spin-wave calculations
including nearest-neighbor interactions, frustrating next-nearest
neighbor  exchange as well as single-ion anisotropy and antisymmetric terms
describe the energy dispersion and the distribution of neutron scattering intensity in
the multiferroic state very well. 
Polarized neutron scattering reveals strong dynamic chirality  of
both signs that may be controlled by external electric fields in the multiferroic phase.
Also above the onset of long-range multiferroic order in zero electric field, a small inelastic
chiral component can be inverted by an electric field.
The microscopic spin-wave calculations fully explain also the dynamic chirality
of magnetic excitations, which is imprinted by the static chirality of the multiferroic phase. 
The ordering of Tb$^{3+}$ moments at lower temperature reduces the broadening of magnons
but also renders the magnon dispersion more complex.

\end{abstract}

\maketitle

\date{\today}




\section{introduction}

The observation of ferroelectricity in \tbmno\ ~\cite{Kimura2003} initiated the discovery of a new group of multiferroic materials, in which a complex magnetic structure directly induces macroscopic electric polarization, so-called type-II multiferroics \cite{Spaldin2019,Fiebig2016}. \tbmno\  is a reference system for this class of materials due to its sizable ferroelectric polarization and its large magnetoelectric coupling~\cite{Goto2004} and is thus well suited for elucidating the magnetic excitations and the multiferroic coupling mechanism.

\begin{figure}[!t]
	\centering
\includegraphics[width=0.95\columnwidth]{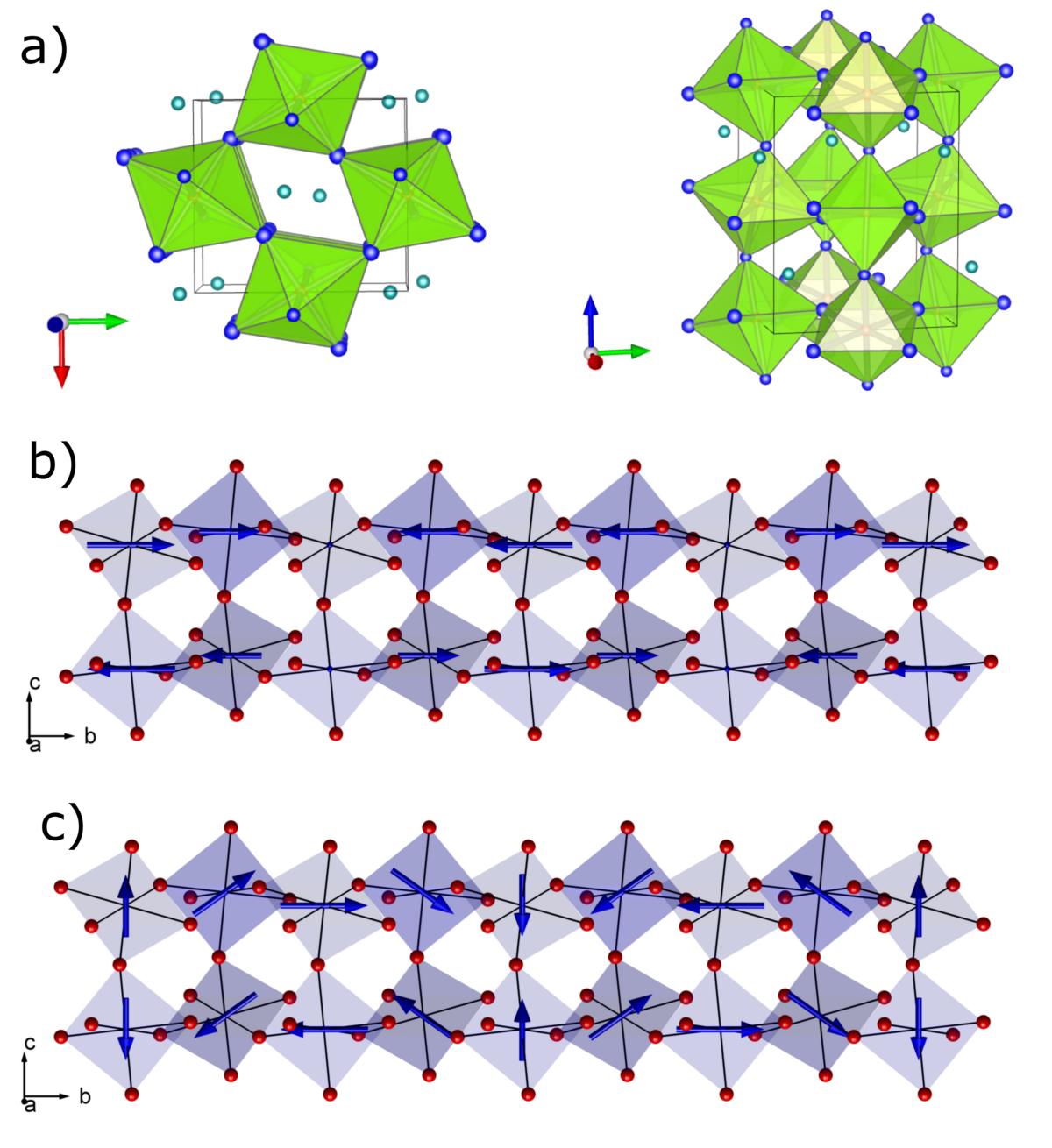}
\caption{ (a) Crystal structure of TbMnO$_3$ with structural parameters obtained by single-crystal X-ray diffraction \cite{struc-baum}. The red, green and blue
arrows at the lower left corners indicate the $a$, $b$, and $c$ directions, respectively.
(b-c) Magnetic structure of manganese moments in \tbmno. The zig-zag chains of MnO$_6$ octahedra are shown along the direction of the propagation vector, with darker oxygen octahedra in the back and lighter in the front. The propagation vector was chosen to be $\vec{k}=(0,0.25,0)$ for a better visibility. The Mn moments order below $T_{N}$ in a spin-density-wave with moments modulated parallel \baxis \ (b), and below $T_{\textrm{MF}}$ in an elliptical spiral in the $bc$-plane (c). }
        	\label{fig:tbmno_mag}
\end{figure}

{\it Magnetic order in} $RE$MnO$_3$ -- \tbmno\ belongs to the series of rare-earth ($RE$) manganates~\cite{Jensen1991}, \REmno , in which the mismatch of ionic radii induces rotations of the MnO$_6$, see Fig.~\ref{fig:tbmno_mag}.
The Mn-O-Mn bond in \tbmno \ strongly deviates from a straight
connection with an angle of only \SI{145}{\degree}~\cite{Kimura2003a,Blasco2000}.
Mn$^{3+}$ has four electrons in the $3d$ orbitals in a $t^3_{2g}e_g^1$ state ~\cite{Rodriguez-Carvajal1998}.
The Jahn-Teller effect in \REmno\ leads to a staggered ordering of single-occupied $d_{3x^2-r^2}$ and $d_{3y^2-r^2}$ orbitals~\cite{Rodriguez-Carvajal1998}, which yields a strong ferromagnetic (FM) nearest-neighbor exchange in the $ab$-plane, $J_{\textrm{FM}}$~\cite{Kimura2003a}. The coupling between the planes along the $c$-axis is antiferromagnetic (AFM)~\cite{Kimura2003a} in agreement with the $A$-type (i.e. AFM stacking of ferromagnetic perovskite layers) AFM order in \lamno ~\cite{Kajimoto2005,Goodenough1955}.
The structural distortion for smaller $RE$ ions weakens $J_{\textrm{FM}}$ and leads to an enhancement of the AFM next-nearest neighbor interaction $J_{\textrm{NN}}$ in the $ab$-plane and thus to frustration \cite{Mochizuki2009a,Fontcuberta2015}.
This $J_{\textrm{NN}}$ exchange is much stronger along \baxis , because the occupied $e_g$ orbitals are rotating towards
the $b$ direction, see Fig.~\ref{fig:tbmno_mag}.
The magnetic structure changes from $A$ type in moderately distorted \lamno \ to $E$-type ordering (i.e. an up-up-down-down stacking along orthorhombic $b$ direction) with $\vec{k}_E=(0,0.5,0)$~\cite{Kimura2003a,Ishiwata2010,Mochizuki2009a,Fontcuberta2015} in orthorhombic HoMnO$_3$, which is also multiferroic.
In the region of intermediate $RE$ ion size, an incommensurate structure with propagation vector $\vec{k}$=(0,$k_{\textrm{inc}}$,0)  and $0<k_{\textrm{inc}}<1/2$ emerges~\cite{Kimura2003a,Mochizuki2009a,Fontcuberta2015}.

{\it Magnetic structure in} TbMnO$_3$ -- In \tbmno, the Mn moments order at $T_N=\,$\SI{42}{\kelvin} in an $A$-type longitudinal spin-density wave (SDW)
with a propagation vector of $\vec{k}_{\textrm{inc}}$=(0, $\sim$0.28, 0)~\cite{Quezel1977}. For this propagation vector and considering only Mn moments, there are four irreducible representations, which just correspond to those at the commensurate $\Gamma$ point \cite{Kajimoto2004}.
These representations can be thus labeled by  the four modes $A$, $C$, $F$ and $G$ known for the commensurate structures~\cite{Bertaut1968,Kajimoto2004,Aliouane2008}. 
The ordering scheme $A$ is explained above, $F$ denotes FM order,
$G$ the AFM order with all nearest neighbors being antiparallel, and $C$ the AFM order in the $a,b$ layer with FM alignment in $c$ direction.

In the SDW phase, the value of the incommensurability amounts to $k_{\textrm{inc}}$$\approx$0.28 and the structure can be described by a single irreducible representation $\Gamma_3=(G_x,A_y,F_z)$~\cite{Aliouane2008}. Neutron diffraction data in this phase could be described using only the dominant $A_y$ mode in the form $\vec{M}^{\textrm{SDW}}=(0,M_b\cos(\vec{k}_{\textrm{inc}}\cdot\vec{r}),0)$ with $M_b=2.9$\mubohr~\cite{Kenzelmann2005}. A model of the magnetic structure is shown in Fig.~\ref{fig:tbmno_mag} (b). Upon further cooling, the incommensurability decreases slightly until about \SI{31}{\kelvin} where a quasi-lock-in at a value of $q_K\approx0.276$ sets in~\cite{Stein2017}.  Macroscopic and diffraction studies show that the lock-in transition is separated from the onset of cycloidal and
multiferroic order~\cite{Meier2007,Stein2017} at $T_{\textrm{MF}}=\SI{27.6}{\kelvin}$.
Magnetic order in the multiferroic phase corresponds to an elliptic cycloid and is described by a coupling of two irreducible representations $\Gamma_2\times\Gamma_3$, with $\Gamma_2=(C_x,F_y,A_z)$~\cite{Kenzelmann2005,Aliouane2008}. Kenzelmann \etal\ described their data by an elliptical cycloid $\vec{M}^{\textrm{MF}}=(0,M_b\cos(\vec{k}_{\textrm{inc}}\cdot\vec{r}),M_c\sin(\vec{k}_{\textrm{inc}}\cdot\vec{r}+\delta))$ with $M_b=3.9$\mubohr\ and $M_c=2.8$\mubohr ~\cite{Kenzelmann2005} that is illustrated in Fig.~\ref{fig:tbmno_mag} (c).
A smaller Mn-spin component emerges along the $a$ axis, which corresponds to a $G_x$-mode~\cite{Aliouane2008}.
The Tb moments order separately below $T_{\textrm{Tb}}=\SI{7}{\kelvin}$ and form an incommensurate SDW with moments mainly oriented parallel to \aaxis\ and a propagation vector of $\vec{k}_{\textrm{Tb}}=(0,0.415,0)$~\cite{Kimura2005,Quezel1977,Meier2007,Goto2004}, but magnetic Bragg peaks are reported to be significantly broadened indicating short-range order.

\begin{figure}[!t]
    \centering
    \includegraphics[width=0.9\columnwidth]{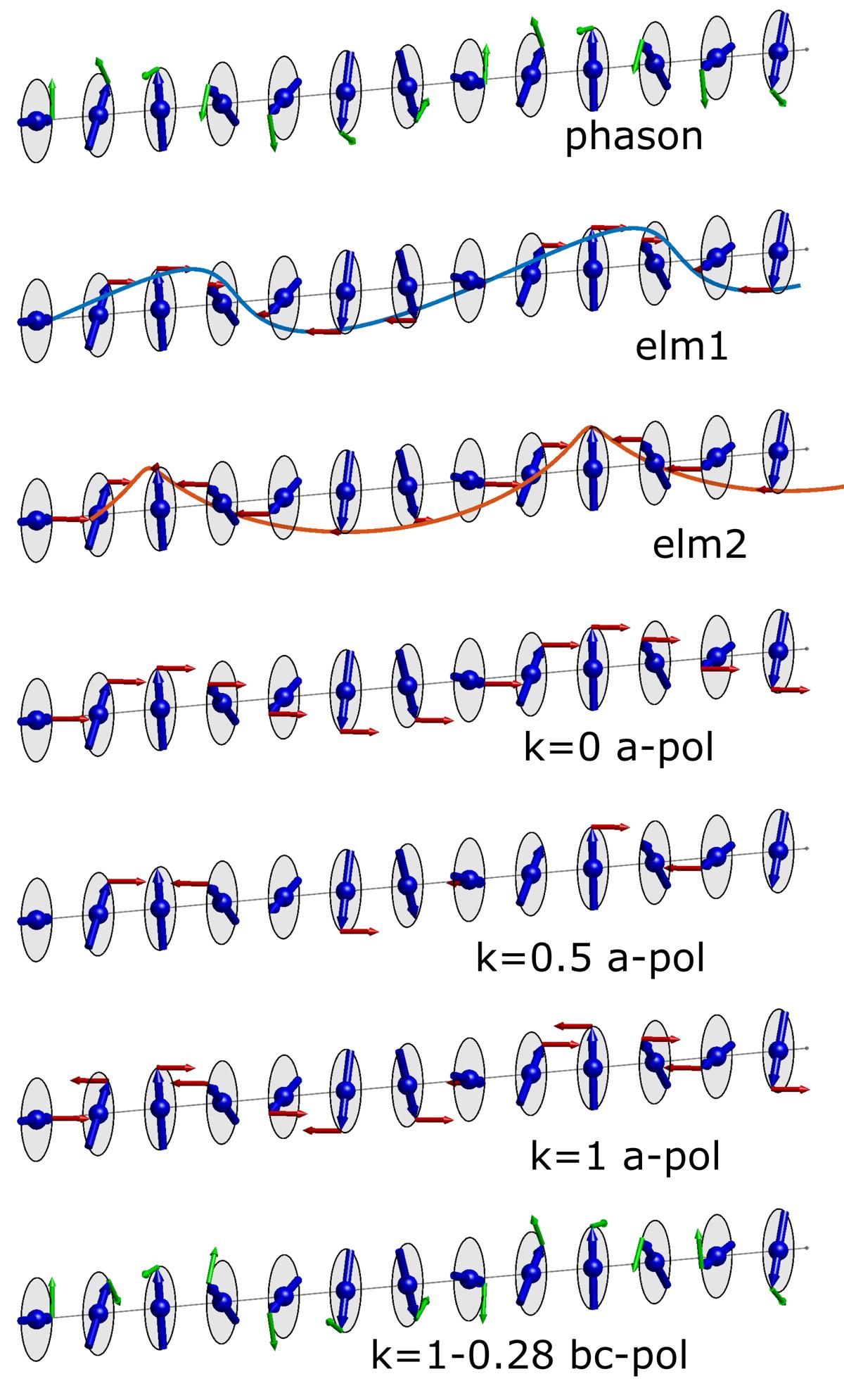}
    \caption{Sketch of the polarization patterns of magnetic excitations at the zone center of the cycloid structure and at characteristic propagation vectors.
    The static cycloid in the incommensurate multiferroic phase is shown for a chain along \baxis ~ by blue arrows (moments) and grey circles (plane in which the
    moments rotate).
    From top to bottom one sees first the zone-center phason and the two electromagnon excitations corresponding to the rotation of the entire cycloid around \baxis \ (elm1) and to the rotation of the cycloid plane around \caxis \ (elm2), see reference \cite{Senff2007}. The phason oscillations are polarized parallel to the cycloid plane (always indicated by green arrows), while for the two electromagnons the oscillation is parallel \aaxis , which is indicated by red arrows. The next pictures present three
    \aaxis -polarized modes for the  wave vector (0,$k$,0) with $k$=0, 0.5, and 1. The lowest pattern corresponds to the in-plane mode at $k$=1-0.28, which
    is also a hybridized mode and even causes the strongest electromagnon signal in IR spectroscopy \cite{ValdesAguilar2009,Finger2014}.
     }
    \label{fig:emagnons}
\end{figure}

Although the Tb moments order at a lower temperature than the Mn subsystem, the two magnetic ions couple. Already at \SI{15}{\kelvin}, the Tb moments contribute to the spiral ordering of Mn moments and had to be taken into account for the structure refinement of the Mn order~\cite{Kenzelmann2005}. At this temperature, the Tb moments align primarily along \aaxis~\cite{Kajimoto2004,Aliouane2009,Voigt2007,Prokhnenko2007}. It was proposed that the Mn and Tb order remain coupled below $T_{\textrm{Tb}}$ through their wave vectors, $3k_{\textrm{Tb}}-k_{\textrm{Mn}}=1$~\cite{Prokhnenko2007}.

{\it Multiferroic order in} TbMnO$_3$ -- The ferroelectricity in TbMnO$_3$ induced through the magnetic order was microscopically explained by the mechanism of the spin current~\cite{Katsura2005}, the inverse \DM\ (DM) interaction~\cite{Sergienko2006} and in a phenomenological approach~\cite{Mostovoy2006} leading to the relation:

\begin{equation}
\vec{P}\propto\vec{r}_{ij}\times(\vec{S}_i\times\vec{S}_j),
\end{equation}

with the neighboring spins $\vec{S}_i$ and $\vec{S}_j$, and their distance vector $\vec{r}_{i,j}$. In the \REmno\ series, the compounds with $RE=$Tb, Dy, Gd and Eu$_{1-x}$Y$_x$ develop such a spin-induced ferroelectric polarization~\cite{Kimura2003,Goto2004,Noda2006,Arima2006,Kuwahara2009,Fontcuberta2015}.
In \tbmno,  electric polarization appears parallel to the orthorhombic \caxis\ axis, $P_c$, because spins rotate in the $bc$ plane and the propagation is along \baxis . The ferroelectric order in TbMnO$_3$ is a secondary effect limiting the size of the polarization to $P_c=\SI{0.08}{\micro\coulomb\cm\tothe{-2}}$ at \SI{10}{\kelvin} in comparison to $P = \SI{26}{\micro\coulomb\cm\tothe{-2}}$ in BaTiO$_3$ \SI{300}{\kelvin}~\cite{Merz1953}. The microscopic model of the multiferroic coupling for \tbmno\ has been corroborated
by the application of magnetic and electric fields~\cite{Kimura2005,Yamasaki2007,Aliouane2009}.

{\it Magnetic excitations  in} TbMnO$_3$ -- The magnon dispersion in \tbmno \ was analyzed using inelastic neutron scattering (INS) by Senff \etal~\cite{Senff2007,Senff2008a}. Three low-energy modes were found at the magnetic zone center: an in-plane mode (with respect to the cycloidal $bc$ plane)  at $\omega_1\approx\SI{0.1}{\milli\eV}$ and two out-of-plane modes at $\omega_2\approx\SI{1.1}{\milli\eV}$ (elm1) and $\omega_3\approx\SI{2.5}{\milli\eV}$ (elm2), respectively, at \SI{17}{\kelvin}. The exchange interactions were estimated by fitting a Heisenberg model to the different magnon branches, but without reflecting the incommensurate character of the order~\cite{Senff2007,Senff2008a}. Milstein and Sushkov described the magnon dispersion in \tbmno\ and \dymno\ using the $\sigma$-model-like effective-field theory~\cite{Milstein2015}. We will discuss this model in Section~V. Magnetic fields along \aaxis ~ or \baxis ~ induce a magnetic transition to a commensurate cycloid with moments in the $ab$ plane (HF-C phase) \cite{Kimura2004,Aliouane2009}. The magnetic excitations in this HF-C phase were studied by INS experiments for both field directions \cite{Senff2008,Holbein2015}.

Multiferroics exhibit hybridized excitations of phonon and magnon character, that are called electromagnons and were predicted in 1982~\cite{Smolenskii1982}. Electromagnons were reported in \tbmno\ or GdMnO$_3$ by infrared (IR) spectroscopy~\cite{Pimenov2006} and by INS~\cite{Senff2007}.
Two such electromagnon modes are present at low frequencies and match the energy of the out-of-plane modes elm1 and elm2 found in INS~\cite{Pimenov2006,Pimenov2009,Senff2007}.
This and the matching temperature dependence of IR and INS mode frequencies~\cite{Shuvaev2011,Holbein2015} strongly support the electromagnon interpretation given in ~\cite{Katsura2007}. Concerning the HF-C phase, only fields \Hparab\ were accessible in IR experiments due to instrument limitations and Shuvaev \textit{et al.} reported the existence of a weak $c$-polarized mode in the HF-C phase above the $H_b$ transition~\cite{Shuvaev2010}, which perfectly agrees with the electromagnon activated by the DM mechanism and INS
measurements \cite{Holbein2015}.
However, the various IR and optical measurements performed on the \REmno\ series yield an additional and even stronger electromagnon signal at a larger energy of about 8\,meV~\cite{ValdesAguilar2009,Mochizuki2011b,Rovillain2011,Finger2014}. This strongest IR signal appears always along the \aaxis\ direction~\cite{Pimenov2009}, even for materials in the HF-C phase associated with an \abplane\ cycloid. A magnetostrictive coupling, in which the Heisenberg interaction is modulated by a variation of the Mn-O-Mn bond angles, perfectly explains the strong $a$-polarized electromagnon~\cite{ValdesAguilar2009}. The corresponding magnon polarization is illustrated in the lowest panel of Fig.~\ref{fig:emagnons} following reference \onlinecite{Finger2014}.

The sign of the ferroelectric polarization follows the chirality of the magnetic structure; therefore an electric field can be used to
pole the magnetic domains and to induce and invert a monodomain system with respect to the chirality \cite{Yamasaki2007,Finger2010,Poole2009,Hearmon2012,Finger2010,Stein2017}.
Neutron scattering directly detects the chiral component by comparing different channels in polarization analysis \cite{Brown2006} and is thus an ideal tool to study chiral domains \cite{Stein2021}, but it can
also be applied to the dynamic chiralities of the magnon modes that were little studied so far. 

{\it New INS experiments and analyses in this work} --
So far the magnon dispersion in \tbmno \ was studied by INS experiments on a triple-axis spectrometer (TAS),
in which a single point in scattering vector ($\vec{Q}$) and energy (E) space is analyzed.
The magnon dispersion of an incommensurate cycloid is however rather complex and exhibits more than just one branch.
Therefore it is important to obtain a full mapping of the intensity distribution in $\vec{Q}$,E  space.
Such a complete picture of the excitations can be obtained with the time-of-flight (TOF) INS
technique, which was the main aim of this new analysis on IN5 at the Institut Laue Langevin (ILL).
We compare these new measurements to comprehensive linear spin-wave theory calculations.
In addition we apply polarization analysis to INS experiments in order to examine
the polarization and the dynamic chirality of selected magnetic modes. This documents the complementarity
of the  INS techniques that is required to obtain the full picture of magnetic correlations in a
multiferroic material.

After presenting the experimental details in Section II, we first discuss the basic microscopic magnetic models
to analyze the magnon dispersion in TbMnO$_3$ in Section III. The new experimental results were obtained  with TOF and neutron-polarization techniques and are presented in Section IV. Sections V and VI discuss possible extensions of the magnetic model and the impact of the Tb-moment ordering, respectively. Finally we conclude
this work.

\section{Experimental}

Single crystals of \tbmno \ were grown by the traveling floating-zone method in an image furnace \cite{Reutler2003,diss_komarek} and characterized by single-crystal X-ray \cite{struc-baum} and neutron diffraction experiments as well as by specific heat and magnetic measurements. Crystals cut from the same batch or from similar growths were already used in references \onlinecite{Stein2017,Stein2021} in which growth and characterization are described. For the INS measurements aiming at the dispersion we studied large
crystals to obtain sufficient intensities in reasonable beam time.

In order to extend the previous INS studies on the magnon dispersion, which focused on single points in $\vec{Q}$,E space, we used  the disk chopper TOF spectrometer IN5 at the ILL \cite{data-4-01-1230}.
This technique measures large parts of $\vec{Q}$,E space, but it does not allow for neutron polarization analysis.
We mounted a large cylindrical single crystal of 25\,mm length and $\sim$6\,mm diameter in a cryostat in $[0,1,0]/[0,0,1]$ scattering geometry.
The instrument is set by construction in direct geometry: a monochromatic neutron pulse arrives at the sample and the position and flight time of the scattered neutrons are measured in a  detector bank covering a large solid angle. The vertical angular range of the detector bank is \SI{\pm20.55}{\degree}, which limits the accessible $\vec{Q}$ range vertically to the scattering plane, i.e. the $a$ direction.
The accessible energy and $Q$ range as well as the resolution can be modulated by the choice of the incident neutron wave length. The frequency of neutron pulses is set so that the highest energy transfer is \SI{70}{\percent} of the incident energy. The experiment was performed at \SI{17}{\kelvin} in order to be sufficiently below the multiferroic transition and to avoid dominant contributions arising from the ordering of Tb moments. We used three different incident neutron wave lengths, $\lambda_{i,1}=\SI{2.0}{\angstrom}$ ($E_{i,1}=\SI{20.5}{\meV}$), $\lambda_{i,2}=\SI{3.75}{\angstrom}$ ($E_{i,2}=\SI{5.8}{\meV}$) and $\lambda_{i,3}=\SI{5.2}{\angstrom}$ ($E_{i,3}=\SI{3.0}{\meV}$). The corresponding energy resolutions determined at the elastic line were $\Delta E_1\approx\SI{0.84}{\meV}$, $\Delta E_2\approx\SI{0.22}{\meV}$ and $\Delta E_3\approx\SI{0.08}{\meV}$. The resolutions are typical for this instrument (incoming energy resolution $\Delta E_i/E_i\approx$\SIrange{1.7}{3.0}{ per cent})~\cite{Ollivier2011}. Intensities were corrected for detector efficiency~\cite{Richard1996}.

Another large single crystal of \tbmno\ was mounted in $[201]/[010]$ scattering geometry on the cold neutron TAS IN14/Thales at the ILL using the \cryopad\ option for neutron polarization analysis. Throughout the experiment, we worked with a fixed final neutron energy of $E_f=\SI{4.98}{\milli\eV}$ ($k_f=\SI{1.55}{\per\angstrom}$) and a Be-filter on $k_f$. The neutrons were polarized by a supermirror bender and analyzed by the Bragg reflection of a Heusler crystal and the CRYOPAD device was used for spherical polarization analysis. The flipping ratio (FR) on a magnetic Bragg peak was FR$\approx$36, which corresponds to a polarization of the neutron beam of approximately \SI{95}{\percent}. The resolution at the fixed final energy was determined on the elastic line, $\Delta E\approx\SI{0.2}{\meV}$. The crystal was placed between two thin aluminum plates in order to apply an electric field along the crystallographic $c$-direction, the direction of the electric polarization in the multiferroic phase. The distance of about \SI{22}{\mm} between the two aluminum plates therefore requires a high voltage to create the electric field needed to pole the large crystal. We were able to apply a voltage of \SI{7.8}{\kilo\volt} corresponding to an electric field of \SI{355}{\volt\per\mm}. In a second run of polarized experiments on IN14/Thales we used the same crystal mounting and instrument configuration besides a set of Helmholtz coils to enable longitudinal polarization analysis \cite{data-4-03-1730}.
In this part of the experiment voltages of plus \SI{10}{\kilo\volt} and minus \SI{5}{\kilo\volt} could be applied to the aluminium plates.
On the thermal neutron TAS IN20 we used the same crystal mounting in a polarized configuration with the CRYOPAD device (Heusler monochromator and analyzer crystals) and $k_f=\SI{1.55}{\per\angstrom}$.
The flipping ratio on a magnetic Bragg peak and on the (0, 2, 0) reflection amounted to  FR$\approx$13.2 and 13.0, respectively.

To study the impact of the ordering of Tb moments on the magnon dispersion the same large crystal used in the IN5 experiment was installed on the cold-neutron TAS IN14/Thales at the ILL in $[010]/[001]$ scattering geometry in an unpolarized configuration. The experiment was performed with a fixed final neutron energy of $E_f=\SI{4.64}{\milli\eV}$ ($k_f=\SI{1.50}{\per\angstrom}$) and a velocity selector on $k_i$. Pyrolytic graphite monochromator and analyzer crystals  were horizontally and vertically focusing. The energy resolution of $\Delta E\approx\SI{0.19}{\meV}$ was determined on the elastic line.

Data obtained at IN5 and at IN14/Thales are available at references \onlinecite{data-4-01-1230} and \onlinecite{data-4-03-1730}, respectively.

\begin{figure} 
    \includegraphics[width=0.9\columnwidth]{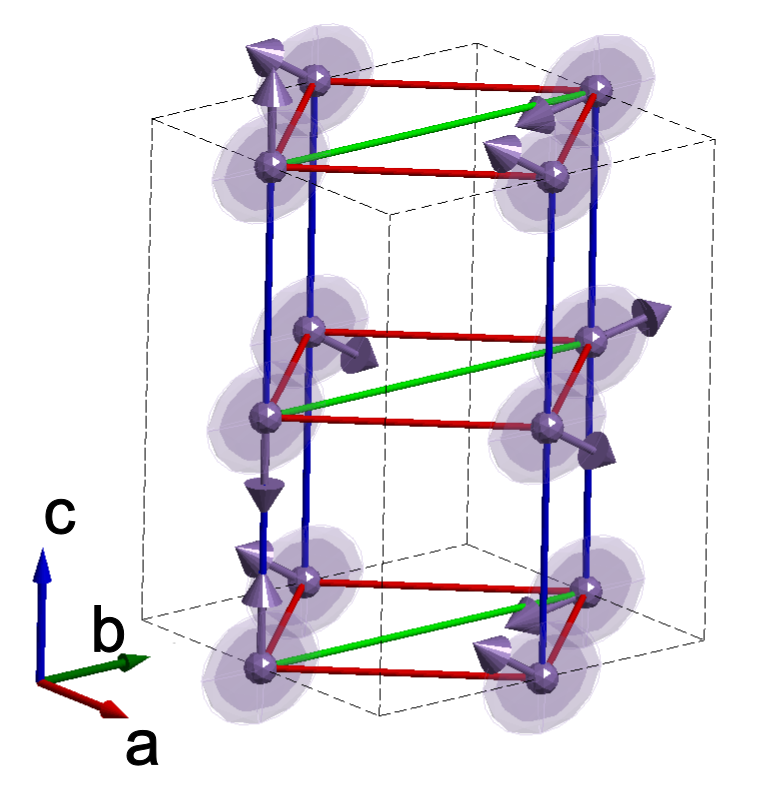}
    \caption{Model of the crystallographic unit cell of \tbmno, showing Mn moments in the spin spiral phase with exchange interactions [(FM $J_{FM}$ (red) and AFM $J_{\textrm{AFM}}$ (blue) and $J_{\textrm{NN}}$ (green)] and elliptic easy-plane anisotropy along \baxis\ and \caxis\ (ellipses).}
    \label{fig:tbmno-spinW_exchanges}
\end{figure}

\begin{figure*}[!t]
    \centering
     \begin{minipage}[b]{1.3\columnwidth}
        \centering
    \includegraphics[width=\columnwidth]{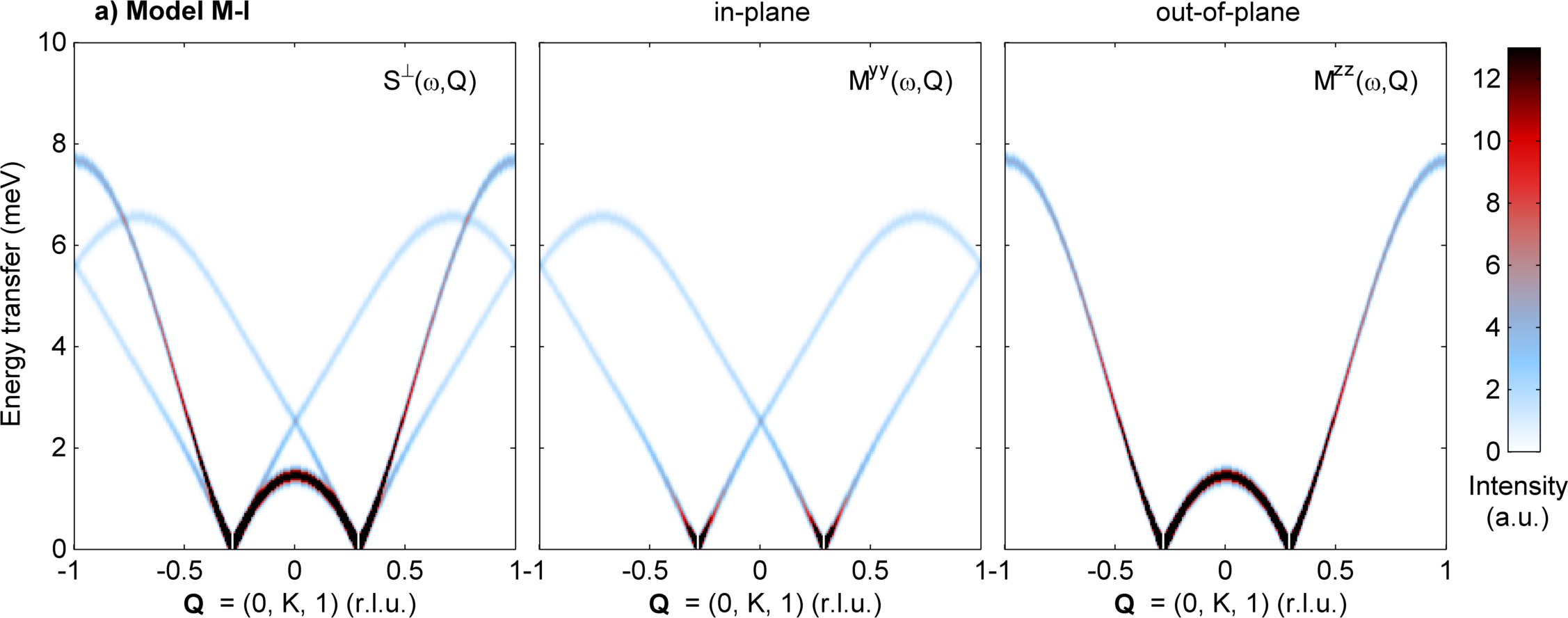}
     \end{minipage}
     \vskip3mm
     \begin{minipage}[b]{1.3\columnwidth}
        \centering
    \includegraphics[width=\columnwidth]{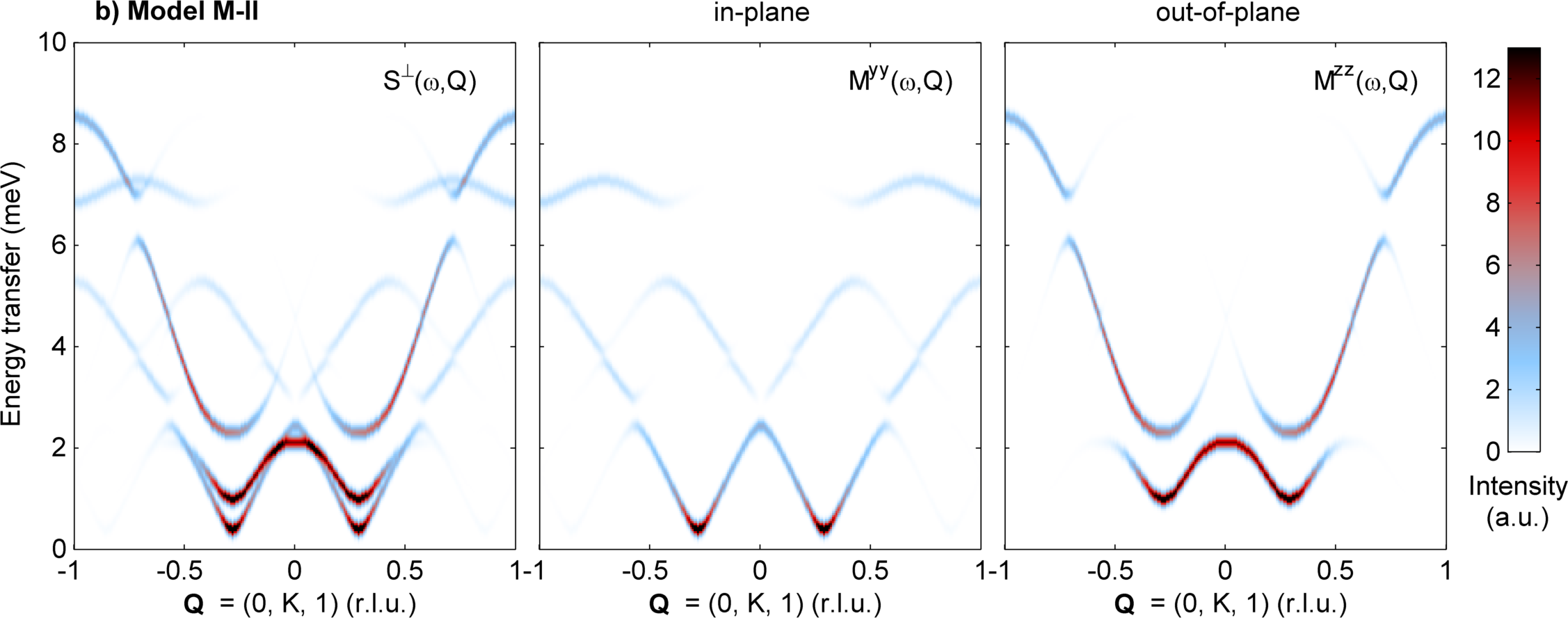}
     \end{minipage}
	\caption{Calculated neutron intensity for \tbmno \ at $\vec{Q}=(0,~k,~1)$ of the total cross section $S^{\perp}(\omega,Q)$ (left) and the components, $M^{yy}(\omega,Q)$ (center) and $M^{zz}(\omega,Q)$ (right). The scattering geometry was chosen in a way that $y$ is in the \bcplane\ plane and $z$ is parallel to \aaxis. Different magnetic models were used for the calculation: (a) Model M-I: spiral magnetic structure with moments in the \bcplane\ plane and $\vec{k}=(0,~2/7,~0)$ and (b) model M-II: elliptic spiral with moments in the \bcplane\ plane, $\vec{k}=(0,~2/7,~0)$ and an elliptic easy-plane anisotropy.}
    \label{fig:tbmno-spinW_easyplane}
\end{figure*}

\section{Models of magnetic interaction}

The magnon dispersion of \tbmno\ in the multiferroic phase was initially described~\cite{Senff2007,Senff2008a} basing on a model for the commensurate order in \lamno\ by Moussa \etal\ and Hirota \etal~\cite{Moussa1996,Hirota1996}. The spin-wave relations were derived for a Hamiltonian including $J_{\textrm{FM}}$ and   $J_{\textrm{NN}}$ within the \abplane\ plane, $J_{\textrm{AFM}}$ along \caxis\ and a single-ion anisotropy and describe reasonably well the dispersion along \aaxis\ and \caxis~\cite{Senff2008a}. However, the incommensurate magnetic structure was not properly taken into account.

In a first attempt to model the magnon dispersion, a linear spin-wave theory calculation was performed with Holstein-Primakoff transformation as described by S\'aenz~\cite{Saenz1962,Ulbrich2012}. This formalism is restricted to collinear magnetic structures with a commensurate propagation vector. The \bcplane\ spiral of \tbmno\ had to be approximated by a sinusoidal modulated spin-density wave in a commensurate lattice, $\vec{k}=(0,~2/7,~0)$ . However, the results determined with this model could not reproduce the experimentally observed dispersion, because the ground state of the magnetic Hamiltonian strongly differs from this collinear magnetic structure.
For describing the excitations with the cycloidal structure we use the program \spinw~\cite{progspinW}. The program bases on linear spin-wave theory and was developed to account for canted and incommensurate spin structures by Toth and Lake~\cite{Toth2015}.

We consider magnetic Mn moments of spin $S=4/2$ and neglect Tb moments, which limits the applicability of the model to temperatures significantly above the onset of Tb-moment order. The simple model of the magnetic moments and interactions is illustrated in Fig.~\ref{fig:tbmno-spinW_exchanges}. It includes the AFM exchange along \caxis\ (blue), FM exchange $J_{\textrm{FM}}$ (red) in the \abplane\ plane and the frustrating AFM nearest-neighbor interaction $J_{\textrm{NN}}$ (green) along \baxis. The single-ion anisotropy \textit{SIA} is indicated as a gray ellipsoid.

\begin{table}
	\begin{ruledtabular}
		\begin{tabular}{c c c c c}
			Model 							& $J_{\textrm{FM}}$   & $J_{\textrm{AFM}}$  & \textit{SIA}  & $DM$ \\
			\hline
			Milstein \etal     				& -0.3     				& 0.9     				& -0.125   			& -0.2 \\ 
			Model M-I   					& -0.34    				& 0.82    				& 0	& \\
			Model M-II   					& -0.38    				& 0.82    				& -(0,0.12,0.09) 	& \\
			Model M-III   					& -0.38    				& 0.82    				& -(0,0.18,0.09) 	&  \\
			Model M-IV   					& -0.38    				& 0.82    				& -0.1    			& (0.64,-0.2,0) \\ 
		\end{tabular}
	\end{ruledtabular}
	\caption{Comparison of exchange interaction parameters (in meV per bond) and anisotropy energies (in meV) in the multiferroic phase of \tbmno\ by Milstein and Sushkov~\cite{Milstein2015} and this work (M-II, M-III, M-IV). The next-nearest neighbor exchange is fixed by the propagation vector $\vec{k}=(0,~k_{\textrm{inc}},~0)$ as $J_{\textrm{NN}}=J_{\textrm{FM}}/2\cos(k_{\textrm{inc}}\pi)$. 
		In model M-III the single-ion anisotropy refers to the local one, that is staggered following the 
		orbital arrangement in TbMnO$_3$. Model M-II is the most simple one that properly describes the
		spin-wave dispersion and its chirality. The results of this model are directly compared to the experimental data in Figures 5 to 9 and 12.
	}
	\label{tab:tbmno-spinWfit}
\end{table}

In a first step we model the excitations along $\vec{Q}=(0,~k,~1)$ for a circular spin-spiral with moments in the \bcplane-plane and a propagation vector of $\vec{k}=(0,~2/7,~0)$ (Model M-I). The value $2/7\approx0.2857$ is close to the value which has been found experimentally in \tbmno~\cite{Kenzelmann2005,Stein2017}. Furthermore, it allows the calculation of a commensurate structure with a magnetic unit cell extended seven times along \baxis\ with respect to the crystallographic cell. 
Thereby we could compare some calculations with those performed in the commensurate collinear model~\cite{Saenz1962,Ulbrich2012}. The calculations presented throughout the paper were performed
with this 7 times extended magnetic structure, but we verified that directly implementing the propagation vector of $\vec{k}=(0,~0.28,~0)$ in the Fourier description of the magnetic structure in \spinw~\cite{progspinW} yields the same results.
The following parameters were used for the calculation: $J_{\textrm{AFM}}=\SI{0.82}{\meV}$, $J_{\textrm{FM}}=\SI{-0.34}{\meV}$, $J_{\textrm{NN}}=\SI{0.24}{\meV}$ and \textit{SIA}$=\SI{0}{\meV}$, see also Table~\ref{tab:tbmno-spinWfit}. The calculated intensities are shown in Fig.~\ref{fig:tbmno-spinW_easyplane}(a). The total cross section $S^{\perp}(\omega,Q)$ (left), and the components $M^{yy}(\omega,Q)$ (center) and $M^{zz}(\omega,Q)$ (right) are given for a $[010]/[001]$ scattering geometry. This setting allows us to distinguish fluctuations of moments in the \bcplane\ plane (in-plane, $M^{yy}(\omega,Q)$) from fluctuations along \aaxis\ (out-of-plane, $M^{zz}(\omega,Q)$). Experimentally, neutron polarization analysis is required in order to
separate these different polarizations of the magnon modes \cite{Senff2007,Senff2008a}.

For a collinear spin chain with FM nearest-neighbor and AFM next-nearest-neighbor interaction ($J_1$-$J_2$ model), $J_{\textrm{FM}}$ and $J_{\textrm{NN}}$ are constrained by the incommensurate pitch (here along \baxis ): $cos(k_{\textrm{inc}}\cdot\pi)=J_{\textrm{FM}}/(2\cdot J_{\textrm{NN}})$. The magnetic zone center splits into two satellites at $\vec{Q}$=(0,$\pm$2/7,~1). Note that the $q_l$=1 component is essential to imply a signal for the $A$-type magnetic order.
In addition, we get a clear separation of an in-plane mode (in $M^{yy}$), which is the phason or sliding mode of the spiral, and an out-of-plane mode (in $M^{zz}$) with moments oscillating along \aaxis , see Fig.~\ref{fig:emagnons}. At the incommensurate zone center, both modes attain zero energy because no anisotropy was inserted in model M-I, and thus there is no preferred direction for the magnetic moments.  The dispersion of the out-of-plane mode can be understood by looking again at the $J_1$-$J_2$ model with cycloidal ground state. At the zone boundary $\vec{Q}=(0,~1,~1)$, the oscillations are out of phase and one finds the polarization with the highest cost in energy $E\sim 4J_{\textrm{FM}}+2J_{\textrm{NN}}$. The $k=0$ mode at $\vec{Q}=(0,~0,~1)$ reveals a much lower energy of $-4J_{\textrm{FM}}$+$2J_{\textrm{NN}}$.
Finally at $\vec{Q}=(0,~0.5,~1)$, one finds an intermediate state whose polarization corresponds to the $E$-type ordering scheme with an energy $-2J_{\textrm{NN}}$.
The polarization of these three out-of-plane modes is illustrated in Fig.~\ref{fig:emagnons}.
For the in-plane modes the mixing of the static and dynamic components severely complicates the interpretation. For $k$=1-$k_{\textrm{inc}}$=0.72, however a
simple picture is obtained that is illustrated in the lowest panel of Fig.~\ref{fig:emagnons}. This mode corresponds to the phason with every second oscillation being inverted
(green arrows in Fig.~\ref{fig:emagnons}). This mode modulates the scalar product of neighboring spin and is therefore coupled to a structural distortion through exchange
striction, which yields a much stronger dynamic magnetoelectric coupling \cite{ValdesAguilar2009}. This mode
exhibits the highest energy of in-plane polarized magnons.
As it can be seen in the middle panels of Fig.~\ref{fig:tbmno-spinW_easyplane}, the maximum in-plane energy appears at $k$=0.72 and belongs to the branch starting at the Bragg in the neighboring zone that is just $\Delta k$=1 away.

The accurate magnetic structure of \tbmno\ corresponds to an elliptic \bcplane\ spiral with moments $M_b$=3.9\,\mubohr\ and $M_c$=2.8\,\mubohr , $\vec{k}=(0,~2/7,~0)$, and to interaction parameters $J_{\textrm{AFM}}=\SI{0.82}{\meV}$, $J_{\textrm{FM}}=\SI{-0.38}{\meV}$, and $J_{\textrm{NN}}=\SI{0.31}{\meV}$, with single-ion anisotropy terms \textit{SIA}$_b=\SI{-0.12}{\meV}$ and \textit{SIA}$_c=\SI{-0.09}{\meV}$, shown in Fig.~\ref{fig:tbmno-spinW_easyplane}(b) (Model M-II).
The values of the ordered magnetic moment along \baxis\ and \caxis\ are taken from the neutron diffraction study~\cite{Kenzelmann2005}.
The elliptic spin spiral is stabilized by an orthorhombic single-ion anisotropy, where the ratio of the major and minor axes is roughly equal to $M_b/M_c$.
Since the ratio $J_{\textrm{FM}}$/$J_{\textrm{NN}}$ is determined by the incommensurability, only three independent parameters needed to be
determined by fitting the magnon energies at characteristic propagation vectors, as it is explained in the next section.
The errors of the independent parameters amount to 0.01, 0.03 and 0.01\,meV for $J_{\textrm{AFM}}$,  $J_{\textrm{FM}}$, and \textit{SIA}$_b$, respectively.
The calculation shows an anti-crossing of in-plane modes  at  $\vec{Q}$=(0, 2$k_{\textrm{inc}}$, 1) and at $\vec{Q}$=(0, 1-$k_{\textrm{inc}}$,1) for out-of-plane modes, with $k_{\textrm{inc}}=2/7\approx 0.28$.
The phason mode at the zone center can attain finite energies due to pinning effects or due the locking of the propagation vector.
These effects appear to be rather small.
The out-of-plane ($a$ polarized) mode splits up into two modes at the zone center which results from the orthorhombic single-ion anisotropy compared to a simple
$b,c$ easy-plane anisotropy.
Modes polarized along \aaxis\ with a static $b$ component are energetically preferred to \aaxis \ modes with a static $c$ component~\cite{note-incom}.

\section{Magnetic excitations in the incommensurate multiferroic phase}

\subsection{Refining interaction parameters with the magnon dispersion}

\begin{figure}[!t]
    \centering
    \includegraphics[width=\columnwidth]{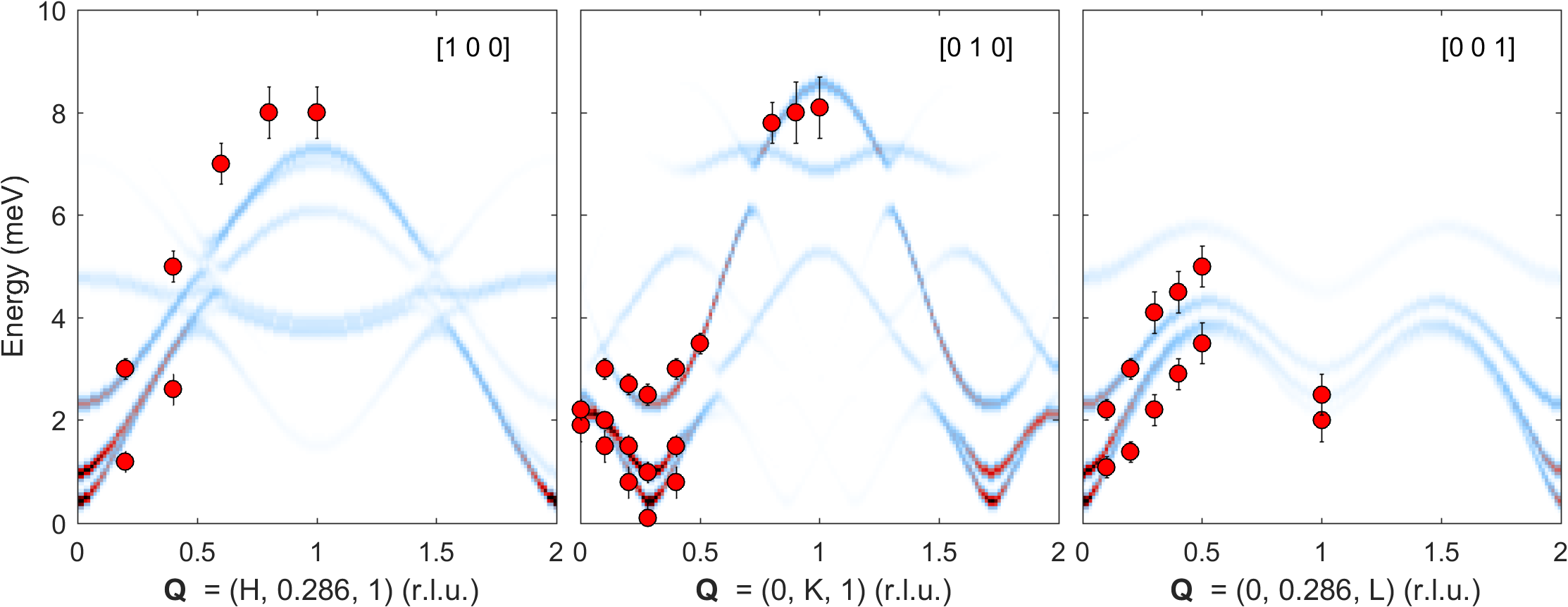}
    \caption{Calculated neutron intensity for \tbmno \ along the paths (a) $[H,0.28,1]$, (b) $[0,K,1]$ and (c) $[0,0.28,L]$. The calculation is based on an elliptic spiral with moments in the \bcplane\ plane, $\vec{k}=(0,~2/7,~0)$ and an elliptic easy-plane anisotropy. The parameters from model M-II were optimized to describe the magnon energies obtained from INS (red points) taken from Ref.~\onlinecite{Senff2008a}.}
    \label{fig:tbmno-spinWfit}
\end{figure}

We used the energies of the different modes extracted from INS data taken with a TAS \cite{Senff2007,Senff2008a} to refine the interaction parameters of our model M-II \cite{note-senff}. Figure~\ref{fig:tbmno-spinWfit} shows the calculated magnon intensity along the principal directions $[H,0.28,1]$, $[0,K,1]$ and $[0,0.28,L]$. The mode energies determined by INS at \SI{17}{\kelvin} taken from Ref.~\cite{Senff2007,Senff2008a} are indicated with red points. 
The exchange and anisotropy parameters of the model have been refined with the experimental data and are given in Table~\ref{tab:tbmno-spinWfit} (model M-II).
The points extracted from INS data are very well described for propagation vectors along the $b$ direction.
The commensurate approximation $k=2/7\approx0.286$ of the incommensurate propagation vector is sufficiently accurate. The dispersion along \aaxis\ and \caxis\ can qualitatively be described and only minor differences are visible.
At this point, we would like to recall the simplicity of the model, which only consists of three exchange parameters and two single-ion anisotropies as described in Section II.
These parameters must be consistent with the magnetic structure yielding constraints, so that only three
independent parameters need to be optimized to describe the experimental dispersion.
In the real system, several other effects may influence the spin dynamics of the Mn moments, such as the DM interactions both along \caxis\ and in the \abplane\ plane~\cite{Mochizuki2009}, the ferroelectric displacements~\cite{Malashevich2009a}, biquadratic~\cite{Mochizuki2010c} and ring exchange interaction~\cite{Fedorova2015} and the influence of the strong Tb moments via direct exchange or crystal field~\cite{Prokhnenko2007,Mostovoy2006}.
We will further discuss such effects in Section V.

The accuracy of the estimated exchange values can be verified by calculating the N\'eel temperature $T_{\textrm{N}}$ and the Weiss temperature using the mean-field equations~\cite{Tovar1999}:
\begin{align}
     T_{\textrm{N,MF}} &=(4/3)S(S+1)(2\cdot J_{\textrm{FM}} - J_{\textrm{AFM}} - J_{\textrm{NN}})/k_b\nonumber\\
      & \approx\SI{30}{\kelvin} \\
\theta_{\textrm{MF}} &=(4/3)S(S+1)(2\cdot J_{\textrm{FM}} + J_{\textrm{AFM}} + J_{\textrm{NN}})/k_b\nonumber\\
      & \approx\SI{-171}{\kelvin}
\end{align}

These values can be compared to experimental observations. The AFM ordering temperature of \tbmno\ is $T_{\textrm{N}}$$\approx\SI{42}{\kelvin}$, where the Mn moments form a spin-density-wave. Yet, the order is incomplete, as only the $b$ component orders and full order sets in below $T_{\textrm{MF}}=\SI{27.6}{\kelvin}$, when the cycloid forms. The mean-field value of the N\'eel temperature agrees well with these experimental values. The Weiss temperature was observed to be $\theta_{\textrm{N}}$=-\SI{21.9\pm0.1}{\kelvin} for a polycrystalline sample of \tbmno~but single-crystal data reveal strong differences along the orthorhombic directions:  $\theta_{a}=\SI{17.6\pm0.1}{\kelvin}$, $\theta_{b}=-\SI{9.3\pm0.5}{\kelvin}$ and $\theta_{c}=-\SI{128\pm1}{\kelvin}$ \cite{OFlynn2014}. Tb moments order below $T_{\textrm{Tb}}$ along \aaxis, and dominate already at higher temperature along this direction, which thus cannot be analyzed with our model for Mn moments. Also the $b$ direction is affected, because the Tb moments are polarized by the ordering of Mn moments in the SDW phase~\cite{Kenzelmann2005}. The $c$-direction primarily sees the Mn moments and the measured value corresponds best to the the mean-field approach, which considers only Mn ordering. The frustration of the Mn system is visible in the enlarged ratio of mean-field values of Weiss and N\'eel temperatures $f=\theta/T_{\textrm{N}}\approx-5.7$.

\subsection{Comparison with zone-center INS and IR data}

In model M-II, there are two $a$-polarized out-of-plane modes at the magnetic zone center, whose energies were fitted to the experimental
results of $\omega_{\perp1}=\SI{1.0}{\meV}$ and $\omega_{\perp2}=\SI{2.5}{\meV}$.
The splitting is due to the distorted easy-plane anisotropy: modes polarized along \aaxis\ with a static $b$ component (rotation around \baxis) are energetically preferred to \aaxis\ modes with a static $c$ component (rotation around \caxis).
These two oscillations correspond to the above-discussed elm1 and  elm2 modes.
When we apply the DM mechanism to both excitations \cite{Katsura2007} only elm1 changes the direction of the induced electric polarization and should thus be electrically active (see Fig.~\ref{fig:emagnons}).

\begin{figure}[!t]
        \includegraphics[width=1\columnwidth]{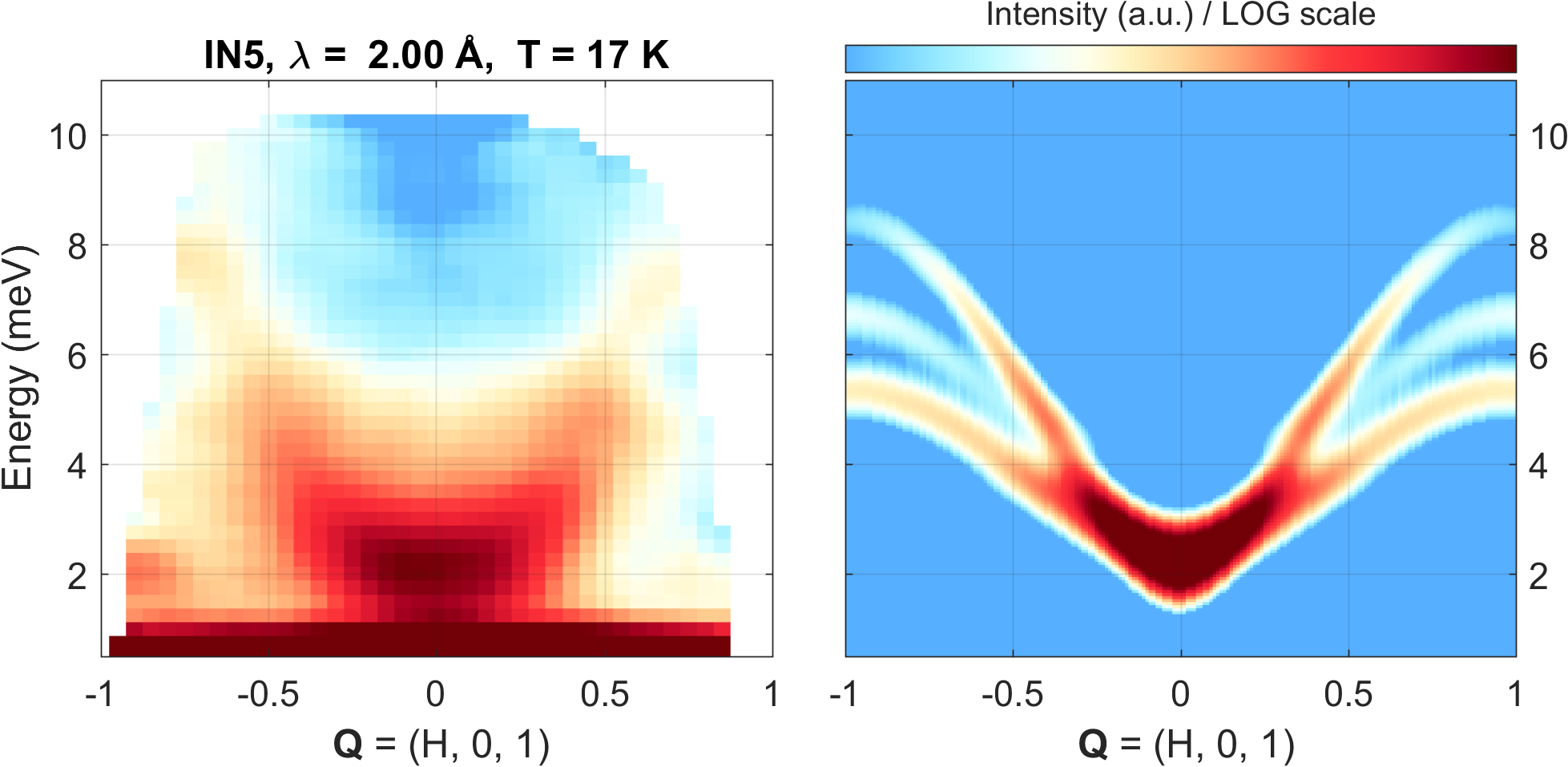}
        \includegraphics[width=1\columnwidth]{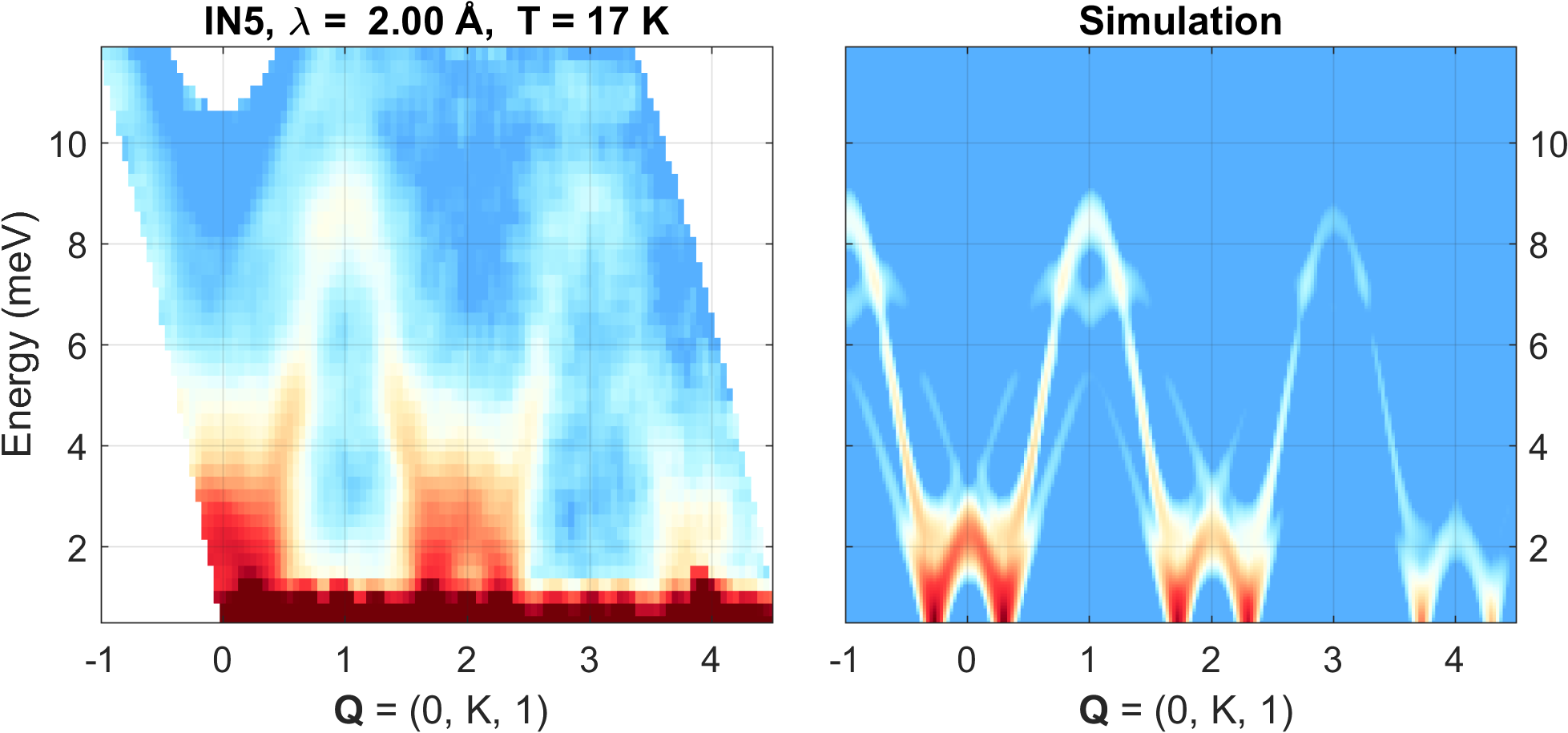}
        \includegraphics[width=1\columnwidth]{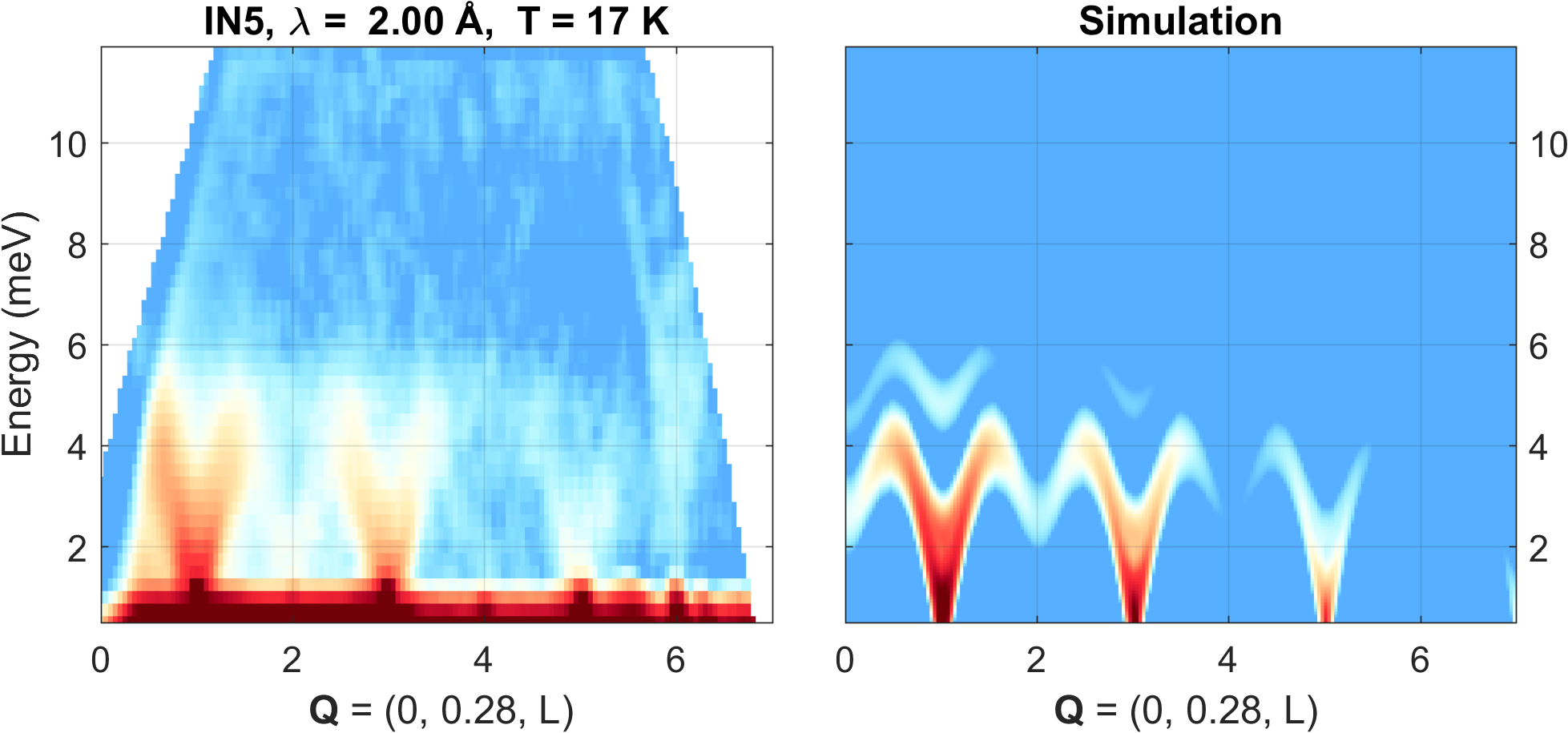}
\caption{Comparison of inelastic scattering maps of \tbmno\ obtained from neutron TOF spectroscopy at IN5 (left) at $T=\SI{17}{\kelvin}$) and simulation (right). The incident neutron wave length and the energy resolution are given in the plot headers. Model M-II (cf. Tab.~\ref{tab:tbmno-spinWfit}) was used for the simulation and the Bose factor as well as the magnetic form factor were taken into account. On top of the upper right panel the logarithmic colorbar of the intensities is shown.}
    \label{fig:tbmno-IN5_scans1}
\end{figure}

\begin{figure}[!t]
        \includegraphics[width=1\columnwidth]{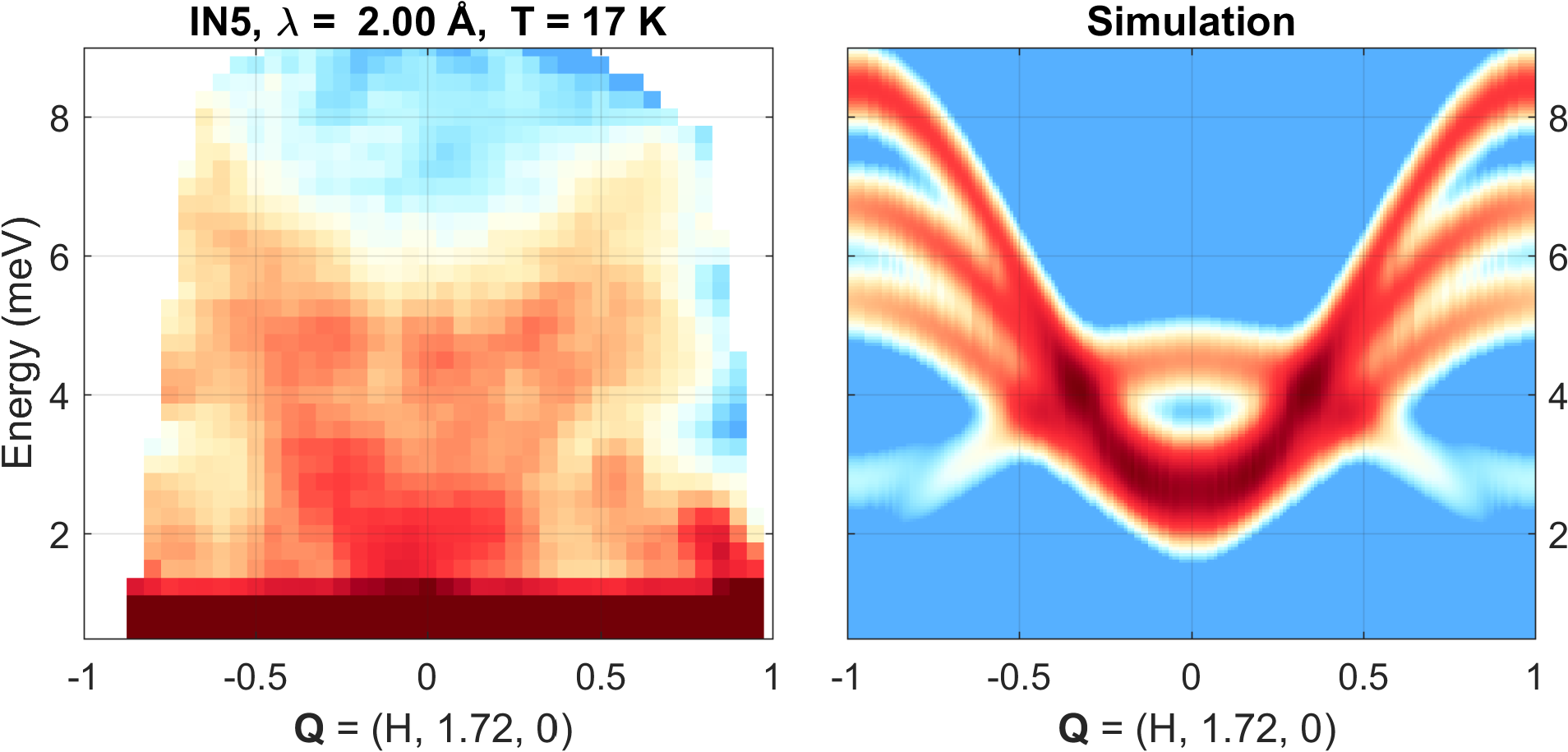}
        \includegraphics[width=1\columnwidth]{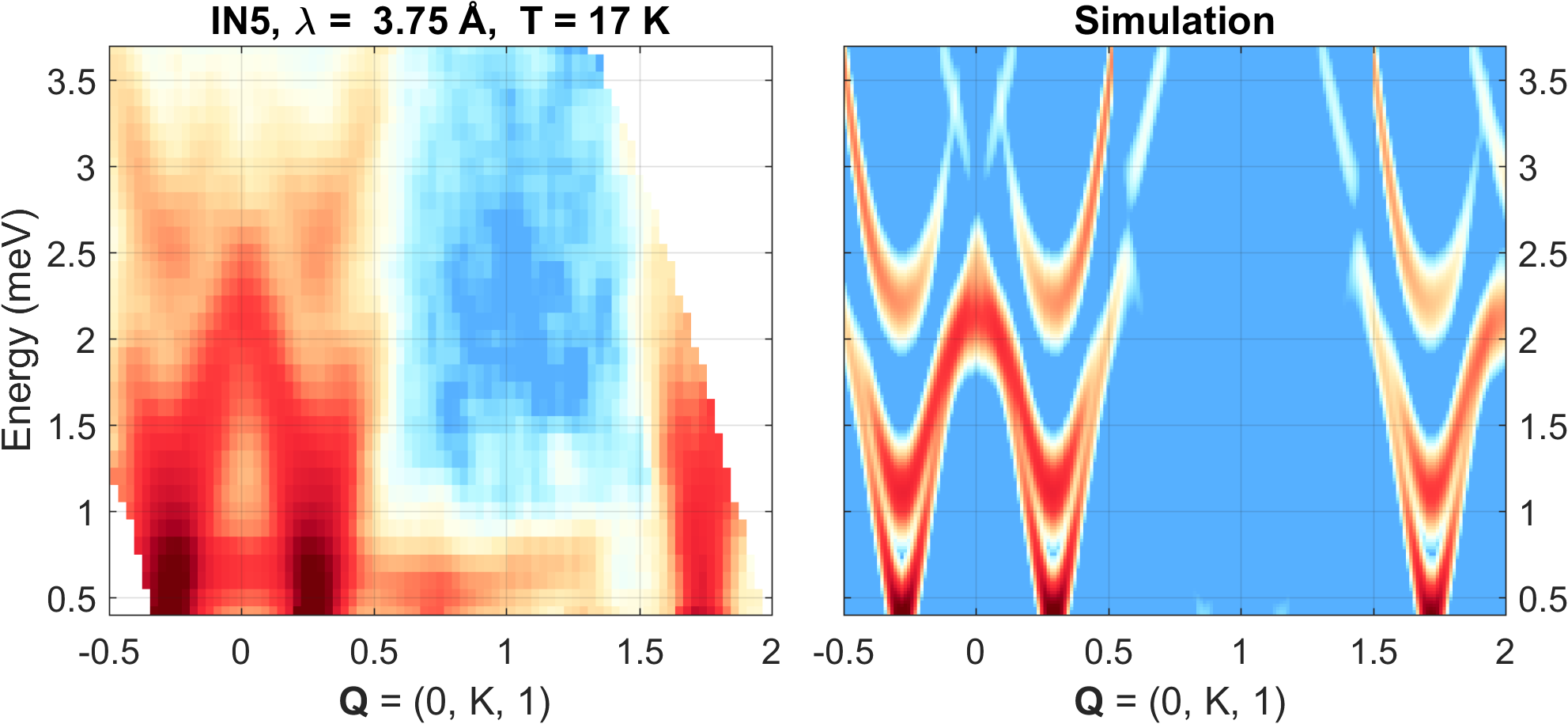}
        \includegraphics[width=1\columnwidth]{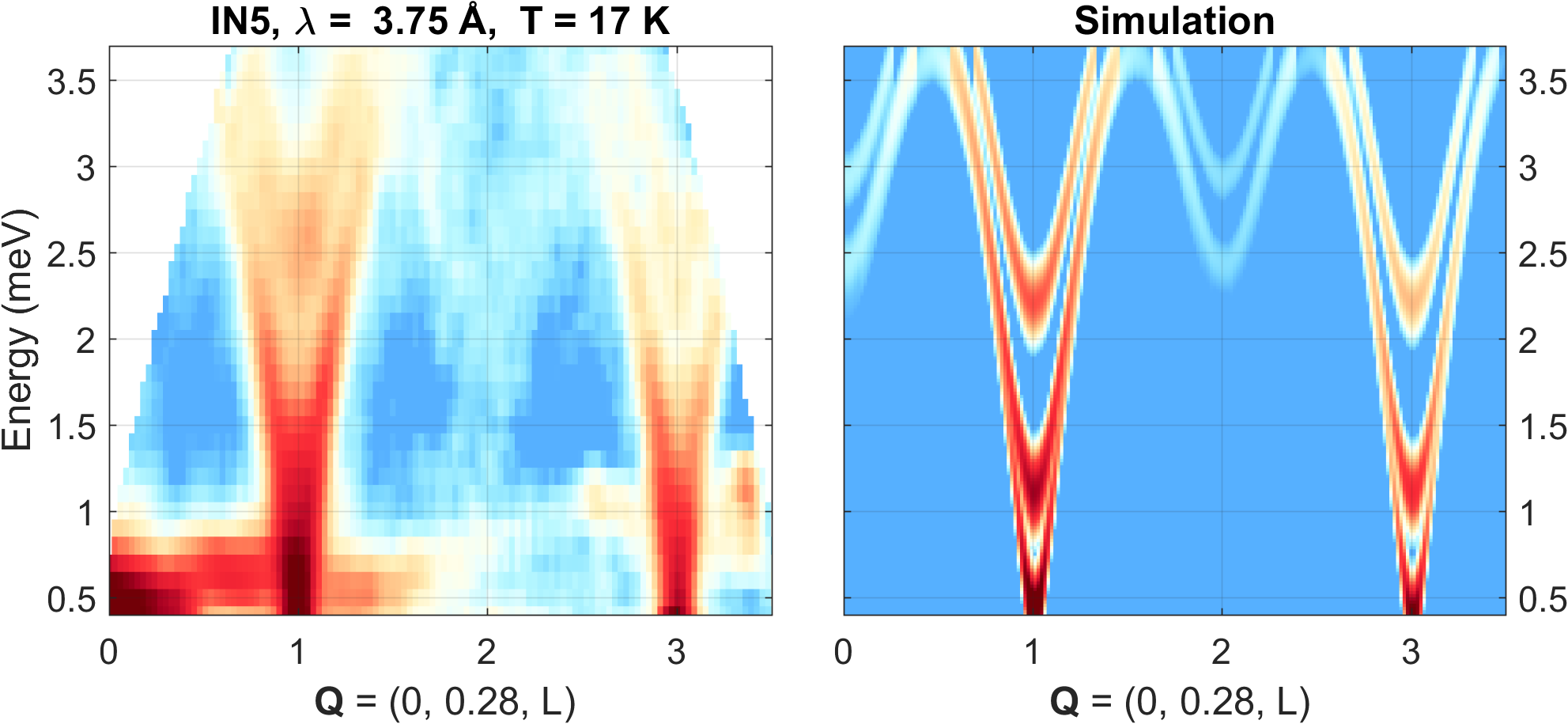}
\caption{Comparison of inelastic scattering maps of \tbmno\ obtained from neutron TOF spectroscopy at IN5 (left) at $T=\SI{17}{\kelvin}$) and simulation (right). The incident neutron wave length and the energy resolution are given in the plot headers. Model M-II (cf. Tab.~\ref{tab:tbmno-spinWfit}) was used for the simulation and the Bose factor as well as the magnetic form factor were taken into account.}
    \label{fig:tbmno-IN5_scans2}
\end{figure}

\begin{figure}[!t]
        \includegraphics[width=1\columnwidth]{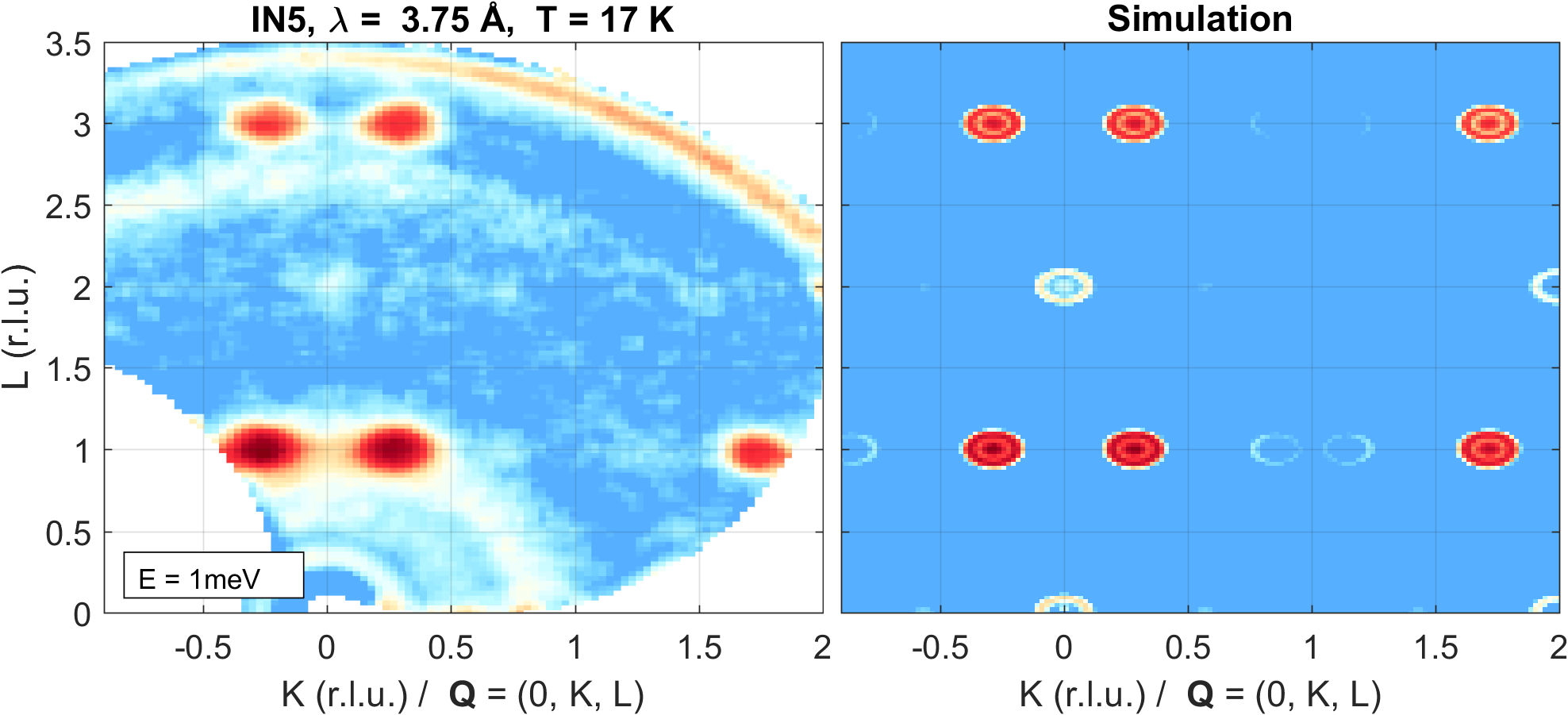}
        \includegraphics[width=1\columnwidth]{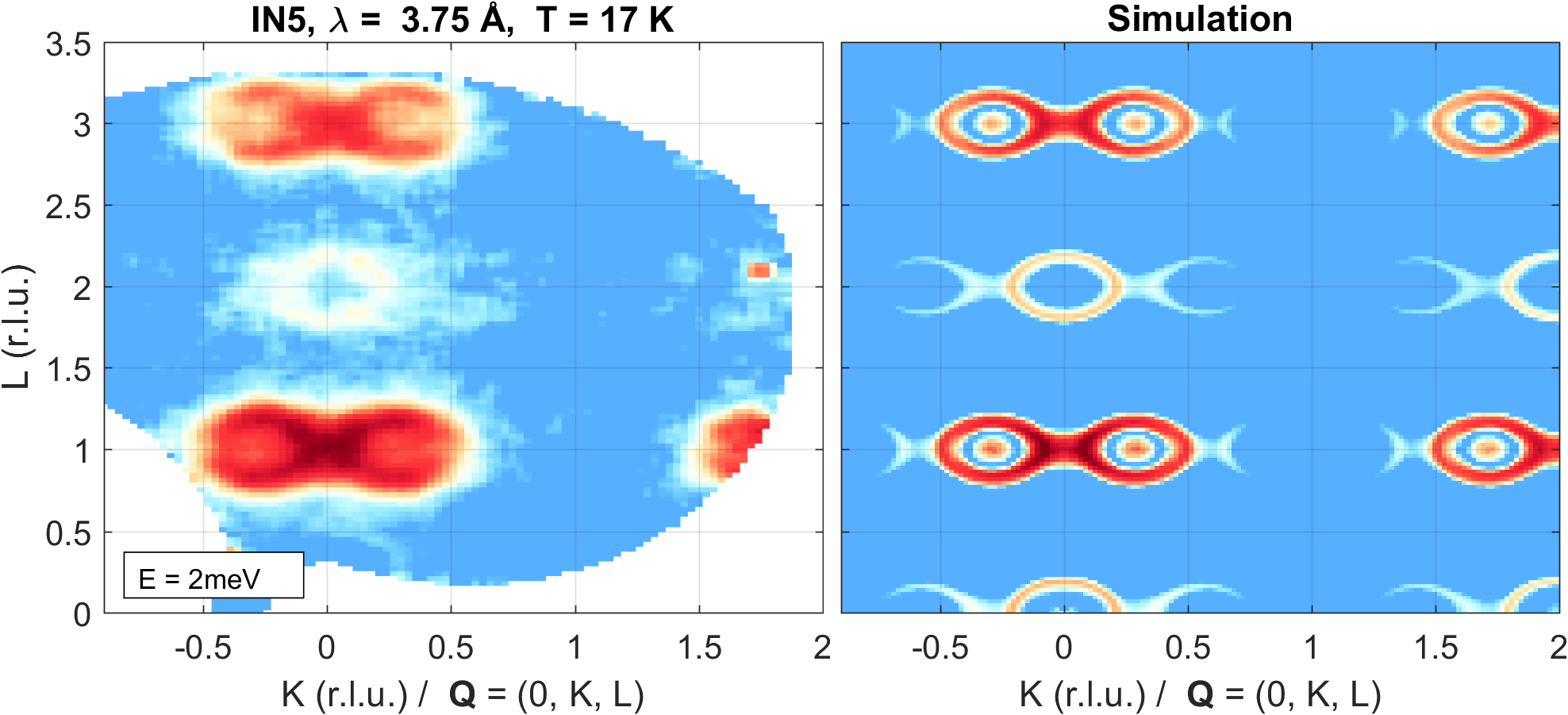}
        \includegraphics[width=1\columnwidth]{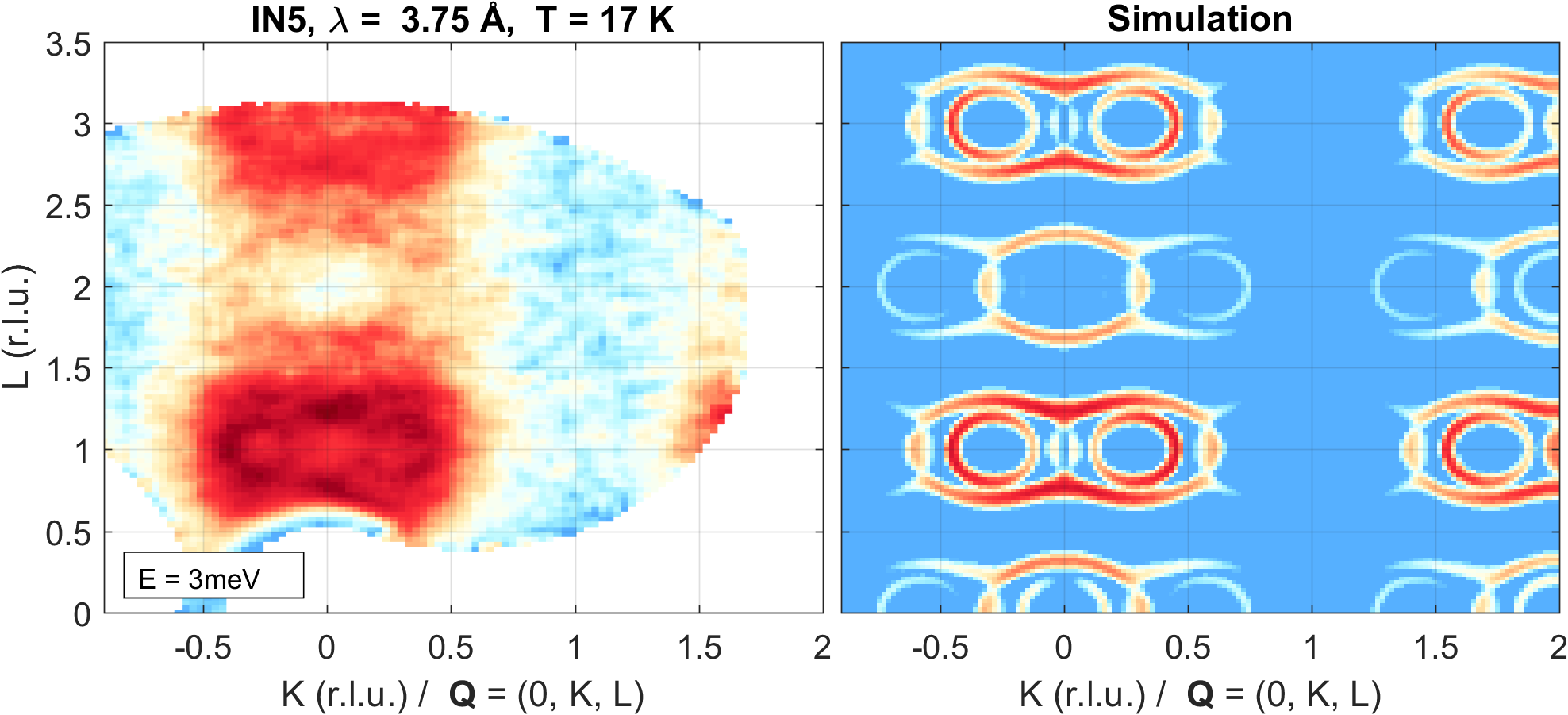}
\caption{Comparison of inelastic scattering maps of \tbmno\ obtained from neutron TOF spectroscopy at IN5 (left) at $T=\SI{17}{\kelvin}$) and simulation (right) in the $[0,K,L]$ plane. The incident neutron wave length and the energy resolution are given in the plot headers. Model M-II (cf. Tab.~\ref{tab:tbmno-spinWfit}) was used for the simulation and the Bose factor as well as the magnetic form factor were taken into account.}
    \label{fig:tbmno-IN5_constE1}
\end{figure}

\begin{figure}[!t]
        \includegraphics[width=1\columnwidth]{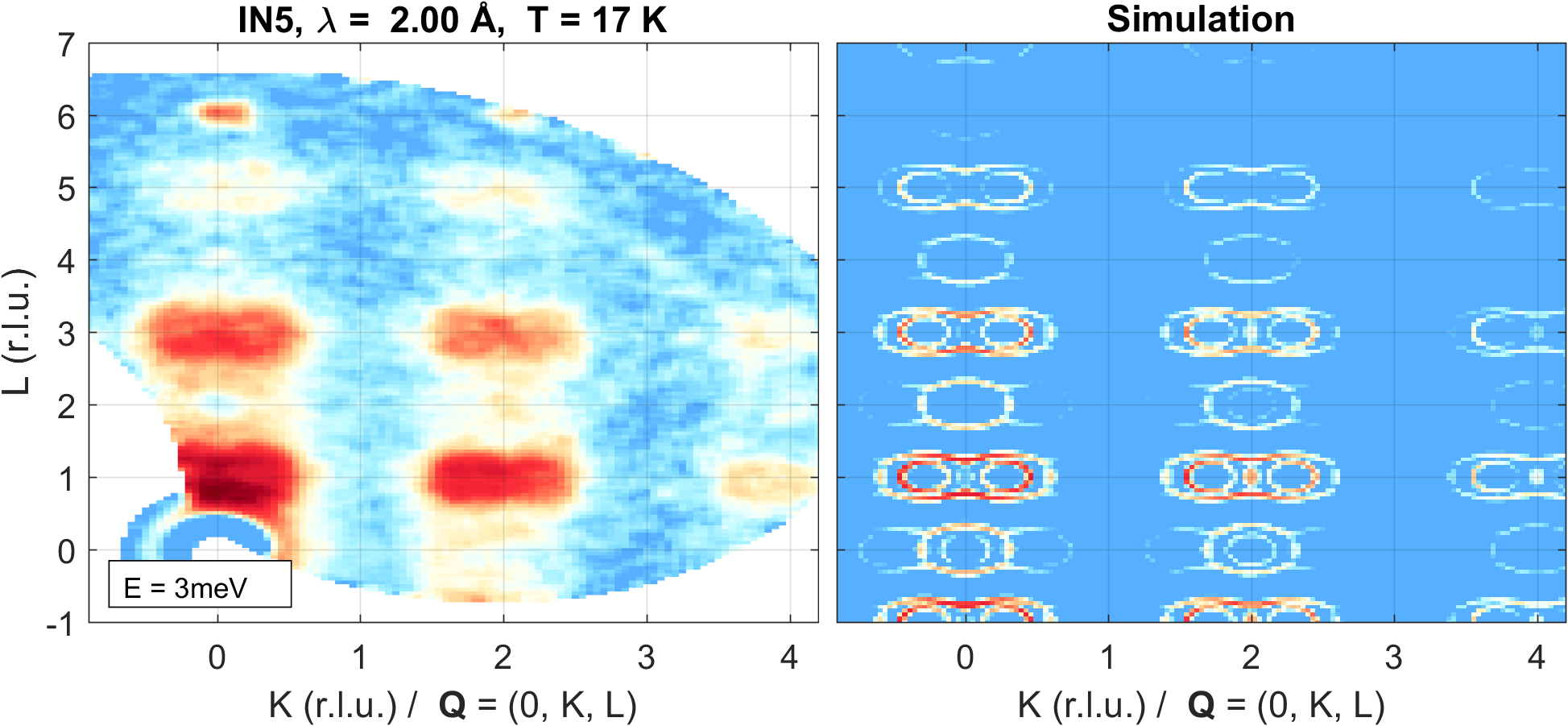}
        \includegraphics[width=1\columnwidth]{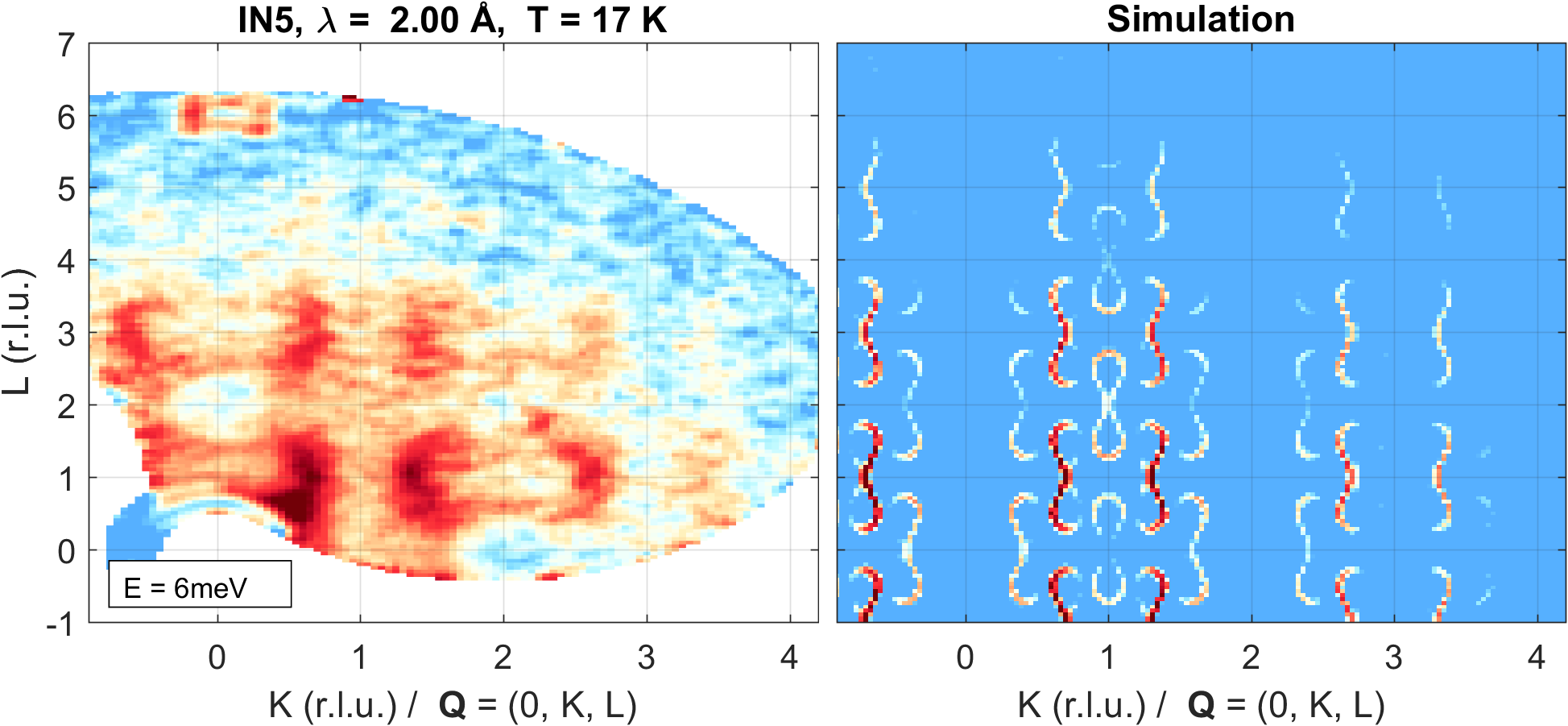}
        \includegraphics[width=1\columnwidth]{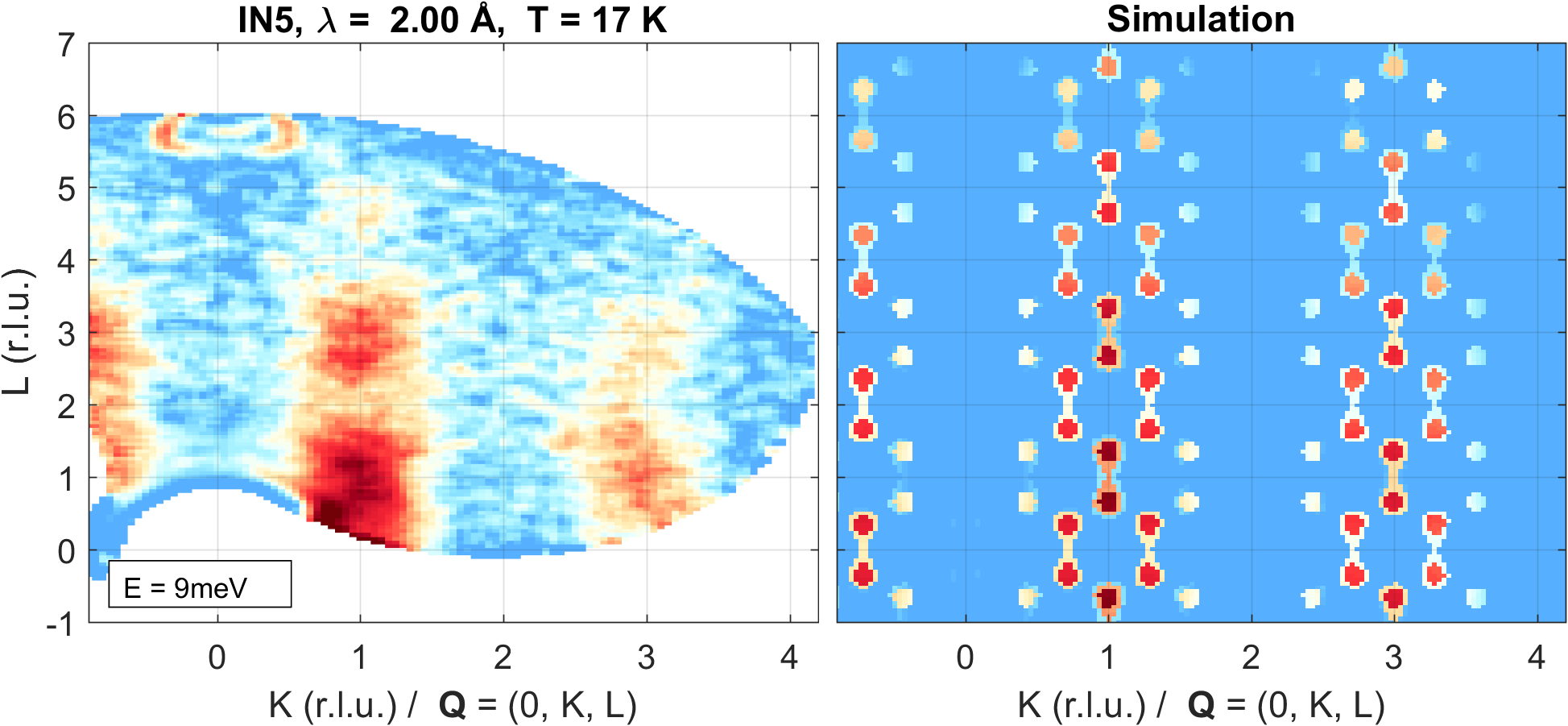}
\caption{Comparison of inelastic scattering maps of \tbmno\ obtained from neutron TOF spectroscopy at IN5 (left) at $T=\SI{17}{\kelvin}$) and simulation (right) in the $[0,K,L]$ plane. The incident neutron wave length and the energy resolution are given in the plot headers. Model M-II (cf. Tab.~\ref{tab:tbmno-spinWfit}) was used for the simulation and the Bose factor as well as the magnetic form factor were taken into account.}
    \label{fig:tbmno-IN5_constE2}
\end{figure}

Senff \etal\ experimentally separated the in-plane mode from the two out-of-plane modes using neutron polarization analysis~\cite{Senff2007}.
The phase of the out-of-plane modes could not be experimentally determined, but taking the magnetic phase diagram into account,
the rotation around \baxis\ must be lower in energy since the cycloid rotates in modest magnetic fields $H_{a,b}$~\cite{Kimura2005,Aliouane2009}.
The strict separation of both out-of-plane modes will not be valid in the case of an elliptic structure, which would allow a mixture of modes, rendering both magnon excitations IR active~\cite{Senff2008a}.
Pimenov \etal\ report two electrically active modes, a stronger broad mode at $\SI{20}{\per\cm}=\SI{2.48}{\meV}$  and a weaker one at $\SI{10}{\per\cm}=\SI{1.24}{\meV}$~\cite{Pimenov2006,Pimenov2009}.
Both positions perfectly correspond to the out-of-plane modes of the model and to the INS experiments, but their spectral weight seems to be exchanged.
Following the DM scenario, one would expect elm1 to be more strongly IR active.
This suggests some additional mechanism involved in the electromagnon at 2.5\,meV, see also references \onlinecite{Stenberg2009,Stenberg2012}, which however is difficult to validate due to the large number of magnon modes with frequencies in this energy range.
The strongest electromagnon response arising from magnetostrictive coupling can be identified with
the in-plane mode at the (0, $k$, 0) propagation vector with $k$=1-$k_{\textrm{inc}}$=0.72, see lowest panel in Fig.~\ref{fig:emagnons}, and is observed at 8\,meV in IR experiments in perfect agreement with the model ~\cite{ValdesAguilar2009,Finger2014,Takahashi2008}.
Even in a single material electromagnons can arise from different mechanisms~\cite{Katsura2007,ValdesAguilar2009,Mochizuki2010c} and the strongest electromagnon signal in \tbmno\ is not caused by the coupling driving the static multiferroic order \cite{ValdesAguilar2009,Finger2014}.

\subsection{Full magnon dispersion and scattering function}

In order to obtain the full picture of the spin-wave dispersion we performed INS experiments on the IN5 TOF spectrometer.
The raw data were first reduced with {\it{Lamp}} \cite{Richard1996} and then processed by Horace under
 \matlab~\cite{progHorace} to generate the  4-dimensional $S(\vec{Q},E)$ datasets and to extract the subsequent cuts.
No background subtraction was performed, because the largest part of it stems from the sample and because the spin-wave signal are sufficiently strong for direct observation. Two-dimensional cuts were generated from the dataset by integrating over a specific range in $\vec{Q}$ and $E$. For an incident neutron wave length $\lambda_{i,1}$=2.0\,\AA \ the integration range was $0.2\,(r.l.u)$ for $Q$ and \SI{1}{\meV} for $E_f$ ($\lambda_{i,2}$=3.75\,\AA : $0.1\,(r.l.u)$ and \SI{0.4}{\meV}, correspondingly). The 2D-cuts were reshaped using the implemented $smooth$ function for $2\times2$ pixels for a better visualization of the data~\cite{progHorace}.

The  TOF data were simulated using the distorted easy-plane model M-II. 2D cuts were produced using \spinw\ by taking into account the magnetic form factor of $Mn^{3+}$ and the Bose-factor for energy-loss scattering $1/(1-\exp(\frac{-E}{k_bT}))$. For the $\vec{Q},E$ maps the energy resolution was adapted to the experimentally determined value. A finite ${\vec{Q}}$ resolution and an integration over the vertical $Q$ component were considered to be dispensable. Accordingly, the simulated constant energy maps were integrated over a finite energy range (corresponding to the experiment) and no ${\vec{Q}}$ resolution was applied.

Figures~\ref{fig:tbmno-IN5_scans1} and \ref{fig:tbmno-IN5_scans2} show the comparison of TOF data and \spinw\ simulations for selected $\vec{Q}$ scans along the principal directions. The intensity is logarithmically color-coded. For all maps, the data could be very well reproduced. The broadening of the modes in the TOF data relative to the calculation is only partly due to the neglected ${\vec{Q}}$ resolution or to the vertical integration of the TOF data. Mainly, it arises from the intrinsic line width of the excitations, which has been found to be significantly larger than the best resolution used here, i.e. higher than \SI{0.2}{\meV}~\cite{Senff2008a}. The datasets using an incident neutron wave length of $\lambda=2.0$ and \SI{3.75}{\angstrom} show spurious signals below $\SI{2}{\meV}$ for wave lengths below the lattice constant of aluminum $\lambda<\SI{4.8}{\angstrom}$ stemming from the aluminum sample holder and the cryostat. Moreover,
there is a strong quasielastic line arising from incoherent scattering and from the Tb moments in addition to
phonons that in particular contribute at large scattering vector.

The comparison of TOF data and \spinw\ simulation for constant energy cuts in the $b^{\star}c^{\star}$ plane is shown in Figures~\ref{fig:tbmno-IN5_constE1} and \ref{fig:tbmno-IN5_constE2}. The difference between the line width of experimental data and simulation can again be attributed to the broadened intrinsic line width of the magnon excitations in \tbmno\ and only partially to the missing $\vec{Q}$ resolution in the simulation.

The comparison of TOF data in the multiferroic phase of \tbmno\ at \SI{17}{\kelvin} with simulated neutron intensity maps using model M-II (parameters given in Table~\ref{tab:tbmno-spinWfit}) is very satisfying, cf. Figs~\ref{fig:tbmno-IN5_scans1},~\ref{fig:tbmno-IN5_scans2},~\ref{fig:tbmno-IN5_constE1} and~\ref{fig:tbmno-IN5_constE2}. The dispersion along all orthorhombic directions as well as the intensities are reproduced quite well.
In particular the various weaker branches visible in the data are correctly reproduced.
The model M-II assumes only the ordering and interaction of Mn moments with a distorted planar anisotropy. A quantitative description of the excitation spectra in this phase is therefore possible without considering a Tb-Mn interaction or the influence of intrinsic and ferroelectric DM interactions. These effects may be hidden in the model in the anisotropy term along \caxis, which stabilizes the spin-spiral in the $bc$ plane.

The comparison between the simulated and experimental maps indicates significant broadening of magnons in TbMnO$_3$, which can be attributed to the coupling with disordered Tb moments. Indeed cooling into the phase with ordered Tb moments
results in sharper features, but yields a more complex dispersion, see Section VI.

\subsection{Chirality of excitations}

Further insight on the complex excitation spectra can be gained by applying experimental techniques to separate the different branches.
Polarized neutrons were already utilized to distinguish the polarization of the three zone-center modes~\cite{Senff2007}.
By the application of an electric field, it is possible to obtain a chiral mono-domain sample~\cite{Finger2010,Poole2009,Stein2015,Stein2017}, which can be studied through the magnetic chiral component $M_{ch}=-i(\vec{M}_\perp\times\vec{M}^*_\perp)_x$, where $\vec{M}_\perp$ is the $Q$th Fourier component of the inelastic magnetization distribution \cite{Brown2006}.

The chiral component can be directly determined by comparing the two neutron-spin-flip
intensities for neutron polarization parallel to the scattering vector, $\sigma_{x\bar{x}}-\sigma_{\bar{x}x}=2M_{ch}$ or by studying the rotation
of the neutron polarization.
Here the indices of the cross section indicate the polarization direction before and after the scattering and the overbar a negative neutron polarization.
The analysis of the neutron polarization rotation (off-diagonal indices) requires however spherical neutron polarization analysis and the corresponding experimental setups~\cite{Brown2006}.
The chirality has been studied in \tbmno\ on magnetic Bragg peaks~\cite{Yamasaki2007,Stein2015} and on the diffuse scattering
above the long-range multiferroic transition ~\cite{Stein2017}.
By using a time-resolved neutron technique it was possible to study the multiferroic relaxation
as a function of temperature and electric field over 8 decades in time ~\cite{Stein2021}.

The aim of our polarized experiments was the investigation of the chiral components of the magnons for scattering vectors $\vec{Q}=(2,~k,~1)$ in the multiferroic phase. In a simple cycloidal incommensurate structure one may expect the phason branch to exhibit a strong chiral component which should vary with the propagation vector as it can be seen in Fig.~\ref{fig:emagnons}. Contrarily, the out-of-plane modes should not carry any chiral signal. An earlier INS study indeed found evidence for
inelastic chiral contributions, but due to statistics limitations it was only possible to count at the expected maximum positions of the phason modes at (2+$\xi$, 0.28, 1) \cite{diss_finger}.

\begin{figure}[!t]
    \centering
    \includegraphics[width=0.9\columnwidth]{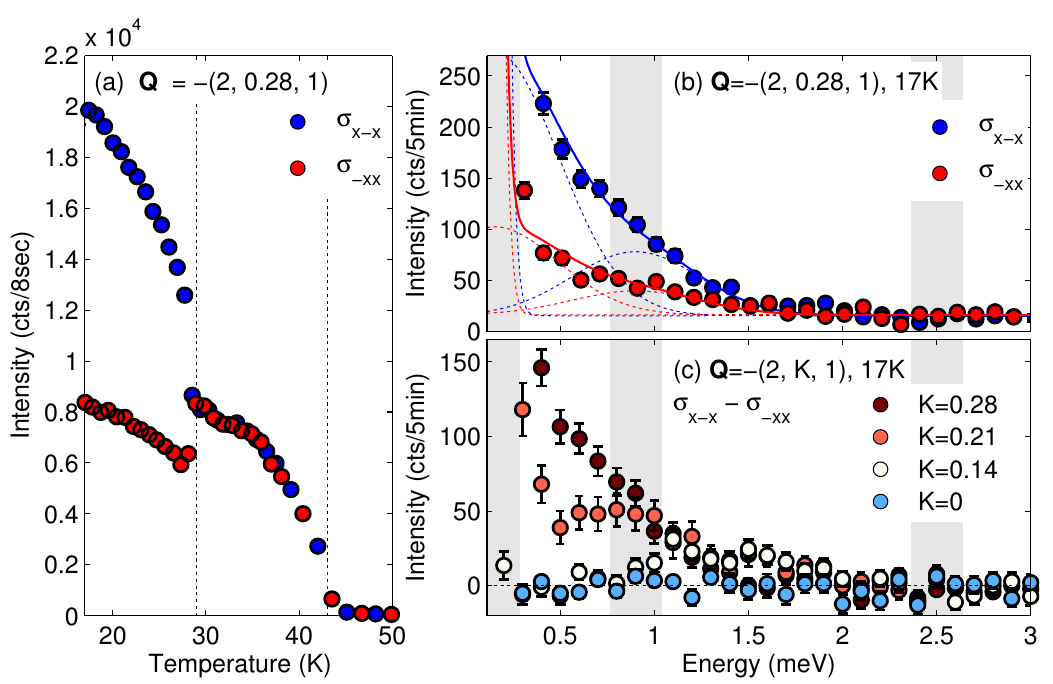}
    \caption{Chiral magnetic scattering of \tbmno\ measured by polarized neutrons. (a) Temperature dependence of the cross sections $\sigma_{x\bar{x}}$ and $\sigma_{\bar{x}x}$ at the magnetic Bragg peak $\vec{Q}=-(2,~0.28,~1)$ at IN14/Thales. An electric voltage of \SI{7.8}{\kilo\volt} ($\approx\SI{300}{\volt\per\mm}$) was applied along \caxis. (b) Energy scan at the magnetic zone center $\vec{Q}$ for both channels at $T=\SI{17}{\kelvin}$. Two Gaussian functions were fitted to each curve to take account of the tail towards large energies. (c) Evolution of the inelastic chiral intensity $\sigma_{x\bar{x}}-\sigma_{\bar{x}x}$ along $-[2,K,1]$ at $T=\SI{17}{\kelvin}$. The gray bars indicate the reported zone center modes~\cite{Senff2007} of which the lowest in energy indicates the phason and the two modes at higher energies are $a$ polarized. The latter two modes are efficiently suppressed by the scattering geometry (factor 1/8) and do not contribute to the
    chiral signal. }
    \label{fig:tbmno-chiral_01}
\end{figure}

To detect the chirality of the magnetic correlations one has to ensure that the scattering vector is nearly vertical to the spin spiral. We performed the experiment around the position $\vec{Q}=-(2,~0.28,~1)$. We recall the coordinate system used in a polarized neutron experiment with $x$ parallel $\vec{Q}$, $z$ vertical to the scattering plane and $y$ perpendicular to $x$ and $z$ \cite{Brown2006}. The corresponding vectors and their angles to the principal crystallographic directions are summarized in Table~\ref{tab:tbmno-chiral} in the Appendix.

The temperature dependence of the two spin-flip channels for neutron polarization along $x$  is shown in Figure~\ref{fig:tbmno-chiral_01}(a)
at the elastic Bragg position $\vec{Q}=-(2,~0.28,~1)$.
The phase transitions at $T_N$ and $T_{\textrm{MF}}$ are indicated by dashed lines. The applied electric field poles the domains in the sample and the polarization analysis is able to distinguish between both vector chiralities. We obtain a ratio between both channels of about $3:1$ at the magnetic Bragg peak which corresponds to a chiral ratio of $r_{\text{chiral}}=\frac{I_{x\bar{x}}-I_{\bar{x}x}}{I_{x\bar{x}}+I_{\bar{x}x}}\approx0.5$.
This chiral ratio can only attain the ideal maximum of $r_{\text{chiral}}=1$ when the scattering vector \qvec\ is perpendicular to the spiral plane.
In TbMnO$_3$ the angle between $\vec{Q}=-(2,~0.28,~1)$ and the spiral plane \bcplane\ amounts to \SI{70}{\degree}, see Table~\ref{tab:tbmno-chiral}.
One also has to take into account the elliptic shape of the spin spiral and the orientation of \qvec\ relative to the ellipse.
This yields a maximum value of the chiral ratio at this position of $r_{\text{chiral,Q,max}}=0.93$.
We can conclude that more than \SI{75}{\percent} of the sample were in one domain.
In previous experiments a higher ratio of up to $r_{\text{chiral}}\approx0.8$ was achieved~\cite{Yamasaki2007,Stein2017,Stein2021,Stein2015}.
This was only possible using smaller samples which increased the maximum applicable electric field.
In our experiment, a large sample was, however, necessary to investigate the  weak inelastic signal. The achieved poling, nevertheless, is sufficient to probe the chiral component of the magnetic excitations. In a consecutive experiment on IN14/Thales a voltage of +10kV could be applied in one direction, but only -5kV in the opposite direction.

\begin{figure}[t]
    \includegraphics[width=0.95\columnwidth]{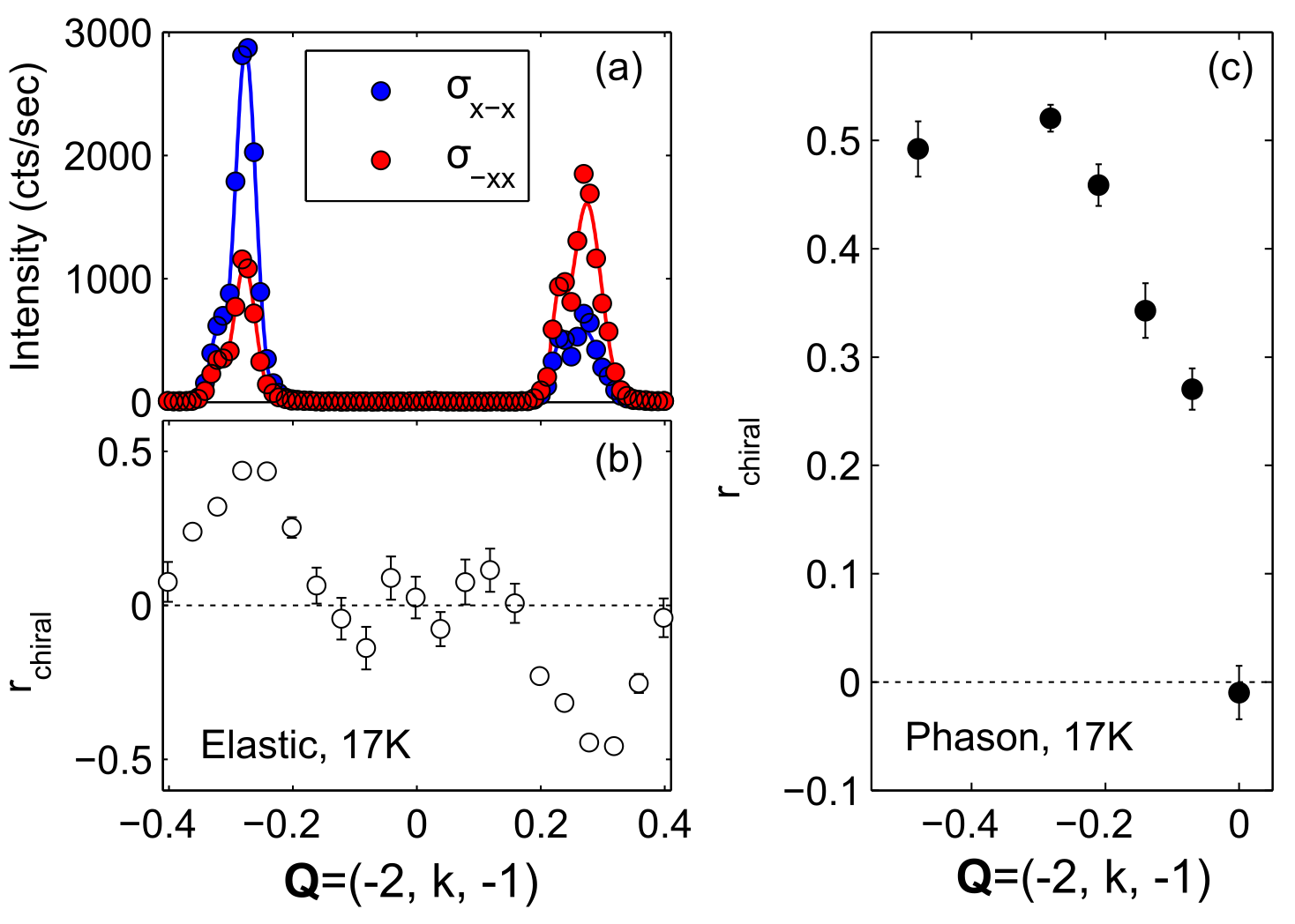}
    \caption{Chirality of elastic and inelastic magnetic scattering in \tbmno\ studied on IN14/Thales. (a)  Intensity of cross sections $\sigma_{x\bar{x}}$ and $\sigma_{\bar{x}x}$ at $\vec{Q}=(-2,~k,~-1)$, (b) Chiral ratio of elastic signal in two Brillouin zones, (c) chiral ratio of low-energy phason mode along $\vec{Q}=(-2,~k,~-1)$.}
    \label{fig:tbmno-chiral_02}
\end{figure}

The magnetic excitations at the magnetic zone center at \SI{17}{\kelvin} inside the multiferroic phase are shown in Fig.~\ref{fig:tbmno-chiral_01}(b). At this scattering vector the contribution of $a$-polarized out-of-plane excitations is strongly suppressed by the sizable component along \aaxis , $\sin^2(\alpha_a)=0.12$, and the scattering intensity stems primarily from in-plane modes. In addition, only the low-energy phason mode carries a chiral component and should be visible in the subtraction of the spin-flip channels along $x$. The positions of the in-plane phason mode ($\omega_1\approx\SI{0.1}{\meV}$) and the two out-of-plane modes elm1 and elm2 are indicated by gray bars. Indeed there is no scattering contribution at these latter two positions. The observed chiral signal is broadened and can only be described by a combination of two Gaussian functions (dotted lines), where the upper Gaussian mimics the high energy tail.

Figures~\ref{fig:tbmno-chiral_01}(c) and 11(c) present the dispersion of the chiral signal $\sigma_{x\bar{x}}-\sigma_{\bar{x}x}$ along $\vec{Q}=(-2,~k,~-1)$.
The intensity difference decreases away from the incommensurate zone center and completely vanishes at $\vec{Q}=(-2,~0,~-1)$.
At all positions, the signal remains broadened with respect to the resolution of $\Delta E\approx\SI{0.24}{\meV}$.
The large crystal exhibits a sizable mosaic spread with a FWHM of about 1.7 degrees, so that
the steep dispersion of the phason implies considerable broadening. The smearing of the $k$ component of
the scattering vector and the resulting sensing of the phason dispersion can, however, only account for a
broadening of about 0.4\,meV, while the chiral scattering clearly extends to above 1\,meV.
As the scattering vector is almost fully aligned along the $a$-direction, an overlapping contribution of the out-of-plane modes can be neglected (see above), and these modes are not expected to carry chirality.
In Figure~\ref{fig:tbmno-chiral_02} the chirality at the two  satellite signals (-2, $\pm$0.28, -1) is compared with the dispersion of the chirality of the phason mode.
The elastic scan along $\vec{Q}=(-2,~k,~-1)$ shows two magnetic Bragg reflections with opposite chiral ratio. This is due to the scattering geometry when $k$ changes sign.

There is no indication that the application of an electric field causes a significant variation of the mode energies.
The energy that the static field implies on the microscopic electric dipoles, is negligible in comparison to the magnetic interactions.
Accordingly, the models of the magnon dispersion do not need to take the ferroelectric ordering in the spiral phase into account.

\begin{figure}[!t]
    \includegraphics[width=0.65\columnwidth]{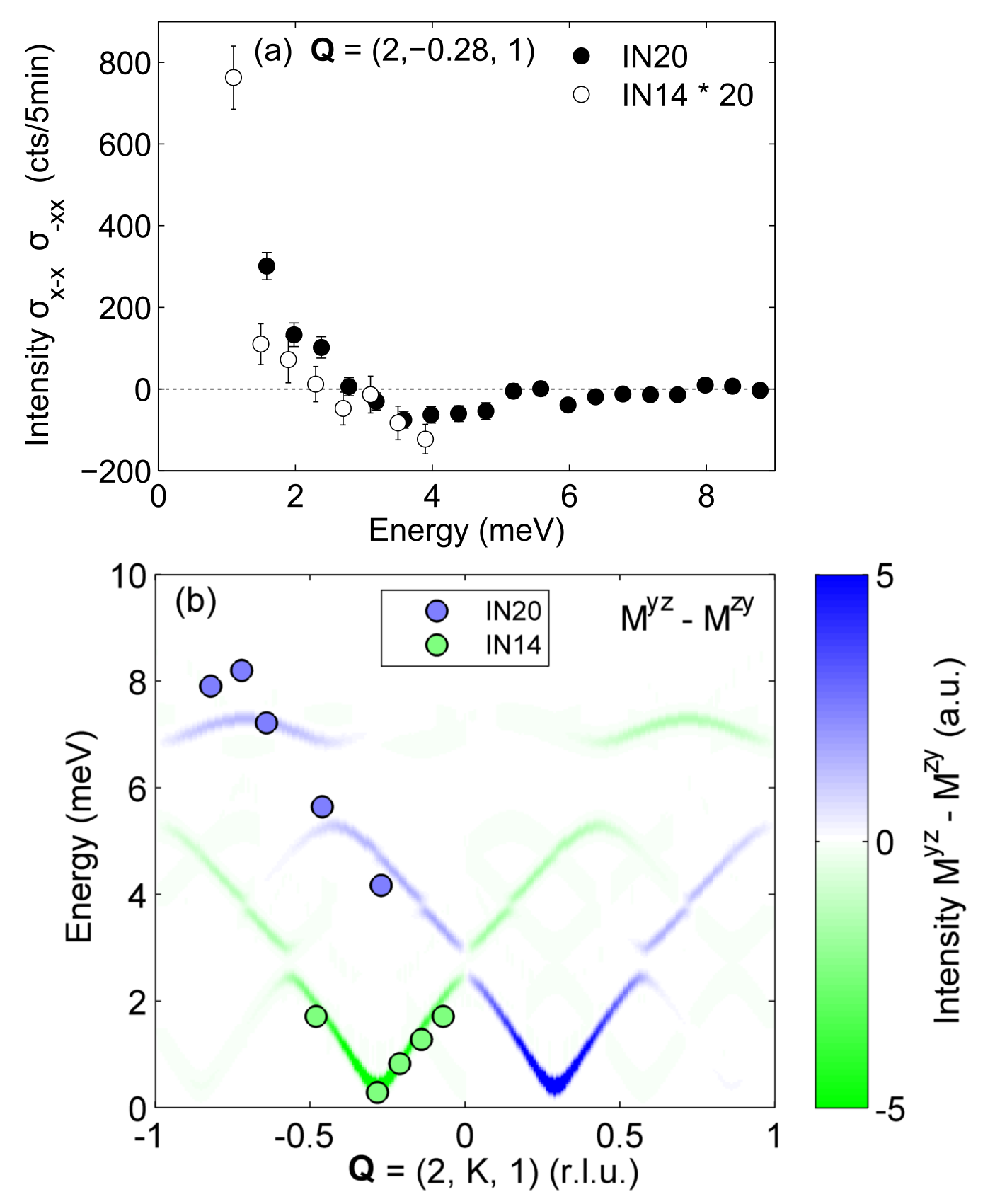}
\caption{Dispersion of chiral magnetic scattering in \tbmno\ at $T=\SI{17}{\kelvin}$, poled by an electric field of $E\approx\SI{355}{\volt\per\mm}$. (a) Chirality component, $\sigma_{x\bar{x}}-\sigma_{\bar{x}x}$, of zone center modes measured on IN14/Thales and IN20. The intensity of the IN14/Thales scan was multiplied by a factor of 20. (b) Simulation of chiral magnetic scattering $\Im(M^{yz}-M^{zy})$ in \tbmno\ along $\vec{Q}=(2,~k,~1)$. The sign of chirality is color-coded. The simulation is compared to data obtained at IN14/Thales and IN20 at $T=\SI{17}{\kelvin}$. The color of the symbols corresponds to the sign of chirality of the modes. }
    \label{fig:tbmno-chiral_03}
\end{figure}

The investigation of the chiral components of the magnons in the spiral phase of \tbmno\ revealed another interesting finding. Figure~\ref{fig:tbmno-chiral_03}(a) shows the chiral component, i.e. the subtraction of both spin-flip channels for neutron polarization along the scattering vector, at the magnetic zone center $\vec{Q}=(2,~-0.28,~1)$ in a larger energy range. The same scan was measured at IN14/Thales, as described above, and at the thermal TAS IN20 using $k_f$=2.66\,\AA$^{-1}$. The latter experimental setting allows one to measure at higher energy transfer, with the cost of a relaxed resolution~\cite{note-IN20}. Both spectra were matched by multiplying the IN14/Thales data by a factor of 20, which accounts for the higher flux on thermal instruments. In spite of the different scattering and polarization geometries in the two experiments, the data can be easily merged with respect to the chirality of static signal. 
In both scans the identical sample was poled by an electric field. At low energies, the intensity of the chiral component is positive and exceeds the vertical scale (see Fig.~\ref{fig:tbmno-chiral_01}(c) for the full scale). Surprisingly, at higher energies, the intensity difference not only decreases, but changes its sign at an energy transfer of approximately \SI{4}{\meV}. This observation is visible in both scans and cannot be explained by statistical variations. At the incommensurate zone center there is
a mode at intermediate energy, which exhibits an opposite chirality with respect to the static Bragg peak.

The mode of opposite chirality could be followed through the Brillouin zone on IN20 (see Appendix for original data).
We simulate the chiral magnetic scattering using the model M-II to investigate the origin of these modes with inverted chiral signs. Figure~\ref{fig:tbmno-chiral_03}(b) shows the calculated neutron intensity $M^{yz}-M^{zy}$ (which corresponds to the dynamic chiral component) along $\vec{Q}=(2,~k,~1)$.
The color scale ranges from negative (green) to positive (blue) values.
The calculation can directly be compared to experimental values.
Green and blue dots denote the position and sign of chirality of the phason modes.
The values have been determined by Gaussian fits to the data (an example is shown in Fig.~\ref{fig:tbmno-chiral_01}).
In addition, we show values of the high-energy modes obtained at IN20 with reversed chirality, with respect to the static Bragg peak.
The data extracted from both experiments perfectly agree with the calculated opposite chirality of the modes associated with the phason branch
starting at the neighboring incommensurate Bragg position, which exhibits an opposite static chirality, see Fig.~\ref{fig:tbmno-chiral_02} (a).  The coexistence of modes with different signs of chirality is just a geometry effect and it has also been reported in Ba$_3$NbFe$_3$Si$_2$O$_{14}$ whose structure is intrinsically chiral~\cite{Jensen2011,Loire2011}.

The dynamic chirality of the in-plane phason modes are thus well described by model M-II.
The high-energy modes of reversed chiral component can be attributed to a phason branch beginning at the neighboring incommensurate Bragg position.
The polarized measurements unambiguously reveal the dispersion of the in-plane modes which otherwise are
superposed with the out-of-plane modes underlining the complementarity of polarized and unpolarized INS.
The experimental data in Fig.~\ref{fig:tbmno-chiral_03} (b) (blue dots) well agree with the calculation predicting a maximum in-plane energy at the wave vector
(0, $(1-k_{\textrm{inc}})$=0.72, 0), which corresponds to the high-energy electromagnon as discussed above.
It may astonish that the splitting between in-plane and out-of-plane modes extends to such high energies,
as one expects the typical anisotropy terms to essentially influence the low-energy modes in an AFM system.
However, the interplay between the static cycloidal moments and the oscillation strongly modifies only the in-plane dispersion, as any longitudinal polarization is suppressed at these moderate energies. The spin-wave calculation with the rotating reference frame perfectly reproduces these chiral aspects.

\begin{figure}[!t]
    \centering
    \includegraphics[width=0.99\columnwidth]{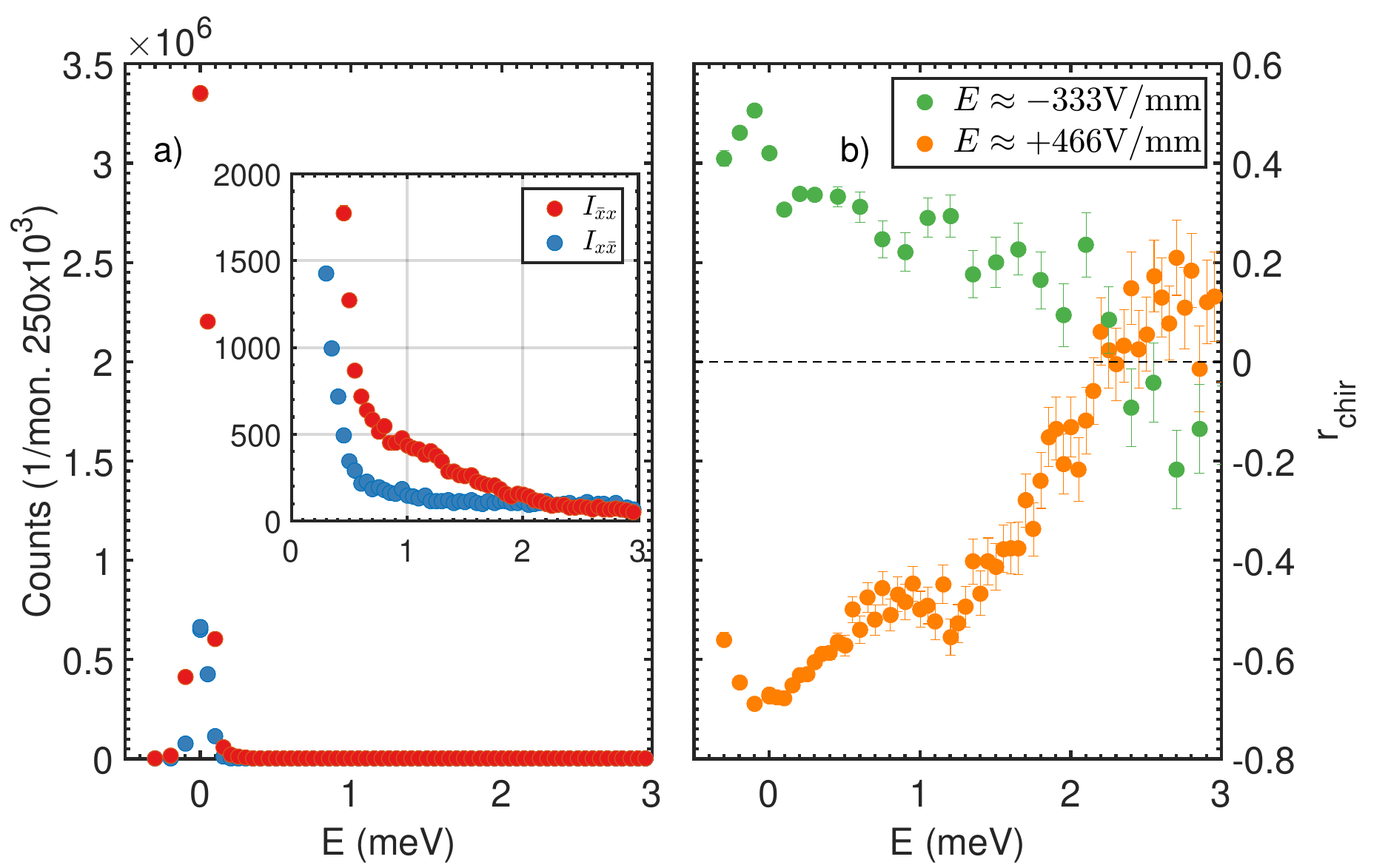}
    \caption{Inelastic neutron intensity  measured  at $\vec{Q}=(0,~0.28,~1)$ on IN14/Thales at the temperature $T=\SI{25}{\kelvin}$ and in a field of +466\,V/mm in the two
    neutron spin-flip channels (a).
     Panel (b) shows the chiral ratio calculated from these data and for data taken at the inverted electric field of -333\,V/mm. The chiral sign is inverted for the elastic response and for the two inelastic features. Also the inelastic signal that exhibits the chirality opposite to the static one gets
     inverted. }
    \label{fig:Thalesnewa}
\end{figure}

In a second set of experiments on IN14/Thales we focused on the possibility to invert the chirality of the inelastic signal by inverting the external electric field. In order to pole the multiferroic phase the sample was cooled through the multiferroic transition to 25\,K in either positive or negative fields.
At higher temperature the plates were short-circuited in order to avoid charging effects.
Fig.~\ref{fig:Thalesnewa} (a) presents the data in the two spin-flip channels taken at the incommensurate zone-center for positive voltage from which the chiral
ratio is calculated in panel (b).
We also show the chiral ratio for the inverted voltage, which indeed changes sign.
Also the chirality of the contribution at higher energy, which was only partially reached in this experiment, is inverted.
In the Appendix we present further data at (0, 0.21, 1) and (0, 0.14, 1), which also show that the inelastic chiralities
are completely reversed by inverting the multiferroic domains. 
These observations unambiguously show that the dynamic
chiralities are fully imprinted by the sign of the static chirality, which is controlled by external electric fields. The dynamic chirality of magnons causes non-reciprocal effects in the neutron
response, see e.g. Fig. 12 (b) for comparing scattering at (2, $\pm$0.28, 1), similar to dichroism effects
in studies with terahertz radiation~\cite{Bordacs2012}.

\begin{figure}[!t]
    \centering
    \includegraphics[width=0.99\columnwidth]{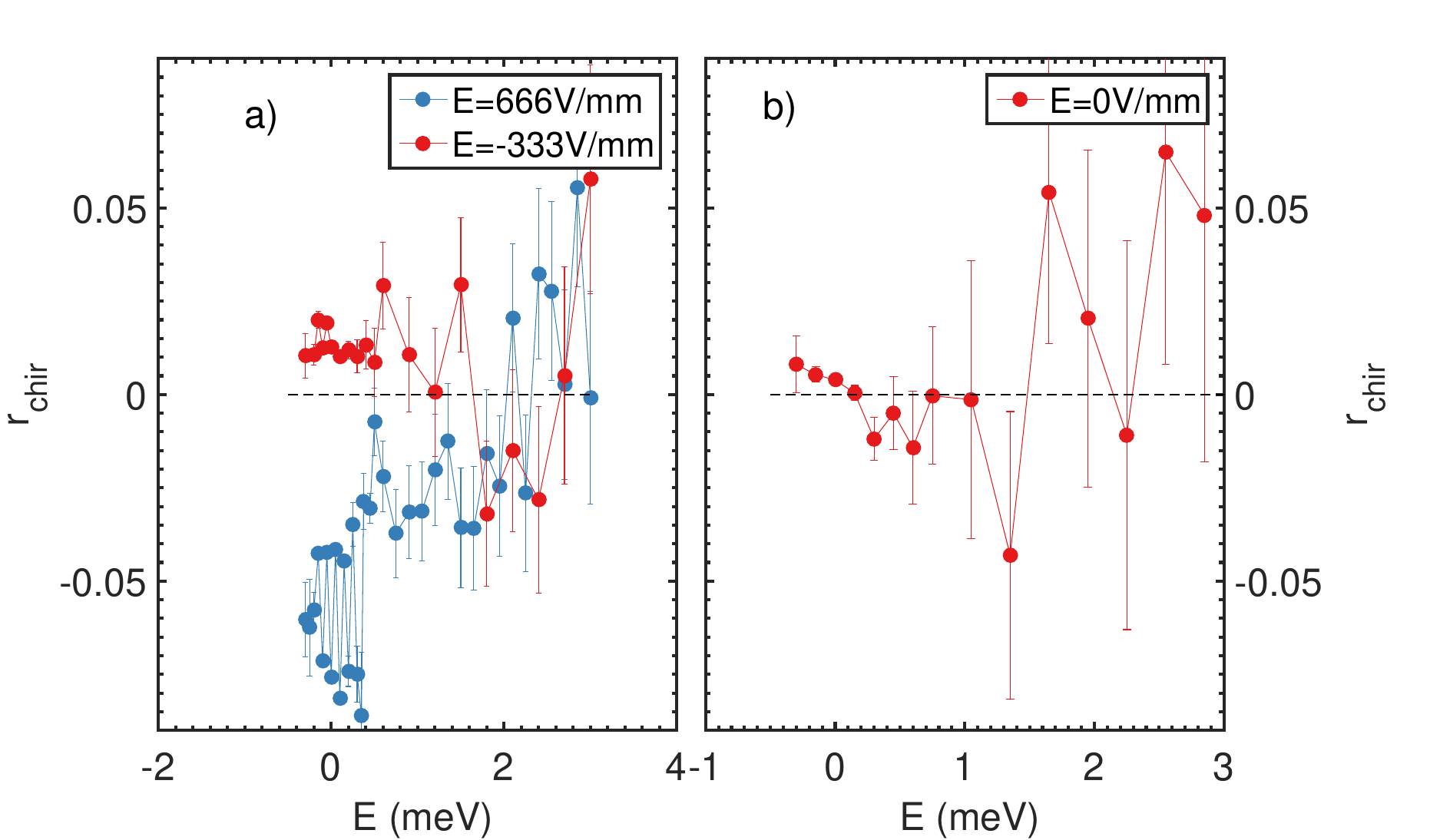}
    \caption{Chiral ratio of polarized INS scans on IN14/Thales taken at $\vec{Q}=(0,~0.28,~1)$ and the temperature $T=\SI{27.93(4)}{\kelvin}$, i.e. above the
    onset of long-range multiferroic order at zero electric field. In panel (a) chiral ratios taken at large electric fields of different signs are shown; panel (b)
    presents the data values of the chiral ratio obtained at zero electric field.  }
    \label{fig:Thalesnewb}
\end{figure}

Polarized neutron diffraction experiments at finite electric fields above the onset of long-range multiferroic order in zero electric field, $T_{\textrm{MF}}$,
revealed that it is still possible to control the chirality of the elastic signal and that this effect exceeds the expectation deduced from Landau theory \cite{Stein2017}.
In order to further analyze the multiferroic poling above the long-range transition, we studied also the inelastic signal slightly above $T_{\textrm{MF}}$.
This scan is comparable to that in the multiferroic phase but
the small amplitude of the chiral contributions considerably hampers the experiments.
Therefore, this new experiment had to be performed on the  large
sample used for the poled inelastic experiments although the control of the applied electric field is more difficult. The results are shown in Fig.~\ref{fig:Thalesnewb} and confirm the control of the elastic signal above $T_{\textrm{MF}}$ reported in reference \cite{Stein2017}.
We emphasize, that the temperature fluctuations during these measurements could be kept below $\pm$0.04\,K.
The inelastic part also exhibits a finite chirality that however remains small, in particular much smaller than
the inelastic chirality seen in the multiferroic poled state.
The applied field is thus not sufficient to fully separate the excitations with different chiral signs in the SDW phase.
The observed inelastic chirality seems to follow the amount of static chirality that can be implied by the external field.
Nevertheless this inelastic chiral poling will
contribute to dielectric measurements \cite{Schrettle2009,Foggetti2020}.

\begin{figure*}[!t]
    \centering
     \begin{minipage}[b]{1.3\columnwidth}
        \centering
        \includegraphics[width=1\columnwidth]{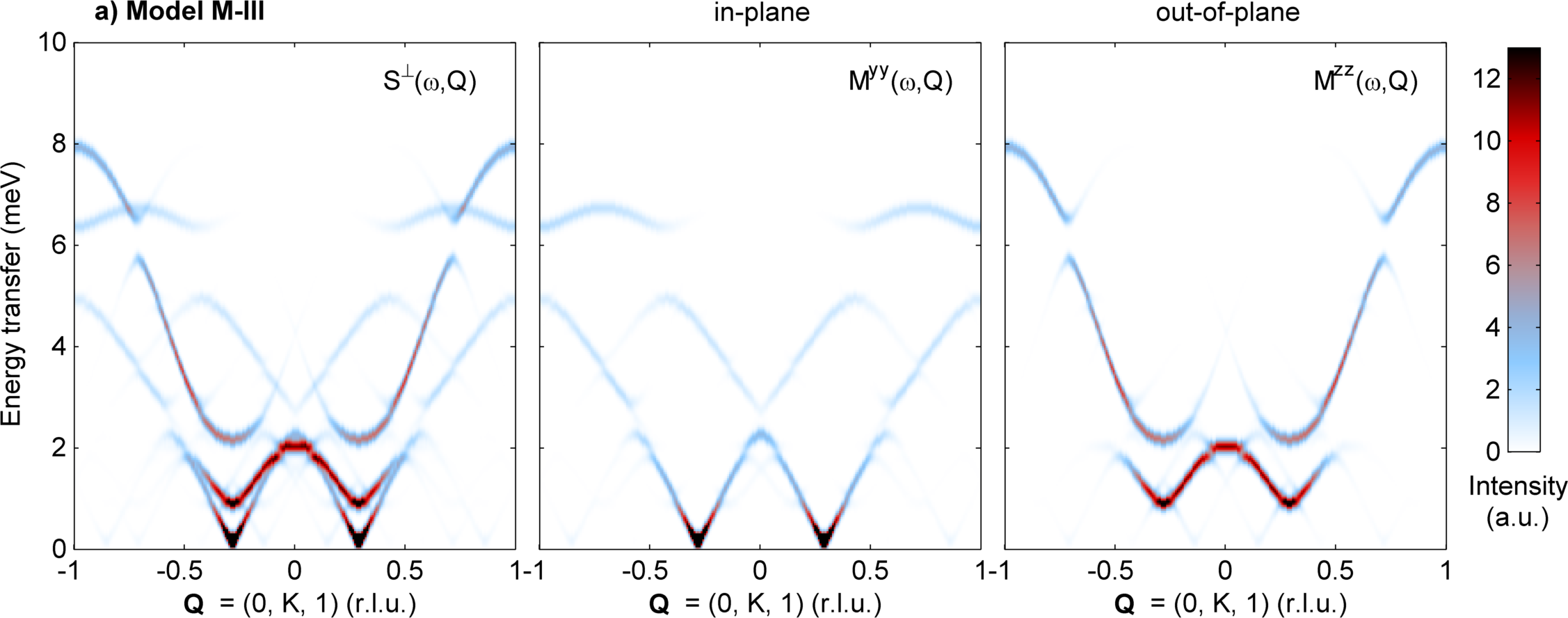}
     \end{minipage}
     \vskip3mm
     \begin{minipage}[b]{1.3\columnwidth}
        \centering
    \includegraphics[width=1\columnwidth]{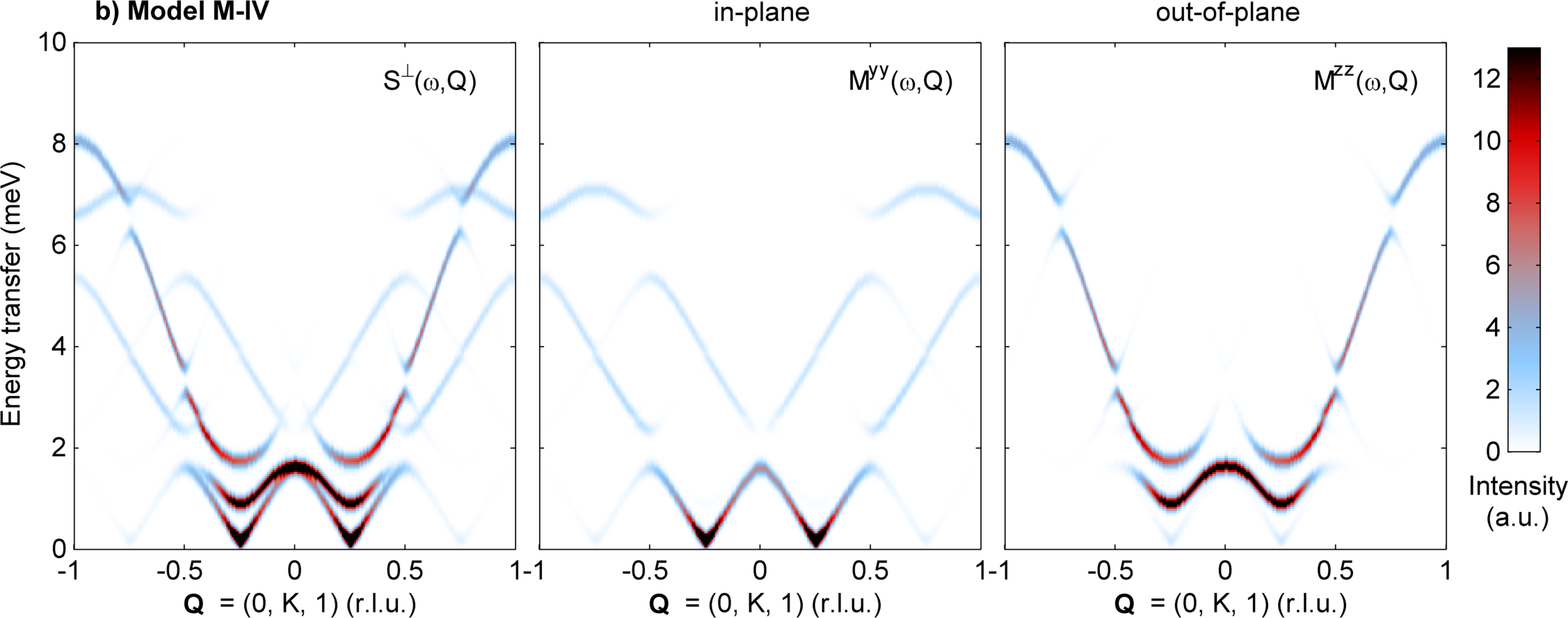}
     \end{minipage}
	\caption{Calculated neutron intensity for \tbmno \ along $[0 K 1]$ of the total cross section $S^{\perp}(\omega,Q)$ and the components $M^{yy}(\omega,Q)$ and $M^{zz}(\omega,Q)$. The scattering geometry was chosen so that $y$ is in the \bcplane\ plane and $z$ is parallel \aaxis. Different magnetic models were used for the calculation: (a) Model M-III: elliptic spiral with moments in the \bcplane\ plane, $\vec{k}=(0,~2/7,~0)$ and a staggered elliptic easy-plane anisotropy and (b) model M-IV: elliptic spiral with moments in the \bcplane\ plane, $\vec{k}=(0,~1/4,~0)$, easy-axis anisotropy along \baxis\ and anisotropic exchange (DM).}
    \label{fig:tbmno-spinW_mochizuki}
\end{figure*}

\section{Extension of the model of magnetic interaction}

Our calculations of the elliptic cycloid dispersion along the direction of the propagation vector match a study by Milstein and Sushkov~\cite{Milstein2015}. Instead of linear spin-wave theory they use effective-field theory to model the magnon branches in \tbmno.
They consider an incommensurate propagation vector of $\vec{k}=(0,~0.28,~0)$ with a deformed elliptic spin spiral and introduce an additional
nearest-neighbor exchange $J_{3a}$ along \aaxis.
The anisotropy-terms consist of an easy-axis along \baxis\ and a DM-type antisymmetric exchange along \aaxis,
which is equivalent to the easy-plane single-ion anisotropy in our model~\cite{Milstein2015}.
Milstein and Sushkov attribute this feature to the dipolar displacement of the FM polarization in the multiferroic phase.
In this model, the DM interaction amounts to $D$=\SI{0.2}{\meV} which appears to be rather strong in view of the relatively weak ionic displacement.
First-principle calculations estimated very small displacements of the order of $10^{-4}$\,\AA~\cite{Xiang2008,Malashevich2008a} in agreement with an experimental X-ray analysis~\cite{Walker2011}. We verified the influence of the additional exchange $J_{3a}$ and of the elliptical deformation on our model. The small deformation proposed by Milstein and Sushkov ~\cite{Milstein2015} only leads to minor variation and $J_{3a}$ did not affect the dispersion along \baxis.

A different calculation of the spin dynamics has been reported by Mochizuki \etal~\cite{Mochizuki2010c}. The underlying Hamiltonian consists of five terms: $H=H_{\textrm{ex}}+H_{\textrm{sia}}^D+H_{\textrm{sia}}^E+H_{\textrm{DM}}+H_{\textrm{biq}}$.
The isotropic exchange parameters in $H_{\textrm{ex}}$ are identical to our model. The introduced anisotropies include a hard axis $H_{\textrm{sia}}^D$ along \caxis\ and alternating local hard and easy axes $H_{\textrm{sia}}^E$ in the \abplane\ plane due to staggering of orbitals. The DM terms along \caxis\ (DM$_c$) and in the \abplane\ plane (DM$_{ab}$) are taken from local spin-density approximation calculations~\cite{Solovyev1996} and experiments on \lamno~\cite{Deisenhofer2002}. They arise from the MnO$_6$-octahedron rotation in \tbmno , which shifts the oxygen ions out of  the high-symmetry position between two Mn ions~\cite{Solovyev1996,Tovar1999}. DM$_c$ is four times stronger than in LaMnO$_3$~\cite{Deisenhofer2002} and plays a crucial role in this model. It is able to stabilize the spin spiral in the \bcplane\ plane and, in competition with $H_{\textrm{sia}}^D$, it can explain the spiral plane flop in HF-C phases~\cite{Mochizuki2009,Mochizuki2009a,Mochizuki2010}. Finally, the biquadratic interaction term $H_{\textrm{biq}}$ in the \abplane\ plane originates from the spin-phonon coupling~\cite{Kaplan2009}. It helped to reproduce optical spectra of \dymno~\cite{Mochizuki2010c}. The calculated magnon dispersion shows very strong anti-crossing and folding of modes. It does not properly reproduce the INS results on the magnon dispersion in \tbmno~\cite{Senff2008a}, as discussed above, or in \dymno~\cite{Finger2014} with only weak anti-crossing features.

An essential difference between the model developed here and the model by Mochizuki \etal\ concerns the anisotropies. Our model M-II stabilizes the \bcplane \ cycloid by a distorted easy-plane anisotropy, while the other model uses the combination of easy-axis anisotropies and DM interactions. We modified our model step-by-step in order to analyze the implications on the dispersion by including a staggered easy-plane anisotropy in the \abplane\ plane and a $DM_c$ anisotropy along \caxis.
The DM interaction in the \abplane\ plane was neglected because of its reported weakness ($DM_c/DM_{ab}\approx4$ in \lamno)~\cite{Deisenhofer2002}.

First, we insert the staggering of the anisotropy direction in the \abplane\ plane. The model consists of an elliptic \bcplane-spiral with moments $M_b$=3.9\,\mubohr\ and $M_c$=2.8\,\mubohr , $\vec{k}$=(0,~2/7,~0), and interaction parameters $J_{\textrm{AFM}}=\SI{0.82}{\meV}$, $J_{\textrm{FM}}=\SI{-0.38}{\meV}$, $J_{\textrm{NN}}=\SI{0.31}{\meV}$, \textit{SIA}$_\parallel$=\SI{-0.18}{\meV} and \textit{SIA}$_c$=\SI{-0.09}{\meV} (Model M-III). The overall dispersion is unchanged by the staggering of the anisotropy. Minor additional features are visible, which may arise due to computational limitations. To introduce the staggered anisotropy, the symmetry of the structure had to be set to $P1$, which strongly increases the number of parameters for the calculation.

In a second step, the anisotropy term along \caxis\ has been removed and the DM interaction activated. The model has the following parameters: elliptic \bcplane \ spiral with moments $M_b=3.9$~\mubohr\ and $M_c=2.8$~\mubohr\ and $\vec{k}=(0,~2/7,~0)$, $J_{\textrm{AFM}}=\SI{0.82}{\meV}$, $J_{\textrm{FM}}=\SI{-0.38}{\meV}$, $J_{\textrm{NN}}=\SI{0.31}{\meV}$, \textit{SIA}$_b=\SI{-0.10}{\meV}$, \textit{SIA}$_c=\SI{0}{\meV}$, $DM_c$\ =\ (0.64,\ -0.2,\ 0)\,meV. In the presence of the $DM_c$ interaction, the moments become alternately modulated~\cite{Mochizuki2009a}. To account for this effect, the moments were tilted \SI{\pm 4}{\degree} around the $a$-axis.
The calculated intensity is shown in Fig.~\ref{fig:tbmno-spinW_mochizuki}(b) (model M-IV).

\begin{figure}[!t]
    \centering
    \includegraphics[width=0.99\columnwidth]{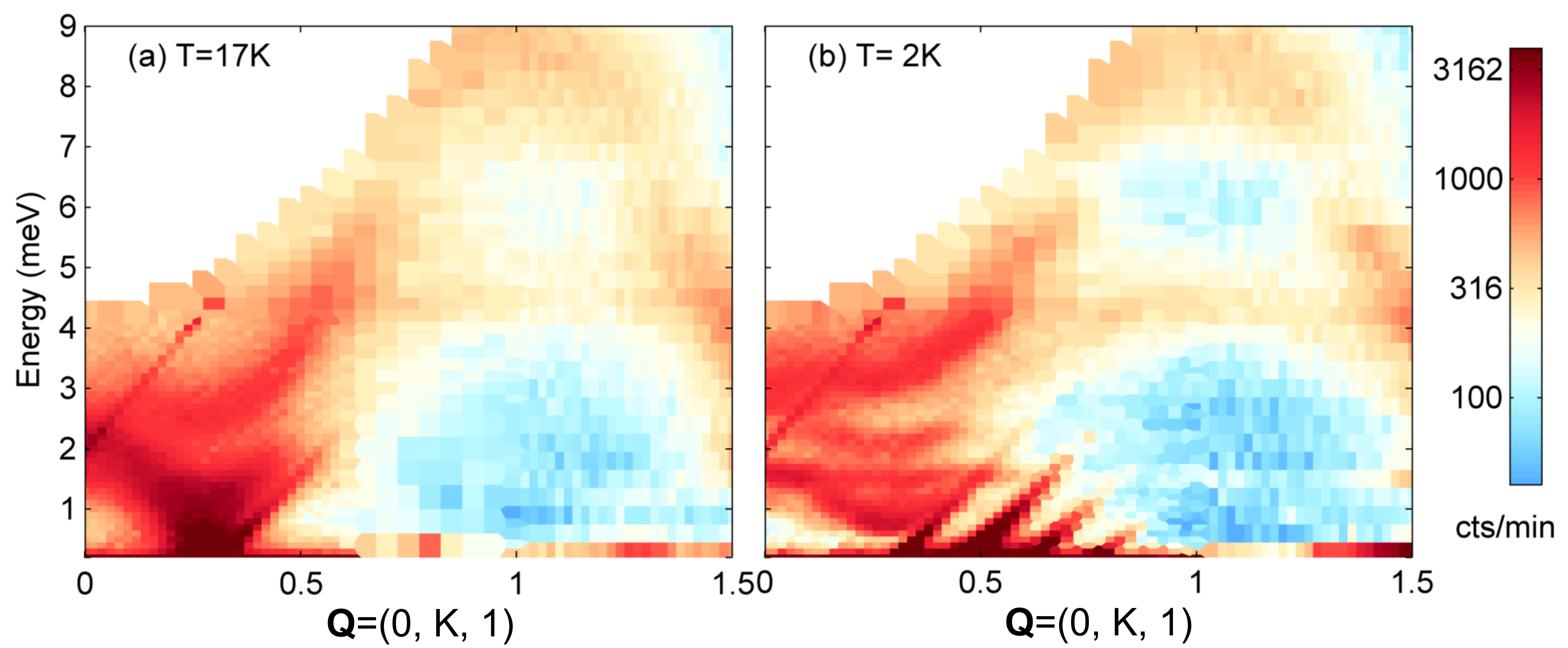}
    \caption{Inelastic neutron intensity of \tbmno\ along $\vec{Q}=(0,~k,~1)$ at $T=\SI{17}{\kelvin}$ (a) and $T=\SI{2}{\kelvin}$ (b) at IN14/Thales. The intensity is logarithmically color-coded. The sharp diagonal features visible at both temperatures can be attributed to spurious signals arising from strong Bragg peak scattering.}
    \label{fig:tbmno-tb_1}
\end{figure}

\begin{figure}[!t]
    \centering
    \includegraphics[width=0.99\columnwidth]{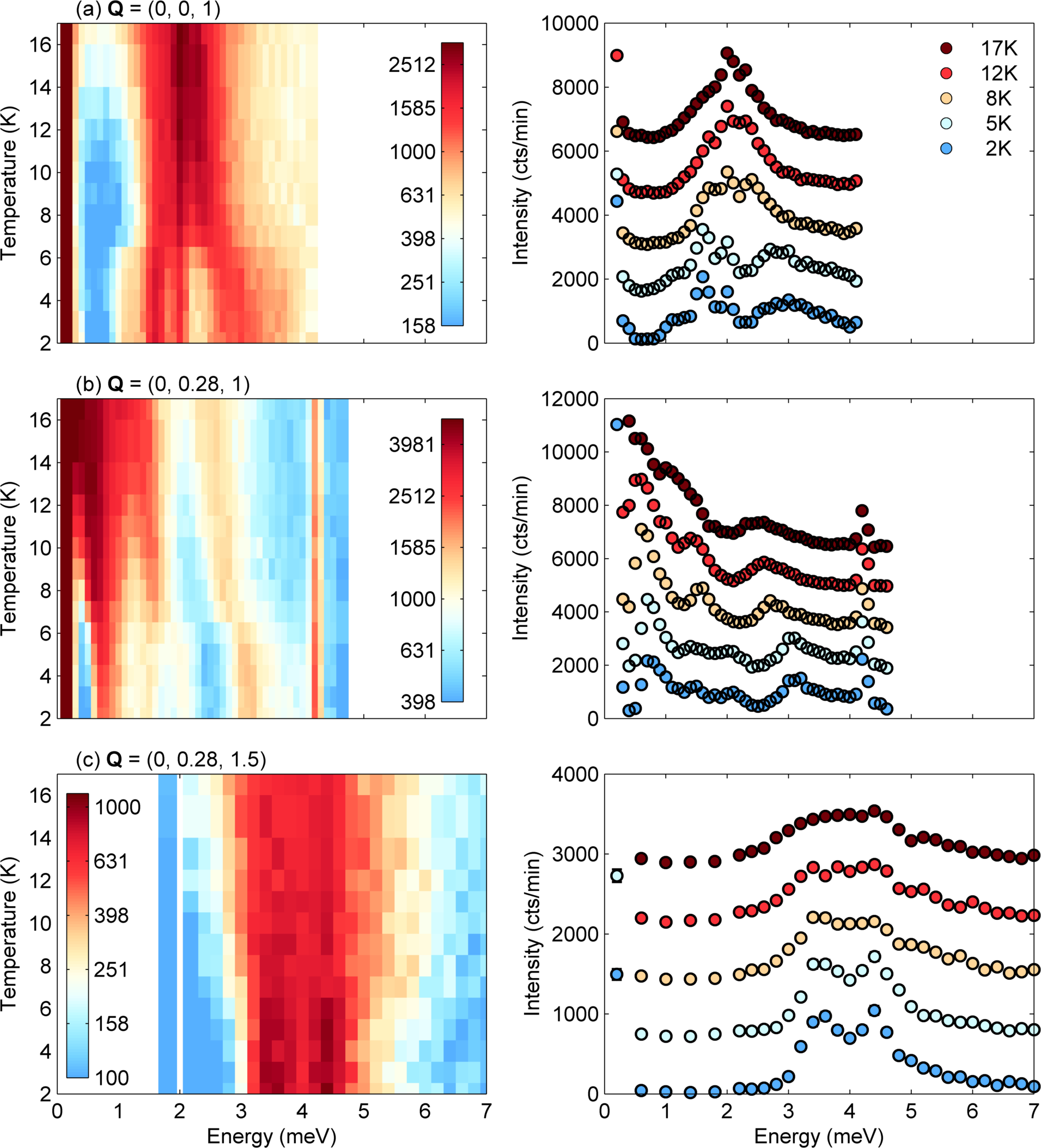}
    \caption{Temperature dependence of INS of \tbmno\ at (a) $\vec{Q}=(0,~0,~1)$, (b) $\vec{Q}=(0,~0.28,~1)$ and (c) $\vec{Q}=(0,~0.28,~1.5)$ at IN14/Thales. The intensity is logarithmically color-coded. Selected scans corresponding to the color map are shown on the right side. }
    \label{fig:tb-impact-2}
\end{figure}

The result again reproduces the main features of the previous calculations. 
The calculation with model M-IV confirms the assumption that the \bcplane-spin cycloid can be stabilized by the combination of an easy-axis anisotropy and a DM interaction. However, this model yields a too low
frequency for the main scattering signal at the scattering vector (0,0,1). 
In the presented calculation, a very strong antisymmetric exchange of $DM_c=(0.64,-0.2,~0)$\,\si{\meV} is required to reproduce the experimental dispersion. The strength of the DM interaction seems to be overestimated for a relativistic effect induced by spin-orbit coupling~\cite[p. 144]{book_khomskii}. The $DM$ values are eight times stronger than the values experimentally determined for \lamno\ ($DM_c\approx(0.08,-0.025,~0)$\,\si{\meV})~\cite{Deisenhofer2002}.

In order to estimate the strength of the DM interaction in TbMnO$_3$, one can analyze the canting of the magnetic structure induced by the antisymmetric terms. As explained in the introduction, the two irreducible representations involved in the cycloidal structure allow for various canting modes in the longitudinal spin-density phase and in the multiferroic state. 
In the SDW phase $G_x$ and $F_z$ modes may occur
and in the multiferroic phase also $C_x$ and $F_y$ become allowed. 
Since all phases are modulated the $F_{y,z}$ modes do not generate macroscopic polarization. The Tb moments can contribute to $F_z$, $C_x$ and $F_y$ modes. 
Evidence for the admixture of such components was found already in the first neutron diffraction experiments \cite{Quezel1977,Blasco2000,Kajimoto2004}, and resonant X-ray diffraction at the Mn $L_{2,3}$ edge clearly confirms such effects \cite{Wilkins2009,Jang2011}. 
However, so far no quantitative determination of the canting angles has been reported.
In the neutron single-crystal diffraction experiments \cite{Kajimoto2004}, it is not possible to compare the dominating $A$-type magnetic scattering with the much weaker components because the strong principal magnetic diffraction intensities seem to suffer from extinction
or detector limitations. Furthermore, it is difficult  in general to quantitatively interpret resonant magnetic X-ray diffraction intensities. 
In the neutron experiments the $G_x$ component seems to be the strongest admixture \cite{Blasco2000,Kajimoto2004} and it furthermore purely arises from the Mn moments.
Taking the Lorentz- and geometry factor into account, the powder neutron-diffraction data indicate a $G_x$ canting of $\sim$6 degrees \cite{Blasco2000}, while the other canting angles must be considerably smaller.
The value of the DM interaction required to stabilize this $a,b$ canting is obtained by classically calculating the exchange and single-ion energy as function of the canting angle
with the parameters of model M-II plus a DM term.
The minimum yields a DM interaction of roughly 10\% \ of $J_{AFM}$, i.e. 0.08\,meV, which agrees with the DM values reported for LaMnO$_3$~\cite{Deisenhofer2002}.
The other DM terms in TbMnO$_3$ should be significantly smaller. 
A more precise determination of all the canted admixtures can be realized with polarized neutron diffraction and is desirable. 
 
In a phenomenological approach  Mostovoy discussed the influence of Tb anisotropies and the coupling to the Mn moments~\cite{Mostovoy2006}, while Tb moments have not been taken into account in any of the models described above. The introduction of Tb moments into the calculations presents a considerable enhancement of complexity, since anisotropic interactions between Tb and Mn moments have to be considered. In view of the good agreement with the data taken at 17\,K we did not extend our model in this direction.

Summarizing these calculations, the spin waves in the spiral phase of \tbmno\ have been modeled assuming three different approaches extending the most simple isotropic model M-I. Their main difference consists in the stabilization of the cycloid along the $bc$ plane: Model M-II assumes an orthorhombic single-ion anisotropy, model M-III bases on staggered anisotropy, and in model M-IV the spiral is stabilized by an antisymmetric exchange along \caxis. The comparison of the models M-II, M-III and M-IV shows, that the principal features of the dispersion can be reproduced equally well by all models, but model M-IV including a large DM term is slightly less suited. 
It may appear astonishing that DM terms are not needed to reproduce the dispersion
and the dynamic chirality in this multiferroic material, whose ferroelectric polarization clearly is implied through such an interaction. However, this polarization is tiny while the dynamic chirality essentially results from the frustrated interaction parameters.

\section{Influence of the Tb subsystem}

The Tb subsystem orders at $T_{\textrm{Tb}}=\SI{7}{\kelvin}$ with a propagation vector of $\vec{k}_{\textrm{Tb}}=(0,~0.42,~0)$ and with moments aligned parallel to the $a$ direction. The $RE$ magnetic ordering has a clear impact on the multiferroic phase~\cite{Goto2005} and Kajimoto \etal\ observed the development of additional Mn structures (G, C and F type) which arise due to the Tb subsystem~\cite{Kajimoto2004}. Furthermore, Voigt \etal\ reported the induced ordering of $4f$ moments in the multiferroic phase~\cite{Voigt2007}. The investigation of the magnon dispersion in \tbmno\ has so far been performed at \SI{17}{\kelvin} or higher temperatures in order to avoid a stronger influence from the $RE$ ions~\cite{Senff2008a}. Nevertheless, even at this temperature Tb moments have an impact on the magnetic excitations.

Figure~\ref{fig:tbmno-tb_1} compares the excitation spectrum in \tbmno\ along $\vec{Q}=(0,~k,~1)$ at \SI{17}{\kelvin}, in the multiferroic phase, and at \SI{2}{\kelvin}, below $T_{\textrm{Tb}}$. The sharp features can be attributed to spurious scattering of the magnetic Bragg reflections. Usually these spurious effects are eliminated using a Be-filter on $k_f$, which was not available during the experiment. At the higher temperature, only two spurions from the Mn Bragg reflections are visible and at the lower temperature also the Tb Bragg reflections are visible. The overall features of the dispersion are recovered below $T_{\textrm{Tb}}$ but especially at the zone center significant changes occur. All modes shift to higher energies and at least one additional mode can be identified at 1.5\,meV. 
In addition modes become sharper at 2\,K supporting our conclusion that the interaction between Mn and Tb moments causes considerable magnon broadening already far above $T_{\textrm{Tb}}$ in the multiferroic phase.
We verified that the measured line width of the magnetic excitations are intrinsic and did not sharpen with higher instrument resolution at $k_f$=1.20\,\AA$^{-1}$.

The changes of the spectra are studied in more detail in Fig.~\ref{fig:tb-impact-2} (a)-(c).
The temperature dependencies in the range of \SIrange{2}{17}{\kelvin} are given for energy scans at (a) $\vec{Q}=(0,~0,~1)$, (b) $\vec{Q}=(0,~0.28,~1)$ and (c) $\vec{Q}=(0,~0.28,~1.5)$. First we look at the magnetic zone center in panel (b).
A peak appears at a finite but still low energy (at \SI{2}{\kelvin}; well below the onset of Tb ordering this peaks lies at about \SI{0.7}{\meV}). The smooth temperature dependence enables us to interpret this low-energy peak as the phason mode which acquires finite energy due to enhanced pinning at low temperatures. Furthermore, the position of the mode at \SI{2}{\kelvin} agrees well to an AFM resonance mode at $\SI{5}{\per\cm}=\SI{0.625}{\meV}$ measured in IR spectroscopy~\cite{Pimenov2009,Shuvaev2011}.
The hardening of the phason energy indicates significant Mn-Tb coupling and suggests that Tb moments also contribute to the pinning of multiferroic domain walls \cite{Stein2021}.
At higher temperatures the phason mode shifts to smaller energies and moves out of the energy range accessible by IR. The two $a$-polarized modes, i.e. the electromagnons with energies of 1.1 and 2.5\,meV at 17\,K, considerably harden upon cooling in perfect agreement with the IR studies and earlier experiments~\cite{Pimenov2009,Shuvaev2011,Holbein2015}.
The upper mode shows a kink at the inset of the Tb order, and near the lower mode there is an additional feature appearing that can be also seen at $\vec{Q}=(0,~0,~1)$ and that reaches 1.5\,meV at 2\,K, cf. Fig.~\ref{fig:tb-impact-2} (a). The temperature and scattering-vector dependencies do not
suggest a crystal-field excitation as the origin, but such an interpretation has been raised for the
feature at 4.5\,meV~\cite{Senff2007,Kajimoto2005}.
The data taken at (0,~0.28,~1.5) also confirm the sharpening of magnons with the onset of Tb order; here a broad signal changes into two well separated peaks with a high-energy shoulder, cf. Fig.~\ref{fig:tb-impact-2} (c).

\section{Conclusion}

The combination of various INS experiments with linear spin-wave theory calculations yields a rather complete picture of the magnetic excitations in multiferroic TbMnO$_3$.
In order to analyze the overall dispersion of magnons, data covering a wide range in reciprocal space and energy are essential, because the magnon dispersion in the incommensurate cycloid phase is complex, even when the Tb moments are fully neglected. The incommensurate character of the order results in a folding of branches and thus in a superposition of many modes.
The spiral magnetic arrangement yields an essential difference between modes polarized perpendicular to the spiral plane and modes polarized within this plane, because static and dynamic moments strongly interfere only in the in-plane case.
This splitting differs from the usual impact of single-ion anisotropy in a simple AFM material that mostly affects low-energy modes.
The experimental analysis of the splitting between in-plane and out-of-plane modes, however, requires neutron polarization analysis to distinguish individual signals.
The spin-wave calculations basing on the rotating reference system yield a very good description of the entire data sets.
The model well describes the dispersion of the different magnons and their
intensity distribution in the INS experiments.

Also the dynamic chirality of the magnon modes is perfectly reproduced by the model. 
Even at the zone-center one finds inelastic chiral contributions
of both signs that however simply arise from geometry effects. 
Measurements with different signs of large external electric fields fully confirm
the picture that the dynamic magnon chirality is imprinted by the chirality of the static order. 
Inelastic chirality can even be controlled above
the onset of long-range multiferroic order in zero field.

The new experiments and the modeling fully agree with the previously discussed electromagnon character of modes at the incommensurate zone center or in the Brillouin zone. Hybridization of phonons and
magnons resulting in electromagnon features can arise from different mechanisms as previously discussed. In particular the polarization of the magnon modes at the incommensurate zone-center is perfectly reproduced by the model.

The theoretical models studied differ in the treatment of single-ion anisotropy and in the inclusion of DM terms, which are essential to stabilize the
different magnetic structures as function of temperature and magnetic field.
However, the overall spin-wave dispersion covering energies up to $\sim$8\,meV is similar in various models
and differences in the calculated spin-wave energies remain small.

The magnetic excitations in the multiferroic phase at T=17\,K are considerably broadened, which can be at least partially attributed to the interaction with disordered Tb moments. Indeed modes
become sharper upon cooling below the onset of ordering of Tb moments, but the overall dispersion also becomes more complex due to the appearance of additional
features. The strong Tb-Mn interaction must also be relevant for the pinning and the relaxation of multiferroic domains~\cite{Stein2021}.

\begin{figure}[!t]
    \includegraphics[width=0.95\columnwidth]{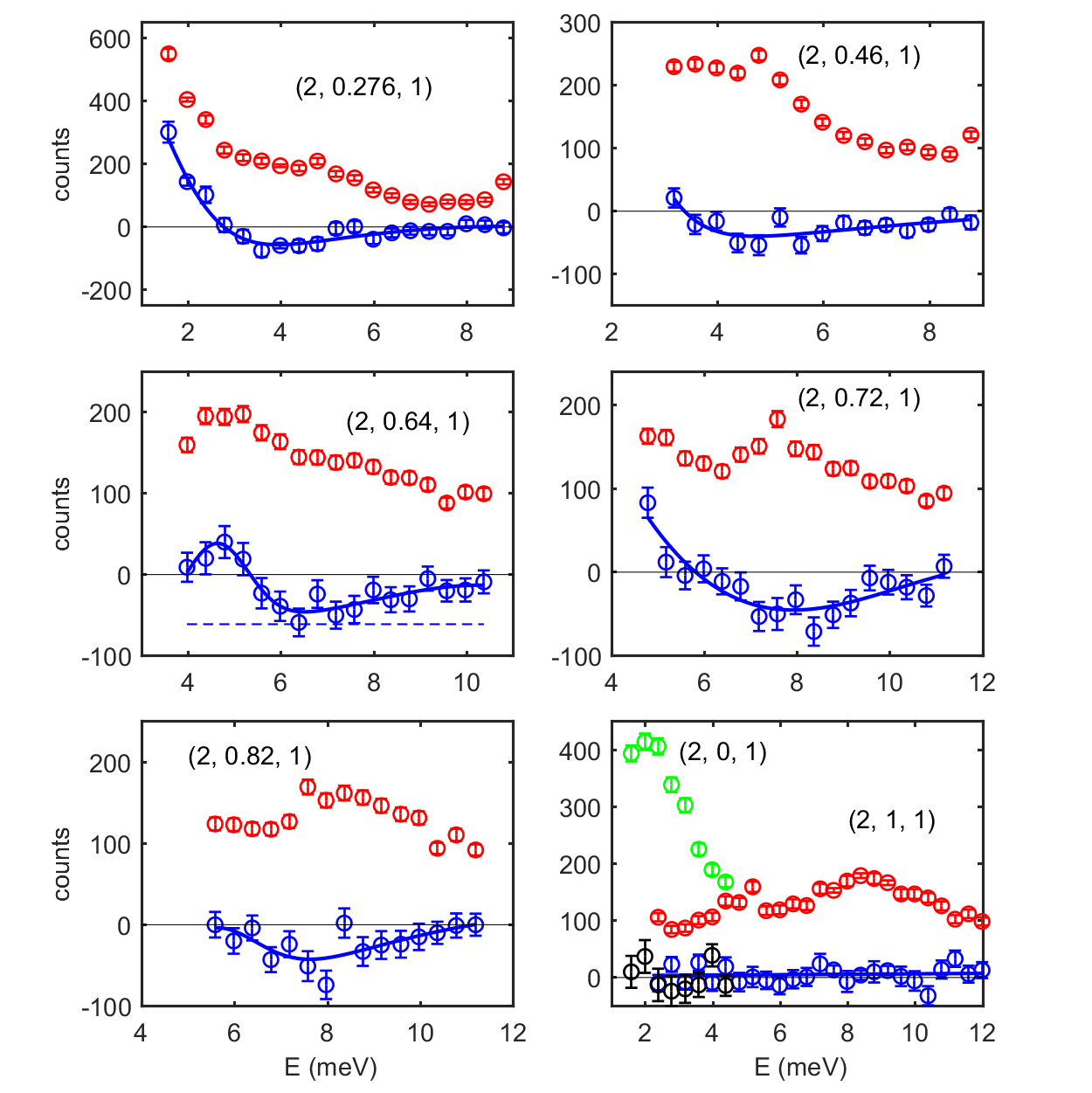}
    \caption{Results of polarized INS experiments on the thermal TAS IN20 performed at a temperature of T=17\,K. The red data points show the
    sum of $I_{x\bar{x}}$ and $I_{\bar{x}x}$ and the blue data points the difference that is proportional
    to the chiral contribution for (2, $\xi$, 1) with $\xi$ varying from 0.27 to 1. The corresponding data obtained at (2, 0, 1) is
    added to the lower right panel in green and black, respectively.}
    \label{fig:in20-all}
\end{figure}

\begin{figure}[!t]
    \includegraphics[width=0.95\columnwidth]{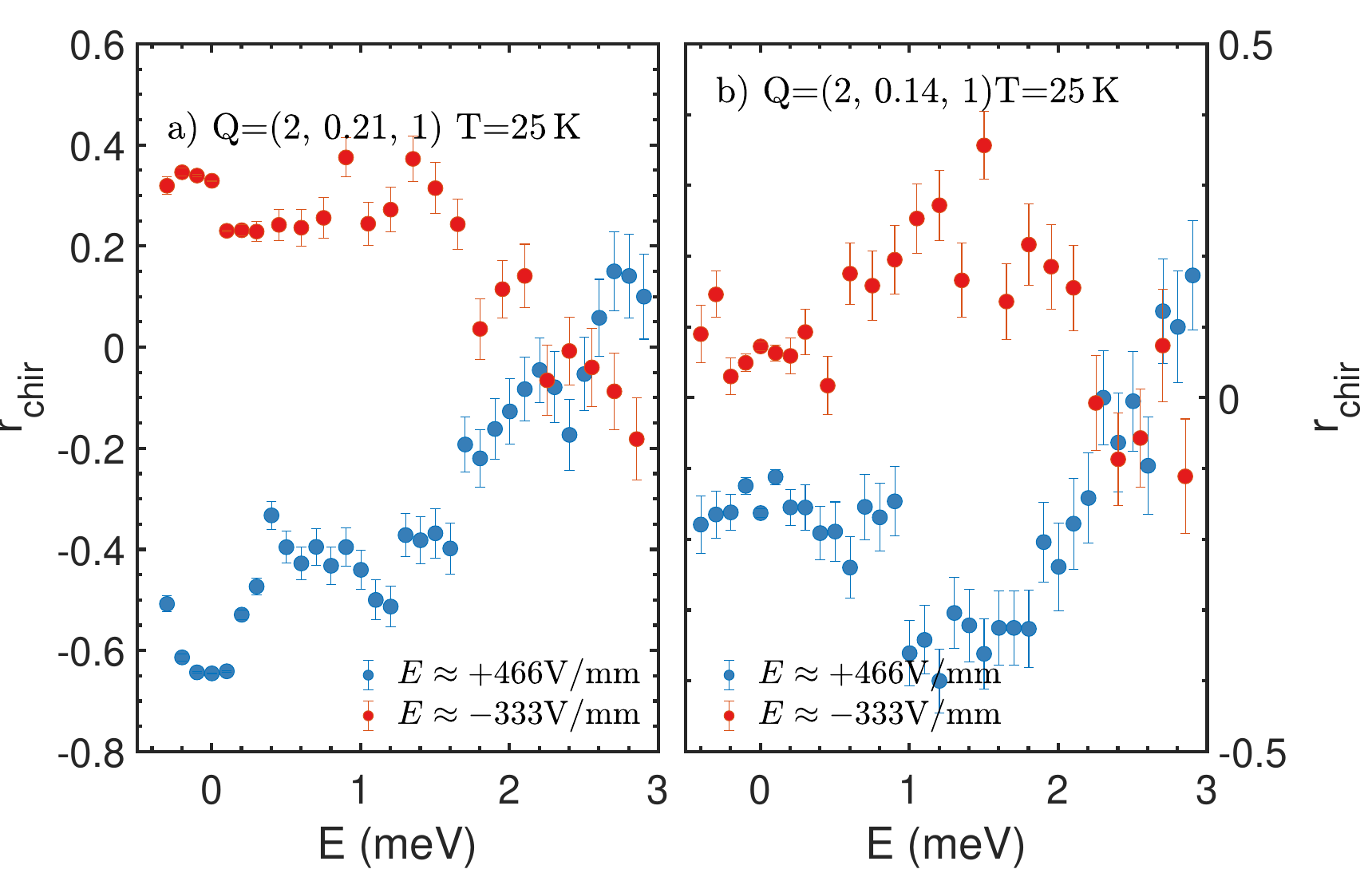}
    \caption{Chiral ratio calculated from polarized INS data taken on IN14/Thales. The neutron polarization was analyzed along the scattering
    vector ($x$ axis) and the crystal was poled in positive and negative electric fields. At both scattering vectors one sees the chiral signal stemming from
    the phason mode and first evidence for the opposite chiral contributions at higher energies. All chiral signs become inverted by the
    external electric field.}
    \label{fig:phason-switch}
\end{figure}

\section{Appendix}

The geometrical angles in the polarized neutron study that are relevant for the interpretation are summarized in Table~\ref{tab:tbmno-chiral}.

\begin{table}[!b]
	\begin{ruledtabular}
	\begin{tabular}{c c c c c }
\toprule %
   & $\vec{Q}$ (r.l.u.) & $\angle$(\qvec,\aaxis) & $\angle$(\qvec,\baxis) & $\angle$(\qvec,\caxis)  \\
\hline
$x$ & $(-2.00,-0.28,-1.00)$ & \SI{20.8}{\degree} & \SI{83.2}{\degree} & \SI{70.4}{\degree} \\
$y$ & $(\phantom{-}0.56,-5.44,\phantom{-}0.28)$ & \SI{83.5}{\degree} & \SI{6.8}{\degree} & \SI{87.7}{\degree} \\
$z$ & $(\phantom{-}1.00,\phantom{-}0.00,-3.90)$ & \SI{70.3}{\degree} & \SI{90}{\degree} & \SI{19.7}{\degree} \\
    \end{tabular}
	\end{ruledtabular}
    \caption{Coordinate system in a polarized neutron experiment at $\vec{Q}=-(2,~0.28,~1)$ in \tbmno\ mounted in $[201]/[010]$ geometry.}
    \label{tab:tbmno-chiral}
\end{table}

On IN20 data similar to those shown in Fig.~\ref{fig:tbmno-chiral_03} were taken for several scattering vectors (2, $k$, 1) covering the entire Brillouin
zone. The results are presented in Fig.~\ref{fig:in20-all}. The red points give the sum of the intensities in both spin-flip INS channels, as one would
observe in an unpolarized experiment. In contrast the difference data (shown in blue) correspond to the chiral component and clearly
illustrate the presence of different signs. Just at the commensurate scattering vectors (2, 0, 1) and (2, 1, 1) there is no chiral component in perfect agreement
with the model calculations. From the chiral signals at larger energy we obtain the dispersion of 
the in-plane polarized modes shown in
Fig.~\ref{fig:tbmno-chiral_03} (b). This dispersion exhibits a maximum at (2, 0.72, 1) and not at the zone boundary (2, 1, 1) due to the interference between the static
and dynamic $bc$ moments. The experiment and the model perfectly agree concerning this peculiar but essentially geometrical effect.

During the second set of experiments on  IN14/Thales  with polarization analysis we studied the 
possibility to invert the
chiral signs of inelastic magnon scattering, as  illustrated in Fig.~\ref{fig:tbmno-chiral_03}. Comparable data in the multiferroic phase were taken away from
the magnetic zone center at (2, 0.21 1) and (2, 0.14 1) and are displayed in Fig.~\ref{fig:phason-switch}.
Also for these scattering vectors the dynamic chirality can be inverted by changing the polarity of the external field.

~


 This work was funded by the Deutsche Forschungsgemeinschaft (DFG,
German Research Foundation) - Project number 277146847 - CRC 1238, projects A02 and B04.


\begin{thebibliography}{98}%
	\makeatletter
	\providecommand \@ifxundefined [1]{%
		\@ifx{#1\undefined}
	}%
	\providecommand \@ifnum [1]{%
		\ifnum #1\expandafter \@firstoftwo
		\else \expandafter \@secondoftwo
		\fi
	}%
	\providecommand \@ifx [1]{%
		\ifx #1\expandafter \@firstoftwo
		\else \expandafter \@secondoftwo
		\fi
	}%
	\providecommand \natexlab [1]{#1}%
	\providecommand \enquote  [1]{``#1''}%
	\providecommand \bibnamefont  [1]{#1}%
	\providecommand \bibfnamefont [1]{#1}%
	\providecommand \citenamefont [1]{#1}%
	\providecommand \href@noop [0]{\@secondoftwo}%
	\providecommand \href [0]{\begingroup \@sanitize@url \@href}%
	\providecommand \@href[1]{\@@startlink{#1}\@@href}%
	\providecommand \@@href[1]{\endgroup#1\@@endlink}%
	\providecommand \@sanitize@url [0]{\catcode `\\12\catcode `\$12\catcode
		`\&12\catcode `\#12\catcode `\^12\catcode `\_12\catcode `\%12\relax}%
	\providecommand \@@startlink[1]{}%
	\providecommand \@@endlink[0]{}%
	\providecommand \url  [0]{\begingroup\@sanitize@url \@url }%
	\providecommand \@url [1]{\endgroup\@href {#1}{\urlprefix }}%
	\providecommand \urlprefix  [0]{URL }%
	\providecommand \Eprint [0]{\href }%
	\providecommand \doibase [0]{https://doi.org/}%
	\providecommand \selectlanguage [0]{\@gobble}%
	\providecommand \bibinfo  [0]{\@secondoftwo}%
	\providecommand \bibfield  [0]{\@secondoftwo}%
	\providecommand \translation [1]{[#1]}%
	\providecommand \BibitemOpen [0]{}%
	\providecommand \bibitemStop [0]{}%
	\providecommand \bibitemNoStop [0]{.\EOS\space}%
	\providecommand \EOS [0]{\spacefactor3000\relax}%
	\providecommand \BibitemShut  [1]{\csname bibitem#1\endcsname}%
	\let\auto@bib@innerbib\@empty
	\bibitem [{\citenamefont {Kimura}\ \emph
		{et~al.}(2003{\natexlab{a}})\citenamefont {Kimura}, \citenamefont {Goto},
		\citenamefont {Shintani}, \citenamefont {Ishizaka}, \citenamefont {Arima},\
		and\ \citenamefont {Tokura}}]{Kimura2003}%
	\BibitemOpen
	\bibfield  {author} {\bibinfo {author} {\bibfnamefont {T.}~\bibnamefont
			{Kimura}}, \bibinfo {author} {\bibfnamefont {T.}~\bibnamefont {Goto}},
		\bibinfo {author} {\bibfnamefont {H.}~\bibnamefont {Shintani}}, \bibinfo
		{author} {\bibfnamefont {K.}~\bibnamefont {Ishizaka}}, \bibinfo {author}
		{\bibfnamefont {T.}~\bibnamefont {Arima}},\ and\ \bibinfo {author}
		{\bibfnamefont {Y.}~\bibnamefont {Tokura}},\ }\bibfield  {title} {\bibinfo
		{title} {{Magnetic control of ferroelectric polarization}},\ }\href
	{https://doi.org/10.1038/nature02018} {\bibfield  {journal} {\bibinfo
			{journal} {Nature}\ }\textbf {\bibinfo {volume} {426}},\ \bibinfo {pages}
		{55} (\bibinfo {year} {2003}{\natexlab{a}})}\BibitemShut {NoStop}%
	\bibitem [{\citenamefont {Spaldin}\ and\ \citenamefont
		{Ramesh}(2019)}]{Spaldin2019}%
	\BibitemOpen
	\bibfield  {author} {\bibinfo {author} {\bibfnamefont {N.~A.}\ \bibnamefont
			{Spaldin}}\ and\ \bibinfo {author} {\bibfnamefont {R.}~\bibnamefont
			{Ramesh}},\ }\bibfield  {title} {\bibinfo {title} {Advances in
			magnetoelectric multiferroics},\ }\href
	{https://doi.org/10.1038/s41563-018-0275-2} {\bibfield  {journal} {\bibinfo
			{journal} {Nature Materials}\ }\textbf {\bibinfo {volume} {18}},\ \bibinfo
		{pages} {203} (\bibinfo {year} {2019})}\BibitemShut {NoStop}%
	\bibitem [{\citenamefont {Fiebig}\ \emph {et~al.}(2016)\citenamefont {Fiebig},
		\citenamefont {Lottermoser}, \citenamefont {Meier},\ and\ \citenamefont
		{Trassin}}]{Fiebig2016}%
	\BibitemOpen
	\bibfield  {author} {\bibinfo {author} {\bibfnamefont {M.}~\bibnamefont
			{Fiebig}}, \bibinfo {author} {\bibfnamefont {T.}~\bibnamefont {Lottermoser}},
		\bibinfo {author} {\bibfnamefont {D.}~\bibnamefont {Meier}},\ and\ \bibinfo
		{author} {\bibfnamefont {M.}~\bibnamefont {Trassin}},\ }\bibfield  {title}
	{\bibinfo {title} {The evolution of multiferroics},\ }\href
	{https://doi.org/10.1038/natrevmats.2016.46} {\bibfield  {journal} {\bibinfo
			{journal} {Nature Reviews Materials}\ }\textbf {\bibinfo {volume} {1}},\
		\bibinfo {pages} {16046} (\bibinfo {year} {2016})}\BibitemShut {NoStop}%
	\bibitem [{\citenamefont {Goto}\ \emph
		{et~al.}(2004{\natexlab{a}})\citenamefont {Goto}, \citenamefont {Kimura},
		\citenamefont {Lawes}, \citenamefont {Ramirez},\ and\ \citenamefont
		{Tokura}}]{Goto2004}%
	\BibitemOpen
	\bibfield  {author} {\bibinfo {author} {\bibfnamefont {T.}~\bibnamefont
			{Goto}}, \bibinfo {author} {\bibfnamefont {T.}~\bibnamefont {Kimura}},
		\bibinfo {author} {\bibfnamefont {G.}~\bibnamefont {Lawes}}, \bibinfo
		{author} {\bibfnamefont {a.~P.}\ \bibnamefont {Ramirez}},\ and\ \bibinfo
		{author} {\bibfnamefont {Y.}~\bibnamefont {Tokura}},\ }\bibfield  {title}
	{\bibinfo {title} {{Ferroelectricity and Giant Magnetocapacitance in
				Perovskite Rare-Earth Manganites}},\ }\href
	{https://doi.org/10.1103/PhysRevLett.92.257201} {\bibfield  {journal}
		{\bibinfo  {journal} {Phys. Rev. Lett.}\ }\textbf {\bibinfo {volume} {92}},\
		\bibinfo {pages} {1} (\bibinfo {year} {2004}{\natexlab{a}})}\BibitemShut
	{NoStop}%
	\bibitem [{str()}]{struc-baum}%
	\BibitemOpen
	\href@noop {} {}\bibinfo {note} {{The crystal structure was studied by
			single-crystal X-ray diffraction on a small crystal cut from the rods grown
			in the mirror furnace.} The lattice parameters are $a$=5.3003(1)\,\AA,
		$b$=5.8532(1)\,\AA, and $c$=7.3987(1)\,\AA (spacegroup $Pbnm$); Tb at
		(0.98349(2),0.08138(2),1/4) $U_{\textrm{iso}}$=0.00342(3)\,\AA$^2$, Mn at
		(1/2,0,0) $U_{\textrm{iso}}$=0.00317(7), O1 at (0.1072(4),0.4660(3),1/4)
		$U_{\textrm{iso}}$=0.0047(4), O2 at (0.7036(3),0.3272(2),0.05145(17))
		$U_{\textrm{iso}}$=0.0048(2). The precision of the structure analysis
		underlines the high quality of the crystals; additional information can be
		found in reference \cite{diss-baum}.}\BibitemShut {Stop}%
	\bibitem [{\citenamefont {Jensen}\ and\ \citenamefont
		{Mackintosh}(1991)}]{Jensen1991}%
	\BibitemOpen
	\bibfield  {author} {\bibinfo {author} {\bibfnamefont {J.}~\bibnamefont
			{Jensen}}\ and\ \bibinfo {author} {\bibfnamefont {A.~R.}\ \bibnamefont
			{Mackintosh}},\ }\href
	{http://www.amazon.com/Rare-Earth-Magnetism-Excitations-International/dp/0198520271%3FSubscriptionId%3D0JYN1NVW651KCA56C102%26tag%3Dtechkie-20%26linkCode%3Dxm2%26camp%3D2025%26creative%3D165953%26creativeASIN%3D0198520271}
	{\emph {\bibinfo {title} {Rare Earth Magnetism: Structures and Excitations
				(The International Series of Monographs on Physics)}}},\ edited by\ \bibinfo
	{editor} {\bibfnamefont {J.}~\bibnamefont {Birman}}, \bibinfo {editor}
	{\bibfnamefont {S.~F.}\ \bibnamefont {Edwards}}, \bibinfo {editor}
	{\bibfnamefont {C.~H.~L.}\ \bibnamefont {Smith}},\ and\ \bibinfo {editor}
	{\bibfnamefont {M.}~\bibnamefont {Rees}}\ (\bibinfo  {publisher} {Clarendon
		Press},\ \bibinfo {year} {1991})\BibitemShut {NoStop}%
	\bibitem [{\citenamefont {Kimura}\ \emph
		{et~al.}(2003{\natexlab{b}})\citenamefont {Kimura}, \citenamefont {Ishihara},
		\citenamefont {Shintani}, \citenamefont {Arima}, \citenamefont {Takahashi},
		\citenamefont {Ishizaka},\ and\ \citenamefont {Tokura}}]{Kimura2003a}%
	\BibitemOpen
	\bibfield  {author} {\bibinfo {author} {\bibfnamefont {T.}~\bibnamefont
			{Kimura}}, \bibinfo {author} {\bibfnamefont {S.}~\bibnamefont {Ishihara}},
		\bibinfo {author} {\bibfnamefont {H.}~\bibnamefont {Shintani}}, \bibinfo
		{author} {\bibfnamefont {T.}~\bibnamefont {Arima}}, \bibinfo {author}
		{\bibfnamefont {K.}~\bibnamefont {Takahashi}}, \bibinfo {author}
		{\bibfnamefont {K.}~\bibnamefont {Ishizaka}},\ and\ \bibinfo {author}
		{\bibfnamefont {Y.}~\bibnamefont {Tokura}},\ }\bibfield  {title} {\bibinfo
		{title} {{Distorted perovskite with e$_{\text{g}}^{\text{1}}$ configuration
				as a frustrated spin system}},\ }\href
	{https://doi.org/10.1103/PhysRevB.68.060403} {\bibfield  {journal} {\bibinfo
			{journal} {Phys. Rev. B}\ }\textbf {\bibinfo {volume} {68}},\ \bibinfo
		{pages} {060403} (\bibinfo {year} {2003}{\natexlab{b}})}\BibitemShut
	{NoStop}%
	\bibitem [{\citenamefont {Blasco}\ \emph {et~al.}(2000)\citenamefont {Blasco},
		\citenamefont {Ritter}, \citenamefont {Garc\'{\i}a}, \citenamefont
		{de~Teresa}, \citenamefont {P\'{e}rez-Cacho},\ and\ \citenamefont
		{Ibarra}}]{Blasco2000}%
	\BibitemOpen
	\bibfield  {author} {\bibinfo {author} {\bibfnamefont {J.}~\bibnamefont
			{Blasco}}, \bibinfo {author} {\bibfnamefont {C.}~\bibnamefont {Ritter}},
		\bibinfo {author} {\bibfnamefont {J.}~\bibnamefont {Garc\'{\i}a}}, \bibinfo
		{author} {\bibfnamefont {J.}~\bibnamefont {de~Teresa}}, \bibinfo {author}
		{\bibfnamefont {J.}~\bibnamefont {P\'{e}rez-Cacho}},\ and\ \bibinfo {author}
		{\bibfnamefont {M.}~\bibnamefont {Ibarra}},\ }\bibfield  {title} {\bibinfo
		{title} {{Structural and magnetic study of
				Tb$_{\text{1-x}}$Ca$_{\text{x}}$MnO$_{\text{3}}$ perovskites}},\ }\href
	{https://doi.org/10.1103/PhysRevB.62.5609} {\bibfield  {journal} {\bibinfo
			{journal} {Phys. Rev. B}\ }\textbf {\bibinfo {volume} {62}},\ \bibinfo
		{pages} {5609} (\bibinfo {year} {2000})}\BibitemShut {NoStop}%
	\bibitem [{\citenamefont {Rodr\'iguez-Carvajal}\ \emph
		{et~al.}(1998)\citenamefont {Rodr\'iguez-Carvajal}, \citenamefont {Hennion},
		\citenamefont {Moussa}, \citenamefont {Moudden}, \citenamefont {Pinsard},\
		and\ \citenamefont {Revcolevschi}}]{Rodriguez-Carvajal1998}%
	\BibitemOpen
	\bibfield  {author} {\bibinfo {author} {\bibfnamefont {J.}~\bibnamefont
			{Rodr\'iguez-Carvajal}}, \bibinfo {author} {\bibfnamefont {M.}~\bibnamefont
			{Hennion}}, \bibinfo {author} {\bibfnamefont {F.}~\bibnamefont {Moussa}},
		\bibinfo {author} {\bibfnamefont {A.~H.}\ \bibnamefont {Moudden}}, \bibinfo
		{author} {\bibfnamefont {L.}~\bibnamefont {Pinsard}},\ and\ \bibinfo {author}
		{\bibfnamefont {A.}~\bibnamefont {Revcolevschi}},\ }\bibfield  {title}
	{\bibinfo {title} {{Neutron-diffraction study of the Jahn-Teller transition
				in stoichiometric LaMnO$_3$}},\ }\href
	{https://doi.org/10.1103/physrevb.57.r3189} {\bibfield  {journal} {\bibinfo
			{journal} {Phys. Rev. B}\ }\textbf {\bibinfo {volume} {57}},\ \bibinfo
		{pages} {R3189} (\bibinfo {year} {1998})}\BibitemShut {NoStop}%
	\bibitem [{\citenamefont {Kajimoto}\ \emph {et~al.}(2005)\citenamefont
		{Kajimoto}, \citenamefont {Mochizuki}, \citenamefont {Yoshizawa},
		\citenamefont {Shintani}, \citenamefont {Kimura},\ and\ \citenamefont
		{Tokura}}]{Kajimoto2005}%
	\BibitemOpen
	\bibfield  {author} {\bibinfo {author} {\bibfnamefont {R.}~\bibnamefont
			{Kajimoto}}, \bibinfo {author} {\bibfnamefont {H.}~\bibnamefont {Mochizuki}},
		\bibinfo {author} {\bibfnamefont {H.}~\bibnamefont {Yoshizawa}}, \bibinfo
		{author} {\bibfnamefont {H.}~\bibnamefont {Shintani}}, \bibinfo {author}
		{\bibfnamefont {T.}~\bibnamefont {Kimura}},\ and\ \bibinfo {author}
		{\bibfnamefont {Y.}~\bibnamefont {Tokura}},\ }\bibfield  {title} {\bibinfo
		{title} {{R -Dependence of Spin Exchange Interactions in RMnO$_{\text{3}}$ (R
				= Rare-Earth Ions)}},\ }\href {https://doi.org/10.1143/JPSJ.74.2430}
	{\bibfield  {journal} {\bibinfo  {journal} {J. Phys. Soc. Japan}\ }\textbf
		{\bibinfo {volume} {74}},\ \bibinfo {pages} {2430} (\bibinfo {year}
		{2005})}\BibitemShut {NoStop}%
	\bibitem [{\citenamefont {Goodenough}(1955)}]{Goodenough1955}%
	\BibitemOpen
	\bibfield  {author} {\bibinfo {author} {\bibfnamefont {J.~B.}\ \bibnamefont
			{Goodenough}},\ }\bibfield  {title} {\bibinfo {title} {{Theory of the Role of
				Covalence in the Perovskite-Type Manganites [La, M(II)]MnO$_3$}},\ }\href
	{https://doi.org/10.1103/physrev.100.564} {\bibfield  {journal} {\bibinfo
			{journal} {Phys. Rev.}\ }\textbf {\bibinfo {volume} {100}},\ \bibinfo {pages}
		{564} (\bibinfo {year} {1955})}\BibitemShut {NoStop}%
	\bibitem [{\citenamefont {Mochizuki}\ and\ \citenamefont
		{Furukawa}(2009{\natexlab{a}})}]{Mochizuki2009a}%
	\BibitemOpen
	\bibfield  {author} {\bibinfo {author} {\bibfnamefont {M.}~\bibnamefont
			{Mochizuki}}\ and\ \bibinfo {author} {\bibfnamefont {N.}~\bibnamefont
			{Furukawa}},\ }\bibfield  {title} {\bibinfo {title} {Microscopic model and
			phase diagrams of the multiferroic perovskite manganites},\ }\href
	{https://doi.org/10.1103/physrevb.80.134416} {\bibfield  {journal} {\bibinfo
			{journal} {Phys. Rev. B}\ }\textbf {\bibinfo {volume} {80}},\ \bibinfo
		{pages} {134416} (\bibinfo {year} {2009}{\natexlab{a}})}\BibitemShut
	{NoStop}%
	\bibitem [{Fon(2015)}]{Fontcuberta2015}%
	\BibitemOpen
	\bibfield  {title} {\bibinfo {title} {{Multiferroic $R$MnO$_3$ thin films}},\
	}\href@noop {} {\bibfield  {journal} {\bibinfo  {journal} {Comptes Rendus
				Physique}\ }\textbf {\bibinfo {volume} {16}},\ \bibinfo {pages} {204}
		(\bibinfo {year} {2015})}\BibitemShut {NoStop}%
	\bibitem [{\citenamefont {Ishiwata}\ \emph {et~al.}(2010)\citenamefont
		{Ishiwata}, \citenamefont {Kaneko}, \citenamefont {Tokunaga}, \citenamefont
		{Taguchi}, \citenamefont {Arima},\ and\ \citenamefont
		{Tokura}}]{Ishiwata2010}%
	\BibitemOpen
	\bibfield  {author} {\bibinfo {author} {\bibfnamefont {S.}~\bibnamefont
			{Ishiwata}}, \bibinfo {author} {\bibfnamefont {Y.}~\bibnamefont {Kaneko}},
		\bibinfo {author} {\bibfnamefont {Y.}~\bibnamefont {Tokunaga}}, \bibinfo
		{author} {\bibfnamefont {Y.}~\bibnamefont {Taguchi}}, \bibinfo {author}
		{\bibfnamefont {T.-h.}\ \bibnamefont {Arima}},\ and\ \bibinfo {author}
		{\bibfnamefont {Y.}~\bibnamefont {Tokura}},\ }\bibfield  {title} {\bibinfo
		{title} {Perovskite manganites hosting versatile multiferroic phases with
			symmetric and antisymmetric exchange strictions},\ }\href
	{https://doi.org/10.1103/PhysRevB.81.100411} {\bibfield  {journal} {\bibinfo
			{journal} {Phys. Rev. B}\ }\textbf {\bibinfo {volume} {81}},\ \bibinfo
		{pages} {100411} (\bibinfo {year} {2010})}\BibitemShut {NoStop}%
	\bibitem [{\citenamefont {Quezel}\ \emph {et~al.}(1977)\citenamefont {Quezel},
		\citenamefont {Tcheou}, \citenamefont {Rossat-Mignod}, \citenamefont
		{Quezel},\ and\ \citenamefont {Roudaut}}]{Quezel1977}%
	\BibitemOpen
	\bibfield  {author} {\bibinfo {author} {\bibfnamefont {S.}~\bibnamefont
			{Quezel}}, \bibinfo {author} {\bibfnamefont {F.}~\bibnamefont {Tcheou}},
		\bibinfo {author} {\bibfnamefont {J.}~\bibnamefont {Rossat-Mignod}}, \bibinfo
		{author} {\bibfnamefont {G.}~\bibnamefont {Quezel}},\ and\ \bibinfo {author}
		{\bibfnamefont {E.}~\bibnamefont {Roudaut}},\ }\bibfield  {title} {\bibinfo
		{title} {{Magnetic structure of the perovskite-like compound
				TbMnO$_{\text{3}}$}},\ }\href@noop {} {\bibfield  {journal} {\bibinfo
			{journal} {Physica}\ }\textbf {\bibinfo {volume} {86}},\ \bibinfo {pages}
		{916} (\bibinfo {year} {1977})}\BibitemShut {NoStop}%
	\bibitem [{\citenamefont {Bertaut}(1968)}]{Bertaut1968}%
	\BibitemOpen
	\bibfield  {author} {\bibinfo {author} {\bibfnamefont {E.~F.}\ \bibnamefont
			{Bertaut}},\ }\bibfield  {title} {\bibinfo {title} {Representation analysis
			of magnetic structures},\ }\href {https://doi.org/10.1107/s0567739468000306}
	{\bibfield  {journal} {\bibinfo  {journal} {Acta Cryst.}\ }\textbf {\bibinfo
			{volume} {A24}},\ \bibinfo {pages} {217} (\bibinfo {year}
		{1968})}\BibitemShut {NoStop}%
	\bibitem [{\citenamefont {Aliouane}\ \emph {et~al.}(2008)\citenamefont
		{Aliouane}, \citenamefont {Prokhnenko}, \citenamefont {Feyerherm},
		\citenamefont {Mostovoy}, \citenamefont {Strempfer}, \citenamefont {Habicht},
		\citenamefont {Rule}, \citenamefont {Dudzik}, \citenamefont {Wolter},
		\citenamefont {Maljuk},\ and\ \citenamefont {Argyriou}}]{Aliouane2008}%
	\BibitemOpen
	\bibfield  {author} {\bibinfo {author} {\bibfnamefont {N.}~\bibnamefont
			{Aliouane}}, \bibinfo {author} {\bibfnamefont {O.}~\bibnamefont
			{Prokhnenko}}, \bibinfo {author} {\bibfnamefont {R.}~\bibnamefont
			{Feyerherm}}, \bibinfo {author} {\bibfnamefont {M.}~\bibnamefont {Mostovoy}},
		\bibinfo {author} {\bibfnamefont {J.}~\bibnamefont {Strempfer}}, \bibinfo
		{author} {\bibfnamefont {K.}~\bibnamefont {Habicht}}, \bibinfo {author}
		{\bibfnamefont {K.~C.}\ \bibnamefont {Rule}}, \bibinfo {author}
		{\bibfnamefont {E.}~\bibnamefont {Dudzik}}, \bibinfo {author} {\bibfnamefont
			{A.~U.~B.}\ \bibnamefont {Wolter}}, \bibinfo {author} {\bibfnamefont
			{A.}~\bibnamefont {Maljuk}},\ and\ \bibinfo {author} {\bibfnamefont {D.~N.}\
			\bibnamefont {Argyriou}},\ }\bibfield  {title} {\bibinfo {title} {{Magnetic
				order and ferroelectricity in RMnO$_{\text{3}}$ multiferroic manganites:
				coupling between R- and Mn-spins}},\ }\href
	{https://doi.org/10.1088/0953-8984/20/43/434215} {\bibfield  {journal}
		{\bibinfo  {journal} {J. Phys.: Condens. Matter}\ }\textbf {\bibinfo {volume}
			{20}},\ \bibinfo {pages} {434215} (\bibinfo {year} {2008})}\BibitemShut
	{NoStop}%
	\bibitem [{\citenamefont {Kenzelmann}\ \emph {et~al.}(2005)\citenamefont
		{Kenzelmann}, \citenamefont {Harris}, \citenamefont {Jonas}, \citenamefont
		{Broholm}, \citenamefont {Schefer}, \citenamefont {Kim}, \citenamefont
		{Zhang}, \citenamefont {Cheong}, \citenamefont {Vajk},\ and\ \citenamefont
		{Lynn}}]{Kenzelmann2005}%
	\BibitemOpen
	\bibfield  {author} {\bibinfo {author} {\bibfnamefont {M.}~\bibnamefont
			{Kenzelmann}}, \bibinfo {author} {\bibfnamefont {A.}~\bibnamefont {Harris}},
		\bibinfo {author} {\bibfnamefont {S.}~\bibnamefont {Jonas}}, \bibinfo
		{author} {\bibfnamefont {C.}~\bibnamefont {Broholm}}, \bibinfo {author}
		{\bibfnamefont {J.}~\bibnamefont {Schefer}}, \bibinfo {author} {\bibfnamefont
			{S.}~\bibnamefont {Kim}}, \bibinfo {author} {\bibfnamefont {C.}~\bibnamefont
			{Zhang}}, \bibinfo {author} {\bibfnamefont {S.-W.}\ \bibnamefont {Cheong}},
		\bibinfo {author} {\bibfnamefont {O.}~\bibnamefont {Vajk}},\ and\ \bibinfo
		{author} {\bibfnamefont {J.}~\bibnamefont {Lynn}},\ }\bibfield  {title}
	{\bibinfo {title} {{Magnetic Inversion Symmetry Breaking and Ferroelectricity
				in TbMnO$_{\text{3}}$}},\ }\href
	{https://doi.org/10.1103/PhysRevLett.95.087206} {\bibfield  {journal}
		{\bibinfo  {journal} {Phys. Rev. Lett.}\ }\textbf {\bibinfo {volume} {95}},\
		\bibinfo {pages} {27} (\bibinfo {year} {2005})}\BibitemShut {NoStop}%
	\bibitem [{\citenamefont {Stein}\ \emph {et~al.}(2017)\citenamefont {Stein},
		\citenamefont {Baum}, \citenamefont {Holbein}, \citenamefont {Finger},
		\citenamefont {Cronert}, \citenamefont {T\"olzer}, \citenamefont
		{Fr\"ohlich}, \citenamefont {Biesenkamp}, \citenamefont {Schmalzl},
		\citenamefont {Steffens}, \citenamefont {Lee},\ and\ \citenamefont
		{Braden}}]{Stein2017}%
	\BibitemOpen
	\bibfield  {author} {\bibinfo {author} {\bibfnamefont {J.}~\bibnamefont
			{Stein}}, \bibinfo {author} {\bibfnamefont {M.}~\bibnamefont {Baum}},
		\bibinfo {author} {\bibfnamefont {S.}~\bibnamefont {Holbein}}, \bibinfo
		{author} {\bibfnamefont {T.}~\bibnamefont {Finger}}, \bibinfo {author}
		{\bibfnamefont {T.}~\bibnamefont {Cronert}}, \bibinfo {author} {\bibfnamefont
			{C.}~\bibnamefont {T\"olzer}}, \bibinfo {author} {\bibfnamefont
			{T.}~\bibnamefont {Fr\"ohlich}}, \bibinfo {author} {\bibfnamefont
			{S.}~\bibnamefont {Biesenkamp}}, \bibinfo {author} {\bibfnamefont
			{K.}~\bibnamefont {Schmalzl}}, \bibinfo {author} {\bibfnamefont
			{P.}~\bibnamefont {Steffens}}, \bibinfo {author} {\bibfnamefont {C.~H.}\
			\bibnamefont {Lee}},\ and\ \bibinfo {author} {\bibfnamefont {M.}~\bibnamefont
			{Braden}},\ }\bibfield  {title} {\bibinfo {title} {{Control of Chiral
				Magnetism Through Electric Fields in Multiferroic Compounds above the
				Long-Range Multiferroic Transition}},\ }\href
	{https://doi.org/10.1103/PhysRevLett.119.177201} {\bibfield  {journal}
		{\bibinfo  {journal} {Phys. Rev. Lett.}\ }\textbf {\bibinfo {volume} {119}},\
		\bibinfo {pages} {177201} (\bibinfo {year} {2017})}\BibitemShut {NoStop}%
	\bibitem [{\citenamefont {Meier}\ \emph {et~al.}(2007)\citenamefont {Meier},
		\citenamefont {Aliouane}, \citenamefont {Argyriou}, \citenamefont {Mydosh},\
		and\ \citenamefont {Lorenz}}]{Meier2007}%
	\BibitemOpen
	\bibfield  {author} {\bibinfo {author} {\bibfnamefont {D.}~\bibnamefont
			{Meier}}, \bibinfo {author} {\bibfnamefont {N.}~\bibnamefont {Aliouane}},
		\bibinfo {author} {\bibfnamefont {D.~N.}\ \bibnamefont {Argyriou}}, \bibinfo
		{author} {\bibfnamefont {J.~a.}\ \bibnamefont {Mydosh}},\ and\ \bibinfo
		{author} {\bibfnamefont {T.}~\bibnamefont {Lorenz}},\ }\bibfield  {title}
	{\bibinfo {title} {{New features in the phase diagram of
				TbMnO$_{\text{3}}$}},\ }\href {https://doi.org/10.1088/1367-2630/9/4/100}
	{\bibfield  {journal} {\bibinfo  {journal} {New J. Phys.}\ }\textbf {\bibinfo
			{volume} {9}},\ \bibinfo {pages} {100} (\bibinfo {year} {2007})}\BibitemShut
	{NoStop}%
	\bibitem [{\citenamefont {Kimura}\ \emph {et~al.}(2005)\citenamefont {Kimura},
		\citenamefont {Lawes}, \citenamefont {Goto}, \citenamefont {Tokura},\ and\
		\citenamefont {Ramirez}}]{Kimura2005}%
	\BibitemOpen
	\bibfield  {author} {\bibinfo {author} {\bibfnamefont {T.}~\bibnamefont
			{Kimura}}, \bibinfo {author} {\bibfnamefont {G.}~\bibnamefont {Lawes}},
		\bibinfo {author} {\bibfnamefont {T.}~\bibnamefont {Goto}}, \bibinfo {author}
		{\bibfnamefont {Y.}~\bibnamefont {Tokura}},\ and\ \bibinfo {author}
		{\bibfnamefont {a.}~\bibnamefont {Ramirez}},\ }\bibfield  {title} {\bibinfo
		{title} {{Magnetoelectric phase diagrams of orthorhombic RMnO$_{\text{3}}$
				(R=Gd, Tb, and Dy)}},\ }\href {https://doi.org/10.1103/PhysRevB.71.224425}
	{\bibfield  {journal} {\bibinfo  {journal} {Phys. Rev. B}\ }\textbf {\bibinfo
			{volume} {71}},\ \bibinfo {pages} {1} (\bibinfo {year} {2005})}\BibitemShut
	{NoStop}%
	\bibitem [{\citenamefont {Senff}\ \emph {et~al.}(2007)\citenamefont {Senff},
		\citenamefont {Link}, \citenamefont {Hradil}, \citenamefont {Hiess},
		\citenamefont {Regnault}, \citenamefont {Sidis}, \citenamefont {Aliouane},
		\citenamefont {Argyriou},\ and\ \citenamefont {Braden}}]{Senff2007}%
	\BibitemOpen
	\bibfield  {author} {\bibinfo {author} {\bibfnamefont {D.}~\bibnamefont
			{Senff}}, \bibinfo {author} {\bibfnamefont {P.}~\bibnamefont {Link}},
		\bibinfo {author} {\bibfnamefont {K.}~\bibnamefont {Hradil}}, \bibinfo
		{author} {\bibfnamefont {A.}~\bibnamefont {Hiess}}, \bibinfo {author}
		{\bibfnamefont {L.~P.}\ \bibnamefont {Regnault}}, \bibinfo {author}
		{\bibfnamefont {Y.}~\bibnamefont {Sidis}}, \bibinfo {author} {\bibfnamefont
			{N.}~\bibnamefont {Aliouane}}, \bibinfo {author} {\bibfnamefont {D.~N.}\
			\bibnamefont {Argyriou}},\ and\ \bibinfo {author} {\bibfnamefont
			{M.}~\bibnamefont {Braden}},\ }\bibfield  {title} {\bibinfo {title}
		{{Magnetic Excitations in Multiferroic TbMnO$_{\text{3}}$: Evidence for a
				Hybridized Soft Mode}},\ }\href
	{https://doi.org/10.1103/PhysRevLett.98.137206} {\bibfield  {journal}
		{\bibinfo  {journal} {Phys. Rev. Lett.}\ }\textbf {\bibinfo {volume} {98}},\
		\bibinfo {pages} {28} (\bibinfo {year} {2007})}\BibitemShut {NoStop}%
	\bibitem [{\citenamefont {{Vald\'es Aguilar}}\ \emph
		{et~al.}(2009)\citenamefont {{Vald\'es Aguilar}}, \citenamefont {Mostovoy},
		\citenamefont {Sushkov}, \citenamefont {Zhang}, \citenamefont {Choi},
		\citenamefont {Cheong},\ and\ \citenamefont {Drew}}]{ValdesAguilar2009}%
	\BibitemOpen
	\bibfield  {author} {\bibinfo {author} {\bibfnamefont {R.}~\bibnamefont
			{{Vald\'es Aguilar}}}, \bibinfo {author} {\bibfnamefont {M.}~\bibnamefont
			{Mostovoy}}, \bibinfo {author} {\bibfnamefont {A.}~\bibnamefont {Sushkov}},
		\bibinfo {author} {\bibfnamefont {C.}~\bibnamefont {Zhang}}, \bibinfo
		{author} {\bibfnamefont {Y.}~\bibnamefont {Choi}}, \bibinfo {author}
		{\bibfnamefont {S.-W.}\ \bibnamefont {Cheong}},\ and\ \bibinfo {author}
		{\bibfnamefont {H.}~\bibnamefont {Drew}},\ }\bibfield  {title} {\bibinfo
		{title} {{Origin of Electromagnon Excitations in Multiferroic RMnO$_3$}},\
	}\href {https://doi.org/10.1103/PhysRevLett.102.047203} {\bibfield  {journal}
		{\bibinfo  {journal} {Phys. Rev. Lett.}\ }\textbf {\bibinfo {volume} {102}},\
		\bibinfo {pages} {047203} (\bibinfo {year} {2009})}\BibitemShut {NoStop}%
	\bibitem [{\citenamefont {Finger}\ \emph {et~al.}(2014)\citenamefont {Finger},
		\citenamefont {Binder}, \citenamefont {Sidis}, \citenamefont {Maljuk},
		\citenamefont {Argyriou},\ and\ \citenamefont {Braden}}]{Finger2014}%
	\BibitemOpen
	\bibfield  {author} {\bibinfo {author} {\bibfnamefont {T.}~\bibnamefont
			{Finger}}, \bibinfo {author} {\bibfnamefont {K.}~\bibnamefont {Binder}},
		\bibinfo {author} {\bibfnamefont {Y.}~\bibnamefont {Sidis}}, \bibinfo
		{author} {\bibfnamefont {A.}~\bibnamefont {Maljuk}}, \bibinfo {author}
		{\bibfnamefont {D.~N.}\ \bibnamefont {Argyriou}},\ and\ \bibinfo {author}
		{\bibfnamefont {M.}~\bibnamefont {Braden}},\ }\bibfield  {title} {\bibinfo
		{title} {{Magnetic order and electromagnon excitations in DyMnO$_{\text{3}}$
				studied by neutron scattering experiments}},\ }\href
	{https://doi.org/10.1103/physrevb.90.224418} {\bibfield  {journal} {\bibinfo
			{journal} {Phys. Rev. B}\ }\textbf {\bibinfo {volume} {90}},\ \bibinfo
		{pages} {224418} (\bibinfo {year} {2014})}\BibitemShut {NoStop}%
	\bibitem [{\citenamefont {Kajimoto}\ \emph {et~al.}(2004)\citenamefont
		{Kajimoto}, \citenamefont {Yoshizawa}, \citenamefont {Shintani},
		\citenamefont {Kimura},\ and\ \citenamefont {Tokura}}]{Kajimoto2004}%
	\BibitemOpen
	\bibfield  {author} {\bibinfo {author} {\bibfnamefont {R.}~\bibnamefont
			{Kajimoto}}, \bibinfo {author} {\bibfnamefont {H.}~\bibnamefont {Yoshizawa}},
		\bibinfo {author} {\bibfnamefont {H.}~\bibnamefont {Shintani}}, \bibinfo
		{author} {\bibfnamefont {T.}~\bibnamefont {Kimura}},\ and\ \bibinfo {author}
		{\bibfnamefont {Y.}~\bibnamefont {Tokura}},\ }\bibfield  {title} {\bibinfo
		{title} {{Magnetic structure of TbMnO$_{\text{3}}$ by neutron diffraction}},\
	}\href {https://doi.org/10.1103/PhysRevB.70.012401} {\bibfield  {journal}
		{\bibinfo  {journal} {Phys. Rev. B}\ }\textbf {\bibinfo {volume} {70}},\
		\bibinfo {pages} {012401} (\bibinfo {year} {2004})}\BibitemShut {NoStop}%
	\bibitem [{\citenamefont {Aliouane}\ \emph {et~al.}(2009)\citenamefont
		{Aliouane}, \citenamefont {Schmalzl}, \citenamefont {Senff}, \citenamefont
		{Maljuk}, \citenamefont {Proke\v{s}}, \citenamefont {Braden},\ and\
		\citenamefont {Argyriou}}]{Aliouane2009}%
	\BibitemOpen
	\bibfield  {author} {\bibinfo {author} {\bibfnamefont {N.}~\bibnamefont
			{Aliouane}}, \bibinfo {author} {\bibfnamefont {K.}~\bibnamefont {Schmalzl}},
		\bibinfo {author} {\bibfnamefont {D.}~\bibnamefont {Senff}}, \bibinfo
		{author} {\bibfnamefont {A.}~\bibnamefont {Maljuk}}, \bibinfo {author}
		{\bibfnamefont {K.}~\bibnamefont {Proke\v{s}}}, \bibinfo {author}
		{\bibfnamefont {M.}~\bibnamefont {Braden}},\ and\ \bibinfo {author}
		{\bibfnamefont {D.}~\bibnamefont {Argyriou}},\ }\bibfield  {title} {\bibinfo
		{title} {{Flop of Electric Polarization Driven by the Flop of the Mn Spin
				Cycloid in Multiferroic TbMnO$_{\text{3}}$}},\ }\href
	{https://doi.org/10.1103/PhysRevLett.102.207205} {\bibfield  {journal}
		{\bibinfo  {journal} {Phys. Rev. Lett.}\ }\textbf {\bibinfo {volume} {102}},\
		\bibinfo {pages} {207205} (\bibinfo {year} {2009})}\BibitemShut {NoStop}%
	\bibitem [{\citenamefont {Voigt}\ \emph {et~al.}(2007)\citenamefont {Voigt},
		\citenamefont {Persson}, \citenamefont {Kim}, \citenamefont {Bihlmayer},\
		and\ \citenamefont {Br\"uckel}}]{Voigt2007}%
	\BibitemOpen
	\bibfield  {author} {\bibinfo {author} {\bibfnamefont {J.}~\bibnamefont
			{Voigt}}, \bibinfo {author} {\bibfnamefont {J.}~\bibnamefont {Persson}},
		\bibinfo {author} {\bibfnamefont {J.~W.}\ \bibnamefont {Kim}}, \bibinfo
		{author} {\bibfnamefont {G.}~\bibnamefont {Bihlmayer}},\ and\ \bibinfo
		{author} {\bibfnamefont {T.}~\bibnamefont {Br\"uckel}},\ }\bibfield  {title}
	{\bibinfo {title} {{Strong coupling between the spin polarization of Mn and
				Tb in multiferroic TbMnO$_3$ determined by x-ray resonance exchange
				scattering}},\ }\href {https://doi.org/10.1103/physrevb.76.104431} {\bibfield
		{journal} {\bibinfo  {journal} {Phys. Rev. B}\ }\textbf {\bibinfo {volume}
			{76}},\ \bibinfo {pages} {104431} (\bibinfo {year} {2007})}\BibitemShut
	{NoStop}%
	\bibitem [{\citenamefont {Prokhnenko}\ \emph {et~al.}(2007)\citenamefont
		{Prokhnenko}, \citenamefont {Feyerherm}, \citenamefont {Mostovoy},
		\citenamefont {Aliouane}, \citenamefont {Dudzik}, \citenamefont {Wolter},
		\citenamefont {Maljuk},\ and\ \citenamefont {Argyriou}}]{Prokhnenko2007}%
	\BibitemOpen
	\bibfield  {author} {\bibinfo {author} {\bibfnamefont {O.}~\bibnamefont
			{Prokhnenko}}, \bibinfo {author} {\bibfnamefont {R.}~\bibnamefont
			{Feyerherm}}, \bibinfo {author} {\bibfnamefont {M.}~\bibnamefont {Mostovoy}},
		\bibinfo {author} {\bibfnamefont {N.}~\bibnamefont {Aliouane}}, \bibinfo
		{author} {\bibfnamefont {E.}~\bibnamefont {Dudzik}}, \bibinfo {author}
		{\bibfnamefont {A.~U.~B.}\ \bibnamefont {Wolter}}, \bibinfo {author}
		{\bibfnamefont {A.}~\bibnamefont {Maljuk}},\ and\ \bibinfo {author}
		{\bibfnamefont {D.~N.}\ \bibnamefont {Argyriou}},\ }\bibfield  {title}
	{\bibinfo {title} {{ Coupling of Frustrated Ising Spins to the Magnetic
				Cycloid in Multiferroic TbMnO$_3$ }},\ }\href
	{https://doi.org/10.1103/physrevlett.99.177206} {\bibfield  {journal}
		{\bibinfo  {journal} {Phys. Rev. Lett.}\ }\textbf {\bibinfo {volume} {99}},\
		\bibinfo {pages} {177206} (\bibinfo {year} {2007})}\BibitemShut {NoStop}%
	\bibitem [{\citenamefont {Katsura}\ \emph {et~al.}(2005)\citenamefont
		{Katsura}, \citenamefont {Nagaosa},\ and\ \citenamefont
		{Balatsky}}]{Katsura2005}%
	\BibitemOpen
	\bibfield  {author} {\bibinfo {author} {\bibfnamefont {H.}~\bibnamefont
			{Katsura}}, \bibinfo {author} {\bibfnamefont {N.}~\bibnamefont {Nagaosa}},\
		and\ \bibinfo {author} {\bibfnamefont {A.}~\bibnamefont {Balatsky}},\
	}\bibfield  {title} {\bibinfo {title} {Spin current and magnetoelectric
			effect in noncollinear magnets},\ }\href
	{https://doi.org/10.1103/PhysRevLett.95.057205} {\bibfield  {journal}
		{\bibinfo  {journal} {Phys. Rev. Lett.}\ }\textbf {\bibinfo {volume} {95}},\
		\bibinfo {pages} {57205} (\bibinfo {year} {2005})}\BibitemShut {NoStop}%
	\bibitem [{\citenamefont {Sergienko}\ and\ \citenamefont
		{Dagotto}(2006)}]{Sergienko2006}%
	\BibitemOpen
	\bibfield  {author} {\bibinfo {author} {\bibfnamefont {I.}~\bibnamefont
			{Sergienko}}\ and\ \bibinfo {author} {\bibfnamefont {E.}~\bibnamefont
			{Dagotto}},\ }\bibfield  {title} {\bibinfo {title} {Role of the
			dzyaloshinskii-moriya interaction in multiferroic perovskites},\ }\href
	{https://doi.org/10.1103/PhysRevB.73.094434} {\bibfield  {journal} {\bibinfo
			{journal} {Phys. Rev. B}\ }\textbf {\bibinfo {volume} {73}},\ \bibinfo
		{pages} {094434} (\bibinfo {year} {2006})}\BibitemShut {NoStop}%
	\bibitem [{\citenamefont {Mostovoy}(2006)}]{Mostovoy2006}%
	\BibitemOpen
	\bibfield  {author} {\bibinfo {author} {\bibfnamefont {M.}~\bibnamefont
			{Mostovoy}},\ }\bibfield  {title} {\bibinfo {title} {Ferroelectricity in
			spiral magnets},\ }\href {https://doi.org/10.1103/PhysRevLett.96.067601}
	{\bibfield  {journal} {\bibinfo  {journal} {Phys. Rev. Lett.}\ }\textbf
		{\bibinfo {volume} {96}},\ \bibinfo {pages} {67601} (\bibinfo {year}
		{2006})}\BibitemShut {NoStop}%
	\bibitem [{\citenamefont {Noda}\ \emph {et~al.}(2006)\citenamefont {Noda},
		\citenamefont {Akaki}, \citenamefont {Kikuchi}, \citenamefont {Akahoshi},\
		and\ \citenamefont {Kuwahara}}]{Noda2006}%
	\BibitemOpen
	\bibfield  {author} {\bibinfo {author} {\bibfnamefont {K.}~\bibnamefont
			{Noda}}, \bibinfo {author} {\bibfnamefont {M.}~\bibnamefont {Akaki}},
		\bibinfo {author} {\bibfnamefont {T.}~\bibnamefont {Kikuchi}}, \bibinfo
		{author} {\bibfnamefont {D.}~\bibnamefont {Akahoshi}},\ and\ \bibinfo
		{author} {\bibfnamefont {H.}~\bibnamefont {Kuwahara}},\ }\bibfield  {title}
	{\bibinfo {title} {{Magnetic-field-induced switching between ferroelectric
				phases in orthorhombic-distortion-controlled RMnO$_3$}},\ }\href
	{https://doi.org/10.1063/1.2177207} {\bibfield  {journal} {\bibinfo
			{journal} {Journal of Applied Physics}\ }\textbf {\bibinfo {volume} {99}},\
		\bibinfo {pages} {08S905} (\bibinfo {year} {2006})}\BibitemShut {NoStop}%
	\bibitem [{\citenamefont {Arima}\ \emph {et~al.}(2006)\citenamefont {Arima},
		\citenamefont {Tokunaga}, \citenamefont {Goto}, \citenamefont {Kimura},
		\citenamefont {Noda},\ and\ \citenamefont {Tokura}}]{Arima2006}%
	\BibitemOpen
	\bibfield  {author} {\bibinfo {author} {\bibfnamefont {T.}~\bibnamefont
			{Arima}}, \bibinfo {author} {\bibfnamefont {A.}~\bibnamefont {Tokunaga}},
		\bibinfo {author} {\bibfnamefont {T.}~\bibnamefont {Goto}}, \bibinfo {author}
		{\bibfnamefont {H.}~\bibnamefont {Kimura}}, \bibinfo {author} {\bibfnamefont
			{Y.}~\bibnamefont {Noda}},\ and\ \bibinfo {author} {\bibfnamefont
			{Y.}~\bibnamefont {Tokura}},\ }\bibfield  {title} {\bibinfo {title}
		{{Collinear to Spiral Spin Transformation without Changing the Modulation
				Wavelength upon Ferroelectric Transition in Tb$_{1-x}$Dy$_x$MnO$_3$}},\
	}\href {https://doi.org/10.1103/physrevlett.96.097202} {\bibfield  {journal}
		{\bibinfo  {journal} {Phys. Rev. Lett.}\ }\textbf {\bibinfo {volume} {96}},\
		\bibinfo {pages} {097202} (\bibinfo {year} {2006})}\BibitemShut {NoStop}%
	\bibitem [{\citenamefont {Kuwahara}\ \emph {et~al.}(2009)\citenamefont
		{Kuwahara}, \citenamefont {Akaki}, \citenamefont {Tozawa}, \citenamefont
		{Hitomi}, \citenamefont {Noda},\ and\ \citenamefont
		{Akahoshi}}]{Kuwahara2009}%
	\BibitemOpen
	\bibfield  {author} {\bibinfo {author} {\bibfnamefont {H.}~\bibnamefont
			{Kuwahara}}, \bibinfo {author} {\bibfnamefont {M.}~\bibnamefont {Akaki}},
		\bibinfo {author} {\bibfnamefont {J.}~\bibnamefont {Tozawa}}, \bibinfo
		{author} {\bibfnamefont {M.}~\bibnamefont {Hitomi}}, \bibinfo {author}
		{\bibfnamefont {K.}~\bibnamefont {Noda}},\ and\ \bibinfo {author}
		{\bibfnamefont {D.}~\bibnamefont {Akahoshi}},\ }\bibfield  {title} {\bibinfo
		{title} {Persistent and reversible phase control in gdmno$_3$ near the phase
			boundary},\ }\href {https://doi.org/10.1088/1742-6596/150/4/042106}
	{\bibfield  {journal} {\bibinfo  {journal} {Journal of Physics: Conference
				Series}\ }\textbf {\bibinfo {volume} {150}},\ \bibinfo {pages} {042106}
		(\bibinfo {year} {2009})}\BibitemShut {NoStop}%
	\bibitem [{\citenamefont {Merz}(1953)}]{Merz1953}%
	\BibitemOpen
	\bibfield  {author} {\bibinfo {author} {\bibfnamefont {W.~J.}\ \bibnamefont
			{Merz}},\ }\bibfield  {title} {\bibinfo {title} {{Double Hysteresis Loop of
				BaTiO$_3$ at the Curie Point}},\ }\href
	{https://doi.org/10.1103/physrev.91.513} {\bibfield  {journal} {\bibinfo
			{journal} {Phys. Rev.}\ }\textbf {\bibinfo {volume} {91}},\ \bibinfo {pages}
		{513} (\bibinfo {year} {1953})}\BibitemShut {NoStop}%
	\bibitem [{\citenamefont {Yamasaki}\ \emph {et~al.}(2007)\citenamefont
		{Yamasaki}, \citenamefont {Sagayama}, \citenamefont {Goto}, \citenamefont
		{Matsuura}, \citenamefont {Hirota}, \citenamefont {Arima},\ and\
		\citenamefont {Tokura}}]{Yamasaki2007}%
	\BibitemOpen
	\bibfield  {author} {\bibinfo {author} {\bibfnamefont {Y.}~\bibnamefont
			{Yamasaki}}, \bibinfo {author} {\bibfnamefont {H.}~\bibnamefont {Sagayama}},
		\bibinfo {author} {\bibfnamefont {T.}~\bibnamefont {Goto}}, \bibinfo {author}
		{\bibfnamefont {M.}~\bibnamefont {Matsuura}}, \bibinfo {author}
		{\bibfnamefont {K.}~\bibnamefont {Hirota}}, \bibinfo {author} {\bibfnamefont
			{T.}~\bibnamefont {Arima}},\ and\ \bibinfo {author} {\bibfnamefont
			{Y.}~\bibnamefont {Tokura}},\ }\bibfield  {title} {\bibinfo {title}
		{{Electric Control of Spin Helicity in a Magnetic Ferroelectric}},\ }\href
	{https://doi.org/10.1103/PhysRevLett.98.147204} {\bibfield  {journal}
		{\bibinfo  {journal} {Phys. Rev. Lett.}\ }\textbf {\bibinfo {volume} {98}},\
		\bibinfo {pages} {1} (\bibinfo {year} {2007})}\BibitemShut {NoStop}%
	\bibitem [{\citenamefont {Senff}\ \emph
		{et~al.}(2008{\natexlab{a}})\citenamefont {Senff}, \citenamefont {Aliouane},
		\citenamefont {Argyriou}, \citenamefont {Hiess}, \citenamefont {Regnault},
		\citenamefont {Link}, \citenamefont {Hradil}, \citenamefont {Sidis},\ and\
		\citenamefont {Braden}}]{Senff2008a}%
	\BibitemOpen
	\bibfield  {author} {\bibinfo {author} {\bibfnamefont {D.}~\bibnamefont
			{Senff}}, \bibinfo {author} {\bibfnamefont {N.}~\bibnamefont {Aliouane}},
		\bibinfo {author} {\bibfnamefont {D.~N.}\ \bibnamefont {Argyriou}}, \bibinfo
		{author} {\bibfnamefont {A.}~\bibnamefont {Hiess}}, \bibinfo {author}
		{\bibfnamefont {L.~P.}\ \bibnamefont {Regnault}}, \bibinfo {author}
		{\bibfnamefont {P.}~\bibnamefont {Link}}, \bibinfo {author} {\bibfnamefont
			{K.}~\bibnamefont {Hradil}}, \bibinfo {author} {\bibfnamefont
			{Y.}~\bibnamefont {Sidis}},\ and\ \bibinfo {author} {\bibfnamefont
			{M.}~\bibnamefont {Braden}},\ }\bibfield  {title} {\bibinfo {title}
		{{Magnetic excitations in a cycloidal magnet: the magnon spectrum of
				multiferroic TbMnO$_{\text{3}}$}},\ }\href
	{https://doi.org/10.1088/0953-8984/20/43/434212} {\bibfield  {journal}
		{\bibinfo  {journal} {J. Phys.: Condens. Matter}\ }\textbf {\bibinfo {volume}
			{20}},\ \bibinfo {pages} {434212} (\bibinfo {year}
		{2008}{\natexlab{a}})}\BibitemShut {NoStop}%
	\bibitem [{\citenamefont {Milstein}\ and\ \citenamefont
		{Sushkov}(2015)}]{Milstein2015}%
	\BibitemOpen
	\bibfield  {author} {\bibinfo {author} {\bibfnamefont {A.~I.}\ \bibnamefont
			{Milstein}}\ and\ \bibinfo {author} {\bibfnamefont {O.~P.}\ \bibnamefont
			{Sushkov}},\ }\bibfield  {title} {\bibinfo {title} {{Magnetic excitations in
				the spin-spiral state of TbMnO$_{\text{3}}$ and DyMnO$_{\text{3}}$}},\ }\href
	{https://doi.org/10.1103/physrevb.91.094417} {\bibfield  {journal} {\bibinfo
			{journal} {Phys. Rev. B}\ }\textbf {\bibinfo {volume} {91}},\ \bibinfo
		{pages} {094417} (\bibinfo {year} {2015})}\BibitemShut {NoStop}%
	\bibitem [{\citenamefont {Goto}\ \emph
		{et~al.}(2004{\natexlab{b}})\citenamefont {Goto}, \citenamefont {Kimura},
		\citenamefont {Lawes}, \citenamefont {Ramirez},\ and\ \citenamefont
		{Tokura}}]{Kimura2004}%
	\BibitemOpen
	\bibfield  {author} {\bibinfo {author} {\bibfnamefont {T.}~\bibnamefont
			{Goto}}, \bibinfo {author} {\bibfnamefont {T.}~\bibnamefont {Kimura}},
		\bibinfo {author} {\bibfnamefont {G.}~\bibnamefont {Lawes}}, \bibinfo
		{author} {\bibfnamefont {A.~P.}\ \bibnamefont {Ramirez}},\ and\ \bibinfo
		{author} {\bibfnamefont {Y.}~\bibnamefont {Tokura}},\ }\bibfield  {title}
	{\bibinfo {title} {{Ferroelectricity and Giant Magnetocapacitance in
				Perovskite Rare-Earth Manganites}},\ }\href
	{https://doi.org/10.1103/PhysRevLett.92.257201} {\bibfield  {journal}
		{\bibinfo  {journal} {Phys. Rev. Lett.}\ }\textbf {\bibinfo {volume} {92}},\
		\bibinfo {pages} {257201} (\bibinfo {year} {2004}{\natexlab{b}})}\BibitemShut
	{NoStop}%
	\bibitem [{\citenamefont {Senff}\ \emph
		{et~al.}(2008{\natexlab{b}})\citenamefont {Senff}, \citenamefont {Link},
		\citenamefont {Aliouane}, \citenamefont {Argyriou},\ and\ \citenamefont
		{Braden}}]{Senff2008}%
	\BibitemOpen
	\bibfield  {author} {\bibinfo {author} {\bibfnamefont {D.}~\bibnamefont
			{Senff}}, \bibinfo {author} {\bibfnamefont {P.}~\bibnamefont {Link}},
		\bibinfo {author} {\bibfnamefont {N.}~\bibnamefont {Aliouane}}, \bibinfo
		{author} {\bibfnamefont {D.}~\bibnamefont {Argyriou}},\ and\ \bibinfo
		{author} {\bibfnamefont {M.}~\bibnamefont {Braden}},\ }\bibfield  {title}
	{\bibinfo {title} {{Field dependence of magnetic correlations through the
				polarization flop transition in multiferroic TbMnO$_{\text{3}}$: Evidence for
				a magnetic memory effect}},\ }\href
	{https://doi.org/10.1103/PhysRevB.77.174419} {\bibfield  {journal} {\bibinfo
			{journal} {Phys. Rev. B}\ }\textbf {\bibinfo {volume} {77}},\ \bibinfo
		{pages} {174419} (\bibinfo {year} {2008}{\natexlab{b}})}\BibitemShut
	{NoStop}%
	\bibitem [{\citenamefont {Holbein}\ \emph {et~al.}(2015)\citenamefont
		{Holbein}, \citenamefont {Steffens}, \citenamefont {Finger}, \citenamefont
		{Komarek}, \citenamefont {Sidis}, \citenamefont {Link},\ and\ \citenamefont
		{Braden}}]{Holbein2015}%
	\BibitemOpen
	\bibfield  {author} {\bibinfo {author} {\bibfnamefont {S.}~\bibnamefont
			{Holbein}}, \bibinfo {author} {\bibfnamefont {P.}~\bibnamefont {Steffens}},
		\bibinfo {author} {\bibfnamefont {T.}~\bibnamefont {Finger}}, \bibinfo
		{author} {\bibfnamefont {A.~C.}\ \bibnamefont {Komarek}}, \bibinfo {author}
		{\bibfnamefont {Y.}~\bibnamefont {Sidis}}, \bibinfo {author} {\bibfnamefont
			{P.}~\bibnamefont {Link}},\ and\ \bibinfo {author} {\bibfnamefont
			{M.}~\bibnamefont {Braden}},\ }\bibfield  {title} {\bibinfo {title} {{Field
				and temperature dependence of electromagnon scattering incTbMnO$_{\text{3}}$
				studied by inelastic neutron scattering}},\ }\href
	{https://doi.org/10.1103/physrevb.91.014432} {\bibfield  {journal} {\bibinfo
			{journal} {Phys. Rev. B}\ }\textbf {\bibinfo {volume} {91}},\ \bibinfo
		{pages} {014432} (\bibinfo {year} {2015})}\BibitemShut {NoStop}%
	\bibitem [{\citenamefont {Smolenskii}\ and\ \citenamefont
		{E.}(1982)}]{Smolenskii1982}%
	\BibitemOpen
	\bibfield  {author} {\bibinfo {author} {\bibfnamefont {G.~A.}\ \bibnamefont
			{Smolenskii}}\ and\ \bibinfo {author} {\bibfnamefont {C.~I.}\ \bibnamefont
			{E.}},\ }\bibfield  {title} {\bibinfo {title} {Ferroelectromagnets},\
	}\href@noop {} {\bibfield  {journal} {\bibinfo  {journal} {Soviet Physics
				Uspekhi}\ }\textbf {\bibinfo {volume} {25}},\ \bibinfo {pages} {475}
		(\bibinfo {year} {1982})}\BibitemShut {NoStop}%
	\bibitem [{\citenamefont {Pimenov}\ \emph {et~al.}(2006)\citenamefont
		{Pimenov}, \citenamefont {Mukhin}, \citenamefont {Ivanov}, \citenamefont
		{Travkin}, \citenamefont {Balbashov},\ and\ \citenamefont
		{Loidl}}]{Pimenov2006}%
	\BibitemOpen
	\bibfield  {author} {\bibinfo {author} {\bibfnamefont {A.}~\bibnamefont
			{Pimenov}}, \bibinfo {author} {\bibfnamefont {A.~A.}\ \bibnamefont {Mukhin}},
		\bibinfo {author} {\bibfnamefont {V.~Y.}\ \bibnamefont {Ivanov}}, \bibinfo
		{author} {\bibfnamefont {V.~D.}\ \bibnamefont {Travkin}}, \bibinfo {author}
		{\bibfnamefont {a.~M.}\ \bibnamefont {Balbashov}},\ and\ \bibinfo {author}
		{\bibfnamefont {A.}~\bibnamefont {Loidl}},\ }\bibfield  {title} {\bibinfo
		{title} {{Possible evidence for electromagnons in multiferroic manganites}},\
	}\href {https://doi.org/10.1038/nphys212} {\bibfield  {journal} {\bibinfo
			{journal} {Nat. Phys.}\ }\textbf {\bibinfo {volume} {2}},\ \bibinfo {pages}
		{97} (\bibinfo {year} {2006})}\BibitemShut {NoStop}%
	\bibitem [{\citenamefont {Pimenov}\ \emph {et~al.}(2009)\citenamefont
		{Pimenov}, \citenamefont {Shuvaev}, \citenamefont {Loidl}, \citenamefont
		{Schrettle}, \citenamefont {Mukhin}, \citenamefont {Travkin}, \citenamefont
		{Ivanov},\ and\ \citenamefont {Balbashov}}]{Pimenov2009}%
	\BibitemOpen
	\bibfield  {author} {\bibinfo {author} {\bibfnamefont {A.}~\bibnamefont
			{Pimenov}}, \bibinfo {author} {\bibfnamefont {A.}~\bibnamefont {Shuvaev}},
		\bibinfo {author} {\bibfnamefont {A.}~\bibnamefont {Loidl}}, \bibinfo
		{author} {\bibfnamefont {F.}~\bibnamefont {Schrettle}}, \bibinfo {author}
		{\bibfnamefont {A.}~\bibnamefont {Mukhin}}, \bibinfo {author} {\bibfnamefont
			{V.}~\bibnamefont {Travkin}}, \bibinfo {author} {\bibfnamefont
			{V.}~\bibnamefont {Ivanov}},\ and\ \bibinfo {author} {\bibfnamefont
			{A.}~\bibnamefont {Balbashov}},\ }\bibfield  {title} {\bibinfo {title}
		{{Magnetic and Magnetoelectric Excitations in TbMnO$_{\text{3}}$}},\ }\href
	{https://doi.org/10.1103/PhysRevLett.102.107203} {\bibfield  {journal}
		{\bibinfo  {journal} {Phys. Rev. Lett.}\ }\textbf {\bibinfo {volume} {102}},\
		\bibinfo {pages} {1} (\bibinfo {year} {2009})}\BibitemShut {NoStop}%
	\bibitem [{\citenamefont {Shuvaev}\ \emph {et~al.}(2011)\citenamefont
		{Shuvaev}, \citenamefont {Mukhin},\ and\ \citenamefont
		{Pimenov}}]{Shuvaev2011}%
	\BibitemOpen
	\bibfield  {author} {\bibinfo {author} {\bibfnamefont {A.~M.}\ \bibnamefont
			{Shuvaev}}, \bibinfo {author} {\bibfnamefont {A.~A.}\ \bibnamefont
			{Mukhin}},\ and\ \bibinfo {author} {\bibfnamefont {A.}~\bibnamefont
			{Pimenov}},\ }\bibfield  {title} {\bibinfo {title} {{Magnetic and
				magnetoelectric excitations in multiferroic manganites}},\ }\href
	{https://doi.org/10.1088/0953-8984/23/11/113201} {\bibfield  {journal}
		{\bibinfo  {journal} {J. Phys.: Condens. Matter}\ }\textbf {\bibinfo {volume}
			{23}},\ \bibinfo {pages} {113201} (\bibinfo {year} {2011})}\BibitemShut
	{NoStop}%
	\bibitem [{\citenamefont {Katsura}\ \emph {et~al.}(2007)\citenamefont
		{Katsura}, \citenamefont {Balatsky},\ and\ \citenamefont
		{Nagaosa}}]{Katsura2007}%
	\BibitemOpen
	\bibfield  {author} {\bibinfo {author} {\bibfnamefont {H.}~\bibnamefont
			{Katsura}}, \bibinfo {author} {\bibfnamefont {A.~V.}\ \bibnamefont
			{Balatsky}},\ and\ \bibinfo {author} {\bibfnamefont {N.}~\bibnamefont
			{Nagaosa}},\ }\bibfield  {title} {\bibinfo {title} {Dynamical magnetoelectric
			coupling in helical magnets},\ }\href
	{https://doi.org/10.1103/physrevlett.98.027203} {\bibfield  {journal}
		{\bibinfo  {journal} {Phys. Rev. Lett.}\ }\textbf {\bibinfo {volume} {98}},\
		\bibinfo {pages} {027203} (\bibinfo {year} {2007})}\BibitemShut {NoStop}%
	\bibitem [{\citenamefont {Shuvaev}\ \emph {et~al.}(2010)\citenamefont
		{Shuvaev}, \citenamefont {Travkin}, \citenamefont {Ivanov}, \citenamefont
		{Mukhin},\ and\ \citenamefont {Pimenov}}]{Shuvaev2010}%
	\BibitemOpen
	\bibfield  {author} {\bibinfo {author} {\bibfnamefont {A.~M.}\ \bibnamefont
			{Shuvaev}}, \bibinfo {author} {\bibfnamefont {V.~D.}\ \bibnamefont
			{Travkin}}, \bibinfo {author} {\bibfnamefont {V.~Y.}\ \bibnamefont {Ivanov}},
		\bibinfo {author} {\bibfnamefont {A.~A.}\ \bibnamefont {Mukhin}},\ and\
		\bibinfo {author} {\bibfnamefont {A.}~\bibnamefont {Pimenov}},\ }\bibfield
	{title} {\bibinfo {title} {{Evidence for Electroactive Excitation of the Spin
				Cycloid in TbMnO$_{\text{3}}$}},\ }\href
	{https://doi.org/10.1103/PhysRevLett.104.097202} {\bibfield  {journal}
		{\bibinfo  {journal} {Phys. Rev. Lett.}\ }\textbf {\bibinfo {volume} {104}},\
		\bibinfo {pages} {097202} (\bibinfo {year} {2010})}\BibitemShut {NoStop}%
	\bibitem [{\citenamefont {Mochizuki}\ \emph {et~al.}(2011)\citenamefont
		{Mochizuki}, \citenamefont {Furukawa},\ and\ \citenamefont
		{Nagaosa}}]{Mochizuki2011b}%
	\BibitemOpen
	\bibfield  {author} {\bibinfo {author} {\bibfnamefont {M.}~\bibnamefont
			{Mochizuki}}, \bibinfo {author} {\bibfnamefont {N.}~\bibnamefont
			{Furukawa}},\ and\ \bibinfo {author} {\bibfnamefont {N.}~\bibnamefont
			{Nagaosa}},\ }\bibfield  {title} {\bibinfo {title} {{Theory of spin-phonon
				coupling in multiferroic manganese perovskites RMnO$_{\text{3}}$}},\ }\href
	{https://doi.org/10.1103/PhysRevB.84.144409} {\bibfield  {journal} {\bibinfo
			{journal} {Phys. Rev. B}\ }\textbf {\bibinfo {volume} {84}},\ \bibinfo
		{pages} {144409} (\bibinfo {year} {2011})}\BibitemShut {NoStop}%
	\bibitem [{\citenamefont {Rovillain}\ \emph {et~al.}(2011)\citenamefont
		{Rovillain}, \citenamefont {Cazayous}, \citenamefont {Gallais}, \citenamefont
		{Measson}, \citenamefont {Sacuto}, \citenamefont {Sakata},\ and\
		\citenamefont {Mochizuki}}]{Rovillain2011}%
	\BibitemOpen
	\bibfield  {author} {\bibinfo {author} {\bibfnamefont {P.}~\bibnamefont
			{Rovillain}}, \bibinfo {author} {\bibfnamefont {M.}~\bibnamefont {Cazayous}},
		\bibinfo {author} {\bibfnamefont {Y.}~\bibnamefont {Gallais}}, \bibinfo
		{author} {\bibfnamefont {M.-A.}\ \bibnamefont {Measson}}, \bibinfo {author}
		{\bibfnamefont {A.}~\bibnamefont {Sacuto}}, \bibinfo {author} {\bibfnamefont
			{H.}~\bibnamefont {Sakata}},\ and\ \bibinfo {author} {\bibfnamefont
			{M.}~\bibnamefont {Mochizuki}},\ }\bibfield  {title} {\bibinfo {title}
		{{Magnetic Field Induced Dehybridization of the Electromagnons in
				Multiferroic TbMnO$_{\text{3}}$}},\ }\href
	{https://doi.org/10.1103/PhysRevLett.107.027202} {\bibfield  {journal}
		{\bibinfo  {journal} {Phys. Rev. Lett.}\ }\textbf {\bibinfo {volume} {107}},\
		\bibinfo {pages} {027202} (\bibinfo {year} {2011})}\BibitemShut {NoStop}%
	\bibitem [{\citenamefont {Finger}\ \emph {et~al.}(2010)\citenamefont {Finger},
		\citenamefont {Senff}, \citenamefont {Schmalzl}, \citenamefont {Schmidt},
		\citenamefont {Regnault}, \citenamefont {Becker}, \citenamefont
		{Bohat\'{y}},\ and\ \citenamefont {Braden}}]{Finger2010}%
	\BibitemOpen
	\bibfield  {author} {\bibinfo {author} {\bibfnamefont {T.}~\bibnamefont
			{Finger}}, \bibinfo {author} {\bibfnamefont {D.}~\bibnamefont {Senff}},
		\bibinfo {author} {\bibfnamefont {K.}~\bibnamefont {Schmalzl}}, \bibinfo
		{author} {\bibfnamefont {W.}~\bibnamefont {Schmidt}}, \bibinfo {author}
		{\bibfnamefont {L.~P.}\ \bibnamefont {Regnault}}, \bibinfo {author}
		{\bibfnamefont {P.}~\bibnamefont {Becker}}, \bibinfo {author} {\bibfnamefont
			{L.}~\bibnamefont {Bohat\'{y}}},\ and\ \bibinfo {author} {\bibfnamefont
			{M.}~\bibnamefont {Braden}},\ }\bibfield  {title} {\bibinfo {title}
		{{Electric-field control of the chiral magnetism of multiferroic MnWO$_4$ as
				seen via polarized neutron diffraction}},\ }\href
	{https://doi.org/10.1103/PhysRevB.81.054430} {\bibfield  {journal} {\bibinfo
			{journal} {Phys. Rev. B}\ }\textbf {\bibinfo {volume} {81}},\ \bibinfo
		{pages} {1} (\bibinfo {year} {2010})}\BibitemShut {NoStop}%
	\bibitem [{\citenamefont {Poole}\ \emph {et~al.}(2009)\citenamefont {Poole},
		\citenamefont {Brown},\ and\ \citenamefont {Wills}}]{Poole2009}%
	\BibitemOpen
	\bibfield  {author} {\bibinfo {author} {\bibfnamefont {A.}~\bibnamefont
			{Poole}}, \bibinfo {author} {\bibfnamefont {P.~J.}\ \bibnamefont {Brown}},\
		and\ \bibinfo {author} {\bibfnamefont {A.~S.}\ \bibnamefont {Wills}},\
	}\bibfield  {title} {\bibinfo {title} {{Spherical neutron polarimetry (SNP)
				study of magneto-electric coupling in the multiferroic MnWO$_4$}},\ }\href
	{https://doi.org/10.1088/1742-6596/145/1/012074} {\bibfield  {journal}
		{\bibinfo  {journal} {J. Phys. Conf. Ser.}\ }\textbf {\bibinfo {volume}
			{145}},\ \bibinfo {pages} {012074} (\bibinfo {year} {2009})}\BibitemShut
	{NoStop}%
	\bibitem [{\citenamefont {Hearmon}\ \emph {et~al.}(2012)\citenamefont
		{Hearmon}, \citenamefont {Fabrizi}, \citenamefont {Chapon}, \citenamefont
		{Johnson}, \citenamefont {Prabhakaran}, \citenamefont {Streltsov},
		\citenamefont {Brown},\ and\ \citenamefont {Radaelli}}]{Hearmon2012}%
	\BibitemOpen
	\bibfield  {author} {\bibinfo {author} {\bibfnamefont {A.~J.}\ \bibnamefont
			{Hearmon}}, \bibinfo {author} {\bibfnamefont {F.}~\bibnamefont {Fabrizi}},
		\bibinfo {author} {\bibfnamefont {L.~C.}\ \bibnamefont {Chapon}}, \bibinfo
		{author} {\bibfnamefont {R.~D.}\ \bibnamefont {Johnson}}, \bibinfo {author}
		{\bibfnamefont {D.}~\bibnamefont {Prabhakaran}}, \bibinfo {author}
		{\bibfnamefont {S.~V.}\ \bibnamefont {Streltsov}}, \bibinfo {author}
		{\bibfnamefont {P.~J.}\ \bibnamefont {Brown}},\ and\ \bibinfo {author}
		{\bibfnamefont {P.~G.}\ \bibnamefont {Radaelli}},\ }\bibfield  {title}
	{\bibinfo {title} {{Electric Field Control of the Magnetic Chiralities in
				Ferroaxial Multiferroic RbFe(MoO$_{4}$)$_{2}$}},\ }\href
	{https://doi.org/10.1103/physrevlett.108.237201} {\bibfield  {journal}
		{\bibinfo  {journal} {Phys. Rev. Lett.}\ }\textbf {\bibinfo {volume} {108}},\
		\bibinfo {pages} {237201} (\bibinfo {year} {2012})}\BibitemShut {NoStop}%
	\bibitem [{\citenamefont {Brown}(2006)}]{Brown2006}%
	\BibitemOpen
	\bibfield  {author} {\bibinfo {author} {\bibfnamefont {P.~J.}\ \bibnamefont
			{Brown}},\ }\bibinfo {title} {Neutron scattering from magnetic materials}\
	(\bibinfo  {publisher} {Elsevier B. V.},\ \bibinfo {year} {2006})\ Chap.\
	\bibinfo {chapter} {5 - Spherical Neutron Polarimetry}, pp.\ \bibinfo {pages}
	{215--244}\BibitemShut {NoStop}%
	\bibitem [{\citenamefont {Stein}\ \emph {et~al.}(2021)\citenamefont {Stein},
		\citenamefont {Biesenkamp}, \citenamefont {Cronert}, \citenamefont
		{Fr\"ohlich}, \citenamefont {Leist}, \citenamefont {Schmalzl}, \citenamefont
		{Komarek},\ and\ \citenamefont {Braden}}]{Stein2021}%
	\BibitemOpen
	\bibfield  {author} {\bibinfo {author} {\bibfnamefont {J.}~\bibnamefont
			{Stein}}, \bibinfo {author} {\bibfnamefont {S.}~\bibnamefont {Biesenkamp}},
		\bibinfo {author} {\bibfnamefont {T.}~\bibnamefont {Cronert}}, \bibinfo
		{author} {\bibfnamefont {T.}~\bibnamefont {Fr\"ohlich}}, \bibinfo {author}
		{\bibfnamefont {J.}~\bibnamefont {Leist}}, \bibinfo {author} {\bibfnamefont
			{K.}~\bibnamefont {Schmalzl}}, \bibinfo {author} {\bibfnamefont {A.~C.}\
			\bibnamefont {Komarek}},\ and\ \bibinfo {author} {\bibfnamefont
			{M.}~\bibnamefont {Braden}},\ }\bibfield  {title} {\bibinfo {title}
		{{Combined Arrhenius-Merz Law Describing Domain Relaxation in Type-II
				Multiferroics}},\ }\href {https://doi.org/10.1103/PhysRevLett.127.097601}
	{\bibfield  {journal} {\bibinfo  {journal} {Phys. Rev. Lett.}\ }\textbf
		{\bibinfo {volume} {127}},\ \bibinfo {pages} {097601} (\bibinfo {year}
		{2021})}\BibitemShut {NoStop}%
	\bibitem [{\citenamefont {Reutler}\ \emph {et~al.}(2003)\citenamefont
		{Reutler}, \citenamefont {Friedt}, \citenamefont {B\"uchner}, \citenamefont
		{Braden},\ and\ \citenamefont {Revcolevschi}}]{Reutler2003}%
	\BibitemOpen
	\bibfield  {author} {\bibinfo {author} {\bibfnamefont {P.}~\bibnamefont
			{Reutler}}, \bibinfo {author} {\bibfnamefont {O.}~\bibnamefont {Friedt}},
		\bibinfo {author} {\bibfnamefont {B.}~\bibnamefont {B\"uchner}}, \bibinfo
		{author} {\bibfnamefont {M.}~\bibnamefont {Braden}},\ and\ \bibinfo {author}
		{\bibfnamefont {A.}~\bibnamefont {Revcolevschi}},\ }\bibfield  {title}
	{\bibinfo {title} {{Growth of La$_x$Sr$_{1+x}$MnO$_4$ single crystals and
				characterization by scattering techniques}},\ }\href
	{https://doi.org/https://doi.org/10.1016/S0022-0248(02)02104-8} {\bibfield
		{journal} {\bibinfo  {journal} {Journal of Crystal Growth}\ }\textbf
		{\bibinfo {volume} {249}},\ \bibinfo {pages} {222} (\bibinfo {year}
		{2003})}\BibitemShut {NoStop}%
	\bibitem [{\citenamefont {Komarek}(2009)}]{diss_komarek}%
	\BibitemOpen
	\bibfield  {author} {\bibinfo {author} {\bibfnamefont {A.~C.}\ \bibnamefont
			{Komarek}},\ }\emph {\bibinfo {title} {Complex ordering phenomena in
			transition metal oxides and oxyhalides}},\ \href@noop {} {Ph.D. thesis},\
	\bibinfo  {school} {Universit{\"a}t zu K{\"o}ln} (\bibinfo {year} {2009}),\
	\bibinfo {note} {https://kups.ub.uni-koeln.de/2982/}\BibitemShut {NoStop}%
	\bibitem [{\citenamefont {Holbein}\ \emph {et~al.}\citenamefont {Holbein},
		\citenamefont {Steffens},\ and\ \citenamefont {Braden}}]{data-4-01-1230}%
	\BibitemOpen
	\bibfield  {author} {\bibinfo {author} {\bibfnamefont {S.}~\bibnamefont
			{Holbein}}, \bibinfo {author} {\bibfnamefont {P.}~\bibnamefont {Steffens}},\
		and\ \bibinfo {author} {\bibfnamefont {M.}~\bibnamefont {Braden}},\
	}\href@noop {} {}\bibinfo {note} {"Magnon dispersion in TbMnO$_3$". Institut
		Laue-Langevin (ILL) doi:10.5291/ILL-DATA.4-01-1230}\BibitemShut {NoStop}%
	\bibitem [{\citenamefont {Ollivier}\ and\ \citenamefont
		{Mutka}(2011)}]{Ollivier2011}%
	\BibitemOpen
	\bibfield  {author} {\bibinfo {author} {\bibfnamefont {J.}~\bibnamefont
			{Ollivier}}\ and\ \bibinfo {author} {\bibfnamefont {H.}~\bibnamefont
			{Mutka}},\ }\bibfield  {title} {\bibinfo {title} {In5~{C}old neutron
			time-of-flight spectrometer, prepared to tackle single crystal
			spectroscopy},\ }\href {https://doi.org/10.1143/jpsjs.80sb.sb003} {\bibfield
		{journal} {\bibinfo  {journal} {J. Phys. Soc. Jpn.}\ }\textbf {\bibinfo
			{volume} {80}},\ \bibinfo {pages} {SB003} (\bibinfo {year}
		{2011})}\BibitemShut {NoStop}%
	\bibitem [{\citenamefont {Richard}\ \emph {et~al.}(1996)\citenamefont
		{Richard}, \citenamefont {Ferrand},\ and\ \citenamefont
		{Kearley}}]{Richard1996}%
	\BibitemOpen
	\bibfield  {author} {\bibinfo {author} {\bibfnamefont {D.}~\bibnamefont
			{Richard}}, \bibinfo {author} {\bibfnamefont {M.}~\bibnamefont {Ferrand}},\
		and\ \bibinfo {author} {\bibfnamefont {G.~J.}\ \bibnamefont {Kearley}},\
	}\bibfield  {title} {\bibinfo {title} {Analysis and visualisation of
			neutron-scattering data},\ }\href {https://doi.org/10.1080/10238169608200065}
	{\bibfield  {journal} {\bibinfo  {journal} {Journal of Neutron Research}\
		}\textbf {\bibinfo {volume} {4}},\ \bibinfo {pages} {33} (\bibinfo {year}
		{1996})}\BibitemShut {NoStop}%
	\bibitem [{\citenamefont {Biesenkamp}\ \emph {et~al.}\citenamefont
		{Biesenkamp}, \citenamefont {Steffens},\ and\ \citenamefont
		{Braden}}]{data-4-03-1730}%
	\BibitemOpen
	\bibfield  {author} {\bibinfo {author} {\bibfnamefont {S.}~\bibnamefont
			{Biesenkamp}}, \bibinfo {author} {\bibfnamefont {P.}~\bibnamefont
			{Steffens}},\ and\ \bibinfo {author} {\bibfnamefont {M.}~\bibnamefont
			{Braden}},\ }\href@noop {} {}\bibinfo {note} {"Control of chiral magnetism
		through E fields in multiferroic TbMnO$_3$ above the long-range multiferroic
		transition". Institut Laue-Langevin (ILL)
		doi:10.5291/ILL-DATA.4-03-1730}\BibitemShut {NoStop}%
	\bibitem [{\citenamefont {Moussa}\ \emph {et~al.}(1996)\citenamefont {Moussa},
		\citenamefont {Hennion}, \citenamefont {Rodriguez-Carvajal}, \citenamefont
		{Moudden}, \citenamefont {Pinsard},\ and\ \citenamefont
		{Revcolevschi}}]{Moussa1996}%
	\BibitemOpen
	\bibfield  {author} {\bibinfo {author} {\bibfnamefont {F.}~\bibnamefont
			{Moussa}}, \bibinfo {author} {\bibfnamefont {M.}~\bibnamefont {Hennion}},
		\bibinfo {author} {\bibfnamefont {J.}~\bibnamefont {Rodriguez-Carvajal}},
		\bibinfo {author} {\bibfnamefont {H.}~\bibnamefont {Moudden}}, \bibinfo
		{author} {\bibfnamefont {L.}~\bibnamefont {Pinsard}},\ and\ \bibinfo {author}
		{\bibfnamefont {a.}~\bibnamefont {Revcolevschi}},\ }\bibfield  {title}
	{\bibinfo {title} {{Spin waves in the antiferromagnet perovskite LaMnO3: A
				neutron-scattering study.}},\ }\href
	{http://www.ncbi.nlm.nih.gov/pubmed/9985575} {\bibfield  {journal} {\bibinfo
			{journal} {Phys. Rev. B. Condens. Matter}\ }\textbf {\bibinfo {volume}
			{54}},\ \bibinfo {pages} {15149} (\bibinfo {year} {1996})}\BibitemShut
	{NoStop}%
	\bibitem [{\citenamefont {Hirota}\ \emph {et~al.}(1996)\citenamefont {Hirota},
		\citenamefont {Kaneko}, \citenamefont {Nishizawa},\ and\ \citenamefont
		{Endoh}}]{Hirota1996}%
	\BibitemOpen
	\bibfield  {author} {\bibinfo {author} {\bibfnamefont {K.}~\bibnamefont
			{Hirota}}, \bibinfo {author} {\bibfnamefont {N.}~\bibnamefont {Kaneko}},
		\bibinfo {author} {\bibfnamefont {A.}~\bibnamefont {Nishizawa}},\ and\
		\bibinfo {author} {\bibfnamefont {Y.}~\bibnamefont {Endoh}},\ }\bibfield
	{title} {\bibinfo {title} {{Two-Dimensional Planar Ferromagnetic Coupling in
				LaMnO$_3$}},\ }\href {https://doi.org/10.1143/jpsj.65.3736} {\bibfield
		{journal} {\bibinfo  {journal} {J. Phys. Soc. Jpn.}\ }\textbf {\bibinfo
			{volume} {65}},\ \bibinfo {pages} {3736} (\bibinfo {year}
		{1996})}\BibitemShut {NoStop}%
	\bibitem [{\citenamefont {S\'aenz}(1962)}]{Saenz1962}%
	\BibitemOpen
	\bibfield  {author} {\bibinfo {author} {\bibfnamefont {A.~W.}\ \bibnamefont
			{S\'aenz}},\ }\bibfield  {title} {\bibinfo {title} {Spin waves in
			exchange-coupled complex magnetic structures and neutron scattering},\ }\href
	{https://doi.org/10.1103/physrev.125.1940} {\bibfield  {journal} {\bibinfo
			{journal} {Phys. Rev.}\ }\textbf {\bibinfo {volume} {125}},\ \bibinfo {pages}
		{1940} (\bibinfo {year} {1962})}\BibitemShut {NoStop}%
	\bibitem [{\citenamefont {Ulbrich}\ \emph {et~al.}(2012)\citenamefont
		{Ulbrich}, \citenamefont {Steffens}, \citenamefont {Lamago}, \citenamefont
		{Sidis},\ and\ \citenamefont {Braden}}]{Ulbrich2012}%
	\BibitemOpen
	\bibfield  {author} {\bibinfo {author} {\bibfnamefont {H.}~\bibnamefont
			{Ulbrich}}, \bibinfo {author} {\bibfnamefont {P.}~\bibnamefont {Steffens}},
		\bibinfo {author} {\bibfnamefont {D.}~\bibnamefont {Lamago}}, \bibinfo
		{author} {\bibfnamefont {Y.}~\bibnamefont {Sidis}},\ and\ \bibinfo {author}
		{\bibfnamefont {M.}~\bibnamefont {Braden}},\ }\bibfield  {title} {\bibinfo
		{title} {Hourglass dispersion in overdoped single-layered manganites},\
	}\href {https://doi.org/10.1103/PhysRevLett.108.247209} {\bibfield  {journal}
		{\bibinfo  {journal} {Phys. Rev. Lett.}\ }\textbf {\bibinfo {volume} {108}},\
		\bibinfo {pages} {247209} (\bibinfo {year} {2012})}\BibitemShut {NoStop}%
	\bibitem [{\citenamefont {{SpinW, version 2.1 revision 238, by S.
				Toth}}()}]{progspinW}%
	\BibitemOpen
	\bibfield  {author} {\bibinfo {author} {\bibnamefont {{SpinW, version 2.1
					revision 238, by S. Toth}}},\ }\href@noop {} {}\bibinfo {note} {Available
		online at \url{www.psi.ch/spinw}}\BibitemShut {NoStop}%
	\bibitem [{\citenamefont {Toth}\ and\ \citenamefont {Lake}(2015)}]{Toth2015}%
	\BibitemOpen
	\bibfield  {author} {\bibinfo {author} {\bibfnamefont {S.}~\bibnamefont
			{Toth}}\ and\ \bibinfo {author} {\bibfnamefont {B.}~\bibnamefont {Lake}},\
	}\bibfield  {title} {\bibinfo {title} {Linear spin wave theory for single-q
			incommensurate magnetic structures},\ }\href
	{https://doi.org/10.1088/0953-8984/27/16/166002} {\bibfield  {journal}
		{\bibinfo  {journal} {J. Phys.: Condens. Matter}\ }\textbf {\bibinfo {volume}
			{27}},\ \bibinfo {pages} {166002} (\bibinfo {year} {2015})}\BibitemShut
	{NoStop}%
	\bibitem [{not()}]{note-incom}%
	\BibitemOpen
	\href@noop {} {}\bibinfo {note} {{The same model with an incommensurate
			propagation vector did not show the splitting of out-of-plane modes. We
			attribute this discrepancy to problems of the implementation of
			incommensurate structures.}}\BibitemShut {Stop}%
	\bibitem [{\citenamefont {{The values from Senff \etal\ differ by a factor of
				two due to a different definition of the parameters.}}()}]{note-senff}%
	\BibitemOpen
	\bibfield  {author} {\bibinfo {author} {\bibnamefont {{The values from Senff
					\etal\ differ by a factor of two due to a different definition of the
					parameters.}}},\ }\href@noop {} {}\bibinfo {note} {{Bonds between moments are
			counted either once or twice in different definitions of magnetic
			interaction.}}\BibitemShut {Stop}%
	\bibitem [{\citenamefont {Mochizuki}\ and\ \citenamefont
		{Furukawa}(2009{\natexlab{b}})}]{Mochizuki2009}%
	\BibitemOpen
	\bibfield  {author} {\bibinfo {author} {\bibfnamefont {M.}~\bibnamefont
			{Mochizuki}}\ and\ \bibinfo {author} {\bibfnamefont {N.}~\bibnamefont
			{Furukawa}},\ }\bibfield  {title} {\bibinfo {title} {{Mechanism of
				Lattice-Distortion-Induced Electric-Polarization Flop in the Multiferroic
				Perovskite Manganites}},\ }\href {https://doi.org/10.1143/JPSJ.78.053704}
	{\bibfield  {journal} {\bibinfo  {journal} {J. Phys. Soc. Japan}\ }\textbf
		{\bibinfo {volume} {78}},\ \bibinfo {pages} {053704} (\bibinfo {year}
		{2009}{\natexlab{b}})}\BibitemShut {NoStop}%
	\bibitem [{\citenamefont {Malashevich}\ and\ \citenamefont
		{Vanderbilt}(2009)}]{Malashevich2009a}%
	\BibitemOpen
	\bibfield  {author} {\bibinfo {author} {\bibfnamefont {A.}~\bibnamefont
			{Malashevich}}\ and\ \bibinfo {author} {\bibfnamefont {D.}~\bibnamefont
			{Vanderbilt}},\ }\bibfield  {title} {\bibinfo {title} {{First-principles
				theory of magnetically induced ferroelectricity in TbMnO$_{\text{3}}$}},\
	}\href {https://doi.org/10.1140/epjb/e2009-00208-2} {\bibfield  {journal}
		{\bibinfo  {journal} {Eur. Phys. J. B}\ }\textbf {\bibinfo {volume} {71}},\
		\bibinfo {pages} {345} (\bibinfo {year} {2009})}\BibitemShut {NoStop}%
	\bibitem [{\citenamefont {Mochizuki}\ \emph {et~al.}(2010)\citenamefont
		{Mochizuki}, \citenamefont {Furukawa},\ and\ \citenamefont
		{Nagaosa}}]{Mochizuki2010c}%
	\BibitemOpen
	\bibfield  {author} {\bibinfo {author} {\bibfnamefont {M.}~\bibnamefont
			{Mochizuki}}, \bibinfo {author} {\bibfnamefont {N.}~\bibnamefont
			{Furukawa}},\ and\ \bibinfo {author} {\bibfnamefont {N.}~\bibnamefont
			{Nagaosa}},\ }\bibfield  {title} {\bibinfo {title} {{Theory of Electromagnons
				in the Multiferroic Mn Perovskites: The Vital Role of Higher Harmonic
				Components of the Spiral Spin Order}},\ }\href
	{https://doi.org/10.1103/PhysRevLett.104.177206} {\bibfield  {journal}
		{\bibinfo  {journal} {Phys. Rev. Lett.}\ }\textbf {\bibinfo {volume} {104}},\
		\bibinfo {pages} {177206} (\bibinfo {year} {2010})}\BibitemShut {NoStop}%
	\bibitem [{\citenamefont {Fedorova}\ \emph {et~al.}(2015)\citenamefont
		{Fedorova}, \citenamefont {Ederer}, \citenamefont {Spaldin},\ and\
		\citenamefont {Scaramucci}}]{Fedorova2015}%
	\BibitemOpen
	\bibfield  {author} {\bibinfo {author} {\bibfnamefont {N.~S.}\ \bibnamefont
			{Fedorova}}, \bibinfo {author} {\bibfnamefont {C.}~\bibnamefont {Ederer}},
		\bibinfo {author} {\bibfnamefont {N.~A.}\ \bibnamefont {Spaldin}},\ and\
		\bibinfo {author} {\bibfnamefont {A.}~\bibnamefont {Scaramucci}},\ }\bibfield
	{title} {\bibinfo {title} {Biquadratic and ring exchange interactions in
			orthorhombic perovskite manganites},\ }\href
	{https://doi.org/10.1103/physrevb.91.165122} {\bibfield  {journal} {\bibinfo
			{journal} {Phys. Rev. B}\ }\textbf {\bibinfo {volume} {91}},\ \bibinfo
		{pages} {165122} (\bibinfo {year} {2015})}\BibitemShut {NoStop}%
	\bibitem [{\citenamefont {Tovar}\ \emph {et~al.}(1999)\citenamefont {Tovar},
		\citenamefont {Alejandro}, \citenamefont {Butera}, \citenamefont {Caneiro},
		\citenamefont {Causa}, \citenamefont {Prado},\ and\ \citenamefont
		{S\'anchez}}]{Tovar1999}%
	\BibitemOpen
	\bibfield  {author} {\bibinfo {author} {\bibfnamefont {M.}~\bibnamefont
			{Tovar}}, \bibinfo {author} {\bibfnamefont {G.}~\bibnamefont {Alejandro}},
		\bibinfo {author} {\bibfnamefont {A.}~\bibnamefont {Butera}}, \bibinfo
		{author} {\bibfnamefont {A.}~\bibnamefont {Caneiro}}, \bibinfo {author}
		{\bibfnamefont {M.~T.}\ \bibnamefont {Causa}}, \bibinfo {author}
		{\bibfnamefont {F.}~\bibnamefont {Prado}},\ and\ \bibinfo {author}
		{\bibfnamefont {R.~D.}\ \bibnamefont {S\'anchez}},\ }\bibfield  {title}
	{\bibinfo {title} {{ESR and magnetization in Jahn-Teller-distorted
				LaMnO$_{3+\delta}$: Correlation with crystal structure}},\ }\href
	{https://doi.org/10.1103/physrevb.60.10199} {\bibfield  {journal} {\bibinfo
			{journal} {Phys. Rev. B}\ }\textbf {\bibinfo {volume} {60}},\ \bibinfo
		{pages} {10199} (\bibinfo {year} {1999})}\BibitemShut {NoStop}%
	\bibitem [{\citenamefont {O'Flynn}\ \emph {et~al.}(2014)\citenamefont
		{O'Flynn}, \citenamefont {Lees},\ and\ \citenamefont
		{Balakrishnan}}]{OFlynn2014}%
	\BibitemOpen
	\bibfield  {author} {\bibinfo {author} {\bibfnamefont {D.}~\bibnamefont
			{O'Flynn}}, \bibinfo {author} {\bibfnamefont {M.~R.}\ \bibnamefont {Lees}},\
		and\ \bibinfo {author} {\bibfnamefont {G.}~\bibnamefont {Balakrishnan}},\
	}\bibfield  {title} {\bibinfo {title} {{Magnetic susceptibility and heat
				capacity measurements of single crystal TbMnO$_3$}},\ }\href
	{https://doi.org/10.1088/0953-8984/26/25/256002} {\bibfield  {journal}
		{\bibinfo  {journal} {J. Phys.: Condens. Matter}\ }\textbf {\bibinfo {volume}
			{26}},\ \bibinfo {pages} {256002} (\bibinfo {year} {2014})}\BibitemShut
	{NoStop}%
	\bibitem [{\citenamefont {Stenberg}\ and\ \citenamefont
		{de~Sousa}(2009)}]{Stenberg2009}%
	\BibitemOpen
	\bibfield  {author} {\bibinfo {author} {\bibfnamefont {M.~P.~V.}\
			\bibnamefont {Stenberg}}\ and\ \bibinfo {author} {\bibfnamefont
			{R.}~\bibnamefont {de~Sousa}},\ }\bibfield  {title} {\bibinfo {title} {Model
			for twin electromagnons and magnetically induced oscillatory polarization in
			multiferroic rmno$_{\text{3}}$},\ }\href
	{https://doi.org/10.1103/physrevb.80.094419} {\bibfield  {journal} {\bibinfo
			{journal} {Phys. Rev. B}\ }\textbf {\bibinfo {volume} {80}},\ \bibinfo
		{pages} {094419} (\bibinfo {year} {2009})}\BibitemShut {NoStop}%
	\bibitem [{\citenamefont {Stenberg}\ and\ \citenamefont
		{de~Sousa}(2012)}]{Stenberg2012}%
	\BibitemOpen
	\bibfield  {author} {\bibinfo {author} {\bibfnamefont {M.~P.~V.}\
			\bibnamefont {Stenberg}}\ and\ \bibinfo {author} {\bibfnamefont
			{R.}~\bibnamefont {de~Sousa}},\ }\bibfield  {title} {\bibinfo {title}
		{{Sinusoidal electromagnon in $R$MnO$_{\text{3}}$: Indication of anomalous
				magnetoelectric coupling}},\ }\href
	{https://doi.org/10.1103/physrevb.85.104412} {\bibfield  {journal} {\bibinfo
			{journal} {Phys. Rev. B}\ }\textbf {\bibinfo {volume} {85}},\ \bibinfo
		{pages} {104412} (\bibinfo {year} {2012})}\BibitemShut {NoStop}%
	\bibitem [{\citenamefont {Takahashi}\ \emph {et~al.}(2008)\citenamefont
		{Takahashi}, \citenamefont {Kida}, \citenamefont {Yamasaki}, \citenamefont
		{Fujioka}, \citenamefont {Arima}, \citenamefont {Shimano}, \citenamefont
		{Miyahara}, \citenamefont {Mochizuki}, \citenamefont {Furukawa},\ and\
		\citenamefont {Tokura}}]{Takahashi2008}%
	\BibitemOpen
	\bibfield  {author} {\bibinfo {author} {\bibfnamefont {Y.}~\bibnamefont
			{Takahashi}}, \bibinfo {author} {\bibfnamefont {N.}~\bibnamefont {Kida}},
		\bibinfo {author} {\bibfnamefont {Y.}~\bibnamefont {Yamasaki}}, \bibinfo
		{author} {\bibfnamefont {J.}~\bibnamefont {Fujioka}}, \bibinfo {author}
		{\bibfnamefont {T.}~\bibnamefont {Arima}}, \bibinfo {author} {\bibfnamefont
			{R.}~\bibnamefont {Shimano}}, \bibinfo {author} {\bibfnamefont
			{S.}~\bibnamefont {Miyahara}}, \bibinfo {author} {\bibfnamefont
			{M.}~\bibnamefont {Mochizuki}}, \bibinfo {author} {\bibfnamefont
			{N.}~\bibnamefont {Furukawa}},\ and\ \bibinfo {author} {\bibfnamefont
			{Y.}~\bibnamefont {Tokura}},\ }\bibfield  {title} {\bibinfo {title}
		{{Evidence for an Electric-Dipole Active Continuum Band of Spin Excitations
				in Multiferroic TbMnO$_{\text{3}}$}},\ }\href
	{https://doi.org/10.1103/PhysRevLett.101.187201} {\bibfield  {journal}
		{\bibinfo  {journal} {Phys. Rev. Lett.}\ }\textbf {\bibinfo {volume} {101}},\
		\bibinfo {pages} {187201} (\bibinfo {year} {2008})}\BibitemShut {NoStop}%
	\bibitem [{\citenamefont {{Horace suite for \textit{MATLAB}}}()}]{progHorace}%
	\BibitemOpen
	\bibfield  {author} {\bibinfo {author} {\bibnamefont {{Horace suite for
					\textit{MATLAB}}}},\ }\href@noop {} {}\bibinfo {note} {Available online at
		\url{horace.isis.rl.ac.uk}}\BibitemShut {NoStop}%
	\bibitem [{\citenamefont {Stein}\ \emph {et~al.}(2015)\citenamefont {Stein},
		\citenamefont {Baum}, \citenamefont {Holbein}, \citenamefont {Hutanu},
		\citenamefont {Komarek},\ and\ \citenamefont {Braden}}]{Stein2015}%
	\BibitemOpen
	\bibfield  {author} {\bibinfo {author} {\bibfnamefont {J.}~\bibnamefont
			{Stein}}, \bibinfo {author} {\bibfnamefont {M.}~\bibnamefont {Baum}},
		\bibinfo {author} {\bibfnamefont {S.}~\bibnamefont {Holbein}}, \bibinfo
		{author} {\bibfnamefont {V.}~\bibnamefont {Hutanu}}, \bibinfo {author}
		{\bibfnamefont {A.~C.}\ \bibnamefont {Komarek}},\ and\ \bibinfo {author}
		{\bibfnamefont {M.}~\bibnamefont {Braden}},\ }\bibfield  {title} {\bibinfo
		{title} {{Control of multiferroic domains by external electric fields in
				TbMnO$_3$}},\ }\href {https://doi.org/10.1088/0953-8984/27/44/446001}
	{\bibfield  {journal} {\bibinfo  {journal} {J. Phys.: Condens. matter}\
		}\textbf {\bibinfo {volume} {27}},\ \bibinfo {pages} {446001} (\bibinfo
		{year} {2015})}\BibitemShut {NoStop}%
	\bibitem [{\citenamefont {Finger}(2013)}]{diss_finger}%
	\BibitemOpen
	\bibfield  {author} {\bibinfo {author} {\bibfnamefont {T.}~\bibnamefont
			{Finger}},\ }\emph {\bibinfo {title} {Analyse magnetischer Korrelationen in
			geschichteten oder multiferroischen \"Ubergangsmetalloxiden durch
			Neutronenstreuung}},\ \href@noop {} {Ph.D. thesis},\ \bibinfo  {school}
	{Universit{\"a}t zu K{\"o}ln} (\bibinfo {year} {2013}),\ \bibinfo {note}
	{https://kups.ub.uni-koeln.de/5493/}\BibitemShut {NoStop}%
	\bibitem [{\citenamefont {{An aim of the experiment on IN20 was to search for a
				magnon phonon hybridization of the high-energy magnons near the zone
				boundary.}}()}]{note-IN20}%
	\BibitemOpen
	\bibfield  {author} {\bibinfo {author} {\bibnamefont {{An aim of the
					experiment on IN20 was to search for a magnon phonon hybridization of the
					high-energy magnons near the zone boundary.}}}\ }\href@noop {} {}\bibinfo
	{note} {{but in spite of very long counting times at $\vec{Q}$=(2, -1, 1) and
			energies between 8 and 10 meV no nuclear-magnetic interference could be
			identified.}}\BibitemShut {Stop}%
	\bibitem [{\citenamefont {Jensen}(2011)}]{Jensen2011}%
	\BibitemOpen
	\bibfield  {author} {\bibinfo {author} {\bibfnamefont {J.}~\bibnamefont
			{Jensen}},\ }\bibfield  {title} {\bibinfo {title} {{Chiral spin-wave
				excitations of the spin-$\frac{5}{2}$ trimers in the langasite compound
				Ba$_3$NbFe$_3$Si$_2$O$_{14}$}},\ }\href
	{https://doi.org/10.1103/physrevb.84.104405} {\bibfield  {journal} {\bibinfo
			{journal} {Phys. Rev. B}\ }\textbf {\bibinfo {volume} {84}},\ \bibinfo
		{pages} {104405} (\bibinfo {year} {2011})}\BibitemShut {NoStop}%
	\bibitem [{\citenamefont {Loire}\ \emph {et~al.}(2011)\citenamefont {Loire},
		\citenamefont {Simonet}, \citenamefont {Petit}, \citenamefont {Marty},
		\citenamefont {Bordet}, \citenamefont {Lejay}, \citenamefont {Ollivier},
		\citenamefont {Enderle}, \citenamefont {Steffens}, \citenamefont {Ressouche},
		\citenamefont {Zorko},\ and\ \citenamefont {Ballou}}]{Loire2011}%
	\BibitemOpen
	\bibfield  {author} {\bibinfo {author} {\bibfnamefont {M.}~\bibnamefont
			{Loire}}, \bibinfo {author} {\bibfnamefont {V.}~\bibnamefont {Simonet}},
		\bibinfo {author} {\bibfnamefont {S.}~\bibnamefont {Petit}}, \bibinfo
		{author} {\bibfnamefont {K.}~\bibnamefont {Marty}}, \bibinfo {author}
		{\bibfnamefont {P.}~\bibnamefont {Bordet}}, \bibinfo {author} {\bibfnamefont
			{P.}~\bibnamefont {Lejay}}, \bibinfo {author} {\bibfnamefont
			{J.}~\bibnamefont {Ollivier}}, \bibinfo {author} {\bibfnamefont
			{M.}~\bibnamefont {Enderle}}, \bibinfo {author} {\bibfnamefont
			{P.}~\bibnamefont {Steffens}}, \bibinfo {author} {\bibfnamefont
			{E.}~\bibnamefont {Ressouche}}, \bibinfo {author} {\bibfnamefont
			{A.}~\bibnamefont {Zorko}},\ and\ \bibinfo {author} {\bibfnamefont
			{R.}~\bibnamefont {Ballou}},\ }\bibfield  {title} {\bibinfo {title}
		{{Parity-Broken Chiral Spin Dynamics in $Ba_3NbFe_3Si_2O_{14}$}},\ }\href
	{https://doi.org/10.1103/physrevlett.106.207201} {\bibfield  {journal}
		{\bibinfo  {journal} {Phys. Rev. Lett.}\ }\textbf {\bibinfo {volume} {106}},\
		\bibinfo {pages} {207201} (\bibinfo {year} {2011})}\BibitemShut {NoStop}%
	\bibitem [{\citenamefont {Bord{\'a}cs}\ \emph {et~al.}(2012)\citenamefont
		{Bord{\'a}cs}, \citenamefont {K{\'e}zsm{\'a}rki}, \citenamefont {Szaller},
		\citenamefont {Demk{\'o}}, \citenamefont {Kida}, \citenamefont {Murakawa},
		\citenamefont {Onose}, \citenamefont {Shimano}, \citenamefont {R{\~o}{\~o}m},
		\citenamefont {Nagel}, \citenamefont {Miyahara}, \citenamefont {Furukawa},\
		and\ \citenamefont {Tokura}}]{Bordacs2012}%
	\BibitemOpen
	\bibfield  {author} {\bibinfo {author} {\bibfnamefont {S.}~\bibnamefont
			{Bord{\'a}cs}}, \bibinfo {author} {\bibfnamefont {I.}~\bibnamefont
			{K{\'e}zsm{\'a}rki}}, \bibinfo {author} {\bibfnamefont {D.}~\bibnamefont
			{Szaller}}, \bibinfo {author} {\bibfnamefont {L.}~\bibnamefont {Demk{\'o}}},
		\bibinfo {author} {\bibfnamefont {N.}~\bibnamefont {Kida}}, \bibinfo {author}
		{\bibfnamefont {H.}~\bibnamefont {Murakawa}}, \bibinfo {author}
		{\bibfnamefont {Y.}~\bibnamefont {Onose}}, \bibinfo {author} {\bibfnamefont
			{R.}~\bibnamefont {Shimano}}, \bibinfo {author} {\bibfnamefont
			{T.}~\bibnamefont {R{\~o}{\~o}m}}, \bibinfo {author} {\bibfnamefont
			{U.}~\bibnamefont {Nagel}}, \bibinfo {author} {\bibfnamefont
			{S.}~\bibnamefont {Miyahara}}, \bibinfo {author} {\bibfnamefont
			{N.}~\bibnamefont {Furukawa}},\ and\ \bibinfo {author} {\bibfnamefont
			{Y.}~\bibnamefont {Tokura}},\ }\bibfield  {title} {\bibinfo {title}
		{Chirality of matter shows up via spin excitations},\ }\href
	{https://doi.org/10.1038/nphys2387} {\bibfield  {journal} {\bibinfo
			{journal} {Nature Physics}\ }\textbf {\bibinfo {volume} {8}},\ \bibinfo
		{pages} {734} (\bibinfo {year} {2012})}\BibitemShut {NoStop}%
	\bibitem [{\citenamefont {Schrettle}\ \emph {et~al.}(2009)\citenamefont
		{Schrettle}, \citenamefont {Lunkenheimer}, \citenamefont {Hemberger},
		\citenamefont {Ivanov}, \citenamefont {Mukhin}, \citenamefont {Balbashov},\
		and\ \citenamefont {Loidl}}]{Schrettle2009}%
	\BibitemOpen
	\bibfield  {author} {\bibinfo {author} {\bibfnamefont {F.}~\bibnamefont
			{Schrettle}}, \bibinfo {author} {\bibfnamefont {P.}~\bibnamefont
			{Lunkenheimer}}, \bibinfo {author} {\bibfnamefont {J.}~\bibnamefont
			{Hemberger}}, \bibinfo {author} {\bibfnamefont {V.~Y.}\ \bibnamefont
			{Ivanov}}, \bibinfo {author} {\bibfnamefont {A.~A.}\ \bibnamefont {Mukhin}},
		\bibinfo {author} {\bibfnamefont {A.~M.}\ \bibnamefont {Balbashov}},\ and\
		\bibinfo {author} {\bibfnamefont {A.}~\bibnamefont {Loidl}},\ }\bibfield
	{title} {\bibinfo {title} {Relaxations as key to the magnetocapacitive
			effects in the perovskite manganites},\ }\href
	{https://doi.org/10.1103/PhysRevLett.102.207208} {\bibfield  {journal}
		{\bibinfo  {journal} {Phys. Rev. Lett.}\ }\textbf {\bibinfo {volume} {102}},\
		\bibinfo {pages} {207208} (\bibinfo {year} {2009})}\BibitemShut {NoStop}%
	\bibitem [{\citenamefont {Foggetti}\ and\ \citenamefont
		{Artyukhin}(2020)}]{Foggetti2020}%
	\BibitemOpen
	\bibfield  {author} {\bibinfo {author} {\bibfnamefont {F.}~\bibnamefont
			{Foggetti}}\ and\ \bibinfo {author} {\bibfnamefont {S.}~\bibnamefont
			{Artyukhin}},\ }\href {https://arxiv.org/abs/2012.15383} {\bibinfo {title}
		{Soft magnon contributions to dielectric constant in spiral magnets with
			domain walls}} (\bibinfo {year} {2020}),\ \bibinfo {note}
	{arXiv2012.15383}\BibitemShut {NoStop}%
	\bibitem [{\citenamefont {Xiang}\ \emph {et~al.}(2008)\citenamefont {Xiang},
		\citenamefont {Wei}, \citenamefont {Whangbo},\ and\ \citenamefont {{Da
				Silva}}}]{Xiang2008}%
	\BibitemOpen
	\bibfield  {author} {\bibinfo {author} {\bibfnamefont {H.~J.}\ \bibnamefont
			{Xiang}}, \bibinfo {author} {\bibfnamefont {S.-H.}\ \bibnamefont {Wei}},
		\bibinfo {author} {\bibfnamefont {M.~H.}\ \bibnamefont {Whangbo}},\ and\
		\bibinfo {author} {\bibfnamefont {J.~L.~F.}\ \bibnamefont {{Da Silva}}},\
	}\bibfield  {title} {\bibinfo {title} {{Spin-Orbit Coupling and Ion
				Displacements in Multiferroic TbMnO$_{\text{3}}$}},\ }\href
	{https://doi.org/10.1103/PhysRevLett.101.037209} {\bibfield  {journal}
		{\bibinfo  {journal} {Phys. Rev. Lett.}\ }\textbf {\bibinfo {volume} {101}},\
		\bibinfo {pages} {037209} (\bibinfo {year} {2008})},\ \Eprint
	{https://arxiv.org/abs/0803.2741} {arXiv:0803.2741} \BibitemShut {NoStop}%
	\bibitem [{\citenamefont {Malashevich}\ and\ \citenamefont
		{Vanderbilt}(2008)}]{Malashevich2008a}%
	\BibitemOpen
	\bibfield  {author} {\bibinfo {author} {\bibfnamefont {A.}~\bibnamefont
			{Malashevich}}\ and\ \bibinfo {author} {\bibfnamefont {D.}~\bibnamefont
			{Vanderbilt}},\ }\bibfield  {title} {\bibinfo {title} {{First Principles
				Study of Improper Ferroelectricity in TbMnO$_{\text{3}}$}},\ }\href
	{https://doi.org/10.1103/PhysRevLett.101.037210} {\bibfield  {journal}
		{\bibinfo  {journal} {Phys. Rev. Lett.}\ }\textbf {\bibinfo {volume} {101}},\
		\bibinfo {pages} {037210} (\bibinfo {year} {2008})}\BibitemShut {NoStop}%
	\bibitem [{\citenamefont {Walker}\ \emph {et~al.}(2011)\citenamefont {Walker},
		\citenamefont {Fabrizi}, \citenamefont {Paolasini}, \citenamefont
		{de~Bergevin}, \citenamefont {Herrero-Martin}, \citenamefont {Boothroyd},
		\citenamefont {Prabhakaran},\ and\ \citenamefont {McMorrow}}]{Walker2011}%
	\BibitemOpen
	\bibfield  {author} {\bibinfo {author} {\bibfnamefont {H.~C.}\ \bibnamefont
			{Walker}}, \bibinfo {author} {\bibfnamefont {F.}~\bibnamefont {Fabrizi}},
		\bibinfo {author} {\bibfnamefont {L.}~\bibnamefont {Paolasini}}, \bibinfo
		{author} {\bibfnamefont {F.}~\bibnamefont {de~Bergevin}}, \bibinfo {author}
		{\bibfnamefont {J.}~\bibnamefont {Herrero-Martin}}, \bibinfo {author}
		{\bibfnamefont {A.~T.}\ \bibnamefont {Boothroyd}}, \bibinfo {author}
		{\bibfnamefont {D.}~\bibnamefont {Prabhakaran}},\ and\ \bibinfo {author}
		{\bibfnamefont {D.~F.}\ \bibnamefont {McMorrow}},\ }\bibfield  {title}
	{\bibinfo {title} {{Femtoscale magnetically induced lattice distortions in
				multiferroic TbMnO$_{\text{3}}$}},\ }\href
	{https://doi.org/10.1126/science.1208085} {\bibfield  {journal} {\bibinfo
			{journal} {Science}\ }\textbf {\bibinfo {volume} {333}},\ \bibinfo {pages}
		{1273} (\bibinfo {year} {2011})}\BibitemShut {NoStop}%
	\bibitem [{\citenamefont {Solovyev}\ \emph {et~al.}(1996)\citenamefont
		{Solovyev}, \citenamefont {Hamada},\ and\ \citenamefont
		{Terakura}}]{Solovyev1996}%
	\BibitemOpen
	\bibfield  {author} {\bibinfo {author} {\bibfnamefont {I.}~\bibnamefont
			{Solovyev}}, \bibinfo {author} {\bibfnamefont {N.}~\bibnamefont {Hamada}},\
		and\ \bibinfo {author} {\bibfnamefont {K.}~\bibnamefont {Terakura}},\
	}\bibfield  {title} {\bibinfo {title} {{Crucial Role of the Lattice
				Distortion in the Magnetism of LaMnO$_3$}},\ }\href
	{https://doi.org/10.1103/physrevlett.76.4825} {\bibfield  {journal} {\bibinfo
			{journal} {Phys. Rev. Lett.}\ }\textbf {\bibinfo {volume} {76}},\ \bibinfo
		{pages} {4825} (\bibinfo {year} {1996})}\BibitemShut {NoStop}%
	\bibitem [{\citenamefont {Deisenhofer}\ \emph {et~al.}(2002)\citenamefont
		{Deisenhofer}, \citenamefont {Eremin}, \citenamefont {Zakharov},
		\citenamefont {Ivanshin}, \citenamefont {Eremina}, \citenamefont {Krug~von
			Nidda}, \citenamefont {Mukhin}, \citenamefont {Balbashov},\ and\
		\citenamefont {Loidl}}]{Deisenhofer2002}%
	\BibitemOpen
	\bibfield  {author} {\bibinfo {author} {\bibfnamefont {J.}~\bibnamefont
			{Deisenhofer}}, \bibinfo {author} {\bibfnamefont {M.~V.}\ \bibnamefont
			{Eremin}}, \bibinfo {author} {\bibfnamefont {D.~V.}\ \bibnamefont
			{Zakharov}}, \bibinfo {author} {\bibfnamefont {V.~A.}\ \bibnamefont
			{Ivanshin}}, \bibinfo {author} {\bibfnamefont {R.~M.}\ \bibnamefont
			{Eremina}}, \bibinfo {author} {\bibfnamefont {H.-A.}\ \bibnamefont {Krug~von
				Nidda}}, \bibinfo {author} {\bibfnamefont {A.~A.}\ \bibnamefont {Mukhin}},
		\bibinfo {author} {\bibfnamefont {A.~M.}\ \bibnamefont {Balbashov}},\ and\
		\bibinfo {author} {\bibfnamefont {A.}~\bibnamefont {Loidl}},\ }\bibfield
	{title} {\bibinfo {title} {{Crystal field, Dzyaloshinsky-Moriya interaction,
				and orbital order in La$_{0.95}$Sr$_{0.05}$MnO$_3$ probed by ESR}},\ }\href
	{https://doi.org/10.1103/physrevb.65.104440} {\bibfield  {journal} {\bibinfo
			{journal} {Phys. Rev. B}\ }\textbf {\bibinfo {volume} {65}},\ \bibinfo
		{pages} {104440} (\bibinfo {year} {2002})}\BibitemShut {NoStop}%
	\bibitem [{\citenamefont {Mochizuki}\ and\ \citenamefont
		{Furukawa}(2010)}]{Mochizuki2010}%
	\BibitemOpen
	\bibfield  {author} {\bibinfo {author} {\bibfnamefont {M.}~\bibnamefont
			{Mochizuki}}\ and\ \bibinfo {author} {\bibfnamefont {N.}~\bibnamefont
			{Furukawa}},\ }\bibfield  {title} {\bibinfo {title} {{Theory of Magnetic
				Switching of Ferroelectricity in Spiral Magnets}},\ }\href
	{https://doi.org/10.1103/PhysRevLett.105.187601} {\bibfield  {journal}
		{\bibinfo  {journal} {Phys. Rev. Lett.}\ }\textbf {\bibinfo {volume} {105}},\
		\bibinfo {pages} {187601} (\bibinfo {year} {2010})}\BibitemShut {NoStop}%
	\bibitem [{\citenamefont {Kaplan}(2009)}]{Kaplan2009}%
	\BibitemOpen
	\bibfield  {author} {\bibinfo {author} {\bibfnamefont {T.~A.}\ \bibnamefont
			{Kaplan}},\ }\bibfield  {title} {\bibinfo {title} {Frustrated classical
			heisenberg model in one dimension with nearest-neighbor biquadratic exchange:
			Exact solution for the ground-state phase diagram},\ }\href
	{https://doi.org/10.1103/physrevb.80.012407} {\bibfield  {journal} {\bibinfo
			{journal} {Phys. Rev. B}\ }\textbf {\bibinfo {volume} {80}},\ \bibinfo
		{pages} {012407} (\bibinfo {year} {2009})}\BibitemShut {NoStop}%
	\bibitem [{\citenamefont {Khomskii}(2014)}]{book_khomskii}%
	\BibitemOpen
	\bibfield  {author} {\bibinfo {author} {\bibfnamefont {D.}~\bibnamefont
			{Khomskii}},\ }\href
	{http://www.cambridge.org/us/academic/subjects/physics/condensed-matter-physics-nanoscience-and-mesoscopic-physics/transition-metal-compounds}
	{\emph {\bibinfo {title} {Transition Metal Compounds}}}\ (\bibinfo
	{publisher} {Cambridge University Press},\ \bibinfo {year}
	{2014})\BibitemShut {NoStop}%
	\bibitem [{\citenamefont {Wilkins}\ \emph {et~al.}(2009)\citenamefont
		{Wilkins}, \citenamefont {Forrest}, \citenamefont {Beale}, \citenamefont
		{Bland}, \citenamefont {Walker}, \citenamefont {Mannix}, \citenamefont
		{Yakhou}, \citenamefont {Prabhakaran}, \citenamefont {Boothroyd},
		\citenamefont {Hill}, \citenamefont {Hatton},\ and\ \citenamefont
		{McMorrow}}]{Wilkins2009}%
	\BibitemOpen
	\bibfield  {author} {\bibinfo {author} {\bibfnamefont {S.~B.}\ \bibnamefont
			{Wilkins}}, \bibinfo {author} {\bibfnamefont {T.~R.}\ \bibnamefont
			{Forrest}}, \bibinfo {author} {\bibfnamefont {T.~A.~W.}\ \bibnamefont
			{Beale}}, \bibinfo {author} {\bibfnamefont {S.~R.}\ \bibnamefont {Bland}},
		\bibinfo {author} {\bibfnamefont {H.~C.}\ \bibnamefont {Walker}}, \bibinfo
		{author} {\bibfnamefont {D.}~\bibnamefont {Mannix}}, \bibinfo {author}
		{\bibfnamefont {F.}~\bibnamefont {Yakhou}}, \bibinfo {author} {\bibfnamefont
			{D.}~\bibnamefont {Prabhakaran}}, \bibinfo {author} {\bibfnamefont {A.~T.}\
			\bibnamefont {Boothroyd}}, \bibinfo {author} {\bibfnamefont {J.~P.}\
			\bibnamefont {Hill}}, \bibinfo {author} {\bibfnamefont {P.~D.}\ \bibnamefont
			{Hatton}},\ and\ \bibinfo {author} {\bibfnamefont {D.~F.}\ \bibnamefont
			{McMorrow}},\ }\bibfield  {title} {\bibinfo {title} {{Nature of the Magnetic
				Order and Origin of Induced Ferroelectricity in ${\mathrm{TbMnO}}_{3}$}},\
	}\href {https://doi.org/10.1103/PhysRevLett.103.207602} {\bibfield  {journal}
		{\bibinfo  {journal} {Phys. Rev. Lett.}\ }\textbf {\bibinfo {volume} {103}},\
		\bibinfo {pages} {207602} (\bibinfo {year} {2009})}\BibitemShut {NoStop}%
	\bibitem [{\citenamefont {Jang}\ \emph {et~al.}(2011)\citenamefont {Jang},
		\citenamefont {Lee}, \citenamefont {Ko}, \citenamefont {Noh}, \citenamefont
		{Koo}, \citenamefont {Kim}, \citenamefont {Lee}, \citenamefont {Park},
		\citenamefont {Zhang}, \citenamefont {Kim},\ and\ \citenamefont
		{Cheong}}]{Jang2011}%
	\BibitemOpen
	\bibfield  {author} {\bibinfo {author} {\bibfnamefont {H.}~\bibnamefont
			{Jang}}, \bibinfo {author} {\bibfnamefont {J.-S.}\ \bibnamefont {Lee}},
		\bibinfo {author} {\bibfnamefont {K.-T.}\ \bibnamefont {Ko}}, \bibinfo
		{author} {\bibfnamefont {W.-S.}\ \bibnamefont {Noh}}, \bibinfo {author}
		{\bibfnamefont {T.~Y.}\ \bibnamefont {Koo}}, \bibinfo {author} {\bibfnamefont
			{J.-Y.}\ \bibnamefont {Kim}}, \bibinfo {author} {\bibfnamefont {K.-B.}\
			\bibnamefont {Lee}}, \bibinfo {author} {\bibfnamefont {J.-H.}\ \bibnamefont
			{Park}}, \bibinfo {author} {\bibfnamefont {C.~L.}\ \bibnamefont {Zhang}},
		\bibinfo {author} {\bibfnamefont {S.~B.}\ \bibnamefont {Kim}},\ and\ \bibinfo
		{author} {\bibfnamefont {S.-W.}\ \bibnamefont {Cheong}},\ }\bibfield  {title}
	{\bibinfo {title} {Coupled magnetic cycloids in multiferroic
			${\mathrm{tbmno}}_{3}$ and
			${\mathrm{eu}}_{3/4}{\mathrm{y}}_{1/4}{\mathrm{mno}}_{3}$},\ }\href
	{https://doi.org/10.1103/PhysRevLett.106.047203} {\bibfield  {journal}
		{\bibinfo  {journal} {Phys. Rev. Lett.}\ }\textbf {\bibinfo {volume} {106}},\
		\bibinfo {pages} {047203} (\bibinfo {year} {2011})}\BibitemShut {NoStop}%
	\bibitem [{\citenamefont {Goto}\ \emph {et~al.}(2005)\citenamefont {Goto},
		\citenamefont {Yamasaki}, \citenamefont {Watanabe}, \citenamefont {Kimura},\
		and\ \citenamefont {Tokura}}]{Goto2005}%
	\BibitemOpen
	\bibfield  {author} {\bibinfo {author} {\bibfnamefont {T.}~\bibnamefont
			{Goto}}, \bibinfo {author} {\bibfnamefont {Y.}~\bibnamefont {Yamasaki}},
		\bibinfo {author} {\bibfnamefont {H.}~\bibnamefont {Watanabe}}, \bibinfo
		{author} {\bibfnamefont {T.}~\bibnamefont {Kimura}},\ and\ \bibinfo {author}
		{\bibfnamefont {Y.}~\bibnamefont {Tokura}},\ }\bibfield  {title} {\bibinfo
		{title} {Anticorrelation between ferromagnetism and ferroelectricity in
			perovskite manganites},\ }\href@noop {} {\bibfield  {journal} {\bibinfo
			{journal} {Phys. Rev. B}\ }\textbf {\bibinfo {volume} {72}},\ \bibinfo
		{pages} {220403(R)} (\bibinfo {year} {2005})}\BibitemShut {NoStop}%
	\bibitem [{\citenamefont {Baum}(2013)}]{diss-baum}%
	\BibitemOpen
	\bibfield  {author} {\bibinfo {author} {\bibfnamefont {M.}~\bibnamefont
			{Baum}},\ }\emph {\bibinfo {title} {Neutron-Scattering Studies on Chiral
			Multiferroics}},\ \href@noop {} {Ph.D. thesis},\ \bibinfo  {school}
	{Universit{\"a}t zu K{\"o}ln} (\bibinfo {year} {2013}),\ \bibinfo {note}
	{https://kups.ub.uni-koeln.de/5258/}\BibitemShut {NoStop}%
\end{thebibliography}
%

\end{document}